# Physics of muscle contraction

## M. Caruel

E-mail: `matthieu.caruel@u-pec.fr`
Université Paris Est, Modélisation et Simulation Multi Echelle, MSME CNRS–UMR 8208, 61 Avenue du Général de Gaulle, 94010 Créteil, France

## L. Truskinovsky

E-mail: `lev.truskinovsky@espci.fr`
Physique et Mécanique des Milieux Hétérogènes, CNRS - UMR 7636, ESPCI, PSL, 10 Rue Vauquelin, 75005 Paris, France

**Abstract.** In this paper we report, clarify and broaden various recent efforts to complement the chemistry-centered models of force generation in (skeletal) muscles by mechanics-centered models. The physical mechanisms of interest can be grouped into two classes: passive and active. The main passive effect is the fast force recovery which does not require the detachment of myosin cross-bridges from actin filaments and can operate without a specialized supply of metabolic fuel (ATP). In mechanical terms, it can be viewed as a collective folding-unfolding phenomenon in the system of interacting bi-stable units and modeled by near equilibrium Langevin dynamics. The parallel active force generation mechanism operates at slow time scales, requires detachment and is crucially dependent on ATP hydrolysis. The underlying mechanical processes take place far from equilibrium and are represented by stochastic models with broken time reversal symmetry implying non-potentiality, correlated noise or multiple reservoirs. The modeling approaches reviewed in this paper deal with both active and passive processes and support from the mechanical perspective the biological point of view that phenomena involved in slow (active) and fast (passive) force generation are tightly intertwined. They reveal, however, that biochemical studies in solution, macroscopic physiological measurements and structural analysis do not provide by themselves all the necessary insights into the functioning of the organized contractile system. In particular, the reviewed body of work emphasizes the important role of long-range interactions and criticality in securing the targeted mechanical response in the physiological regime of isometric contractions. The importance of the purely mechanical micro-scale modeling is accentuated at the end of the paper where we address the puzzling issue of the stability of muscle response on the so called "descending limb" of the isometric tetanus.

## Contents

## 1. Introduction

In recent years considerable attention has been focused on the study of the physical behavior of cells and tissues. Outside their direct physiological functionality, these biological systems are viewed as prototypes of new artificially produced materials that can actively generate stresses, adjust their rheology and accommodate loading through remodeling and growth. The intriguing mechanical properties of these systems can be linked to hierarchical structures, which bridge a broad range of scales, and to expressly nonlocal interactions which make these systems reminiscent more of structures and mechanisms than of a homogeneous matter. In contrast with traditional materials, where microscopic dynamics can be enslaved through homogenization and averaging, diverse scales in cells and tissues appear to be linked by complex energy cascades. To complicate matters further, in addition to external loading, cells and tissues are driven internally by endogenous mechanisms supplying energy and maintaining non-equilibrium. The multifaceted nature of the ensuing mechanical responses makes the task of constitutive modeling of such distributed systems rather challenging [1–15].

While general principles of active bio-mechanical response of cells and tissues still remain to be found, physical understanding of some specific sub-systems and regimes has been considerably improved in recent years. An example of a class of distributed biological systems whose functioning has been rather thoroughly characterized on both physiological and bio-chemical levels is provided by skeletal (striated) muscles [16–24]. The narrow functionality of skeletal muscles is behind their relatively simple, almost crystalline geometry which makes them a natural first choice for systematic physical modeling. The main challenge in the representation of the underlying microscopic machinery is to strike the right balance between chemistry and mechanics.

In this review, we address only a very small portion of the huge literature on force generation in muscles and mostly focus on recent efforts to complement the chemistry-centered models by the mechanics-centered models. Other perspectives on muscle contraction can be found in a number of comprehensive reviews [25–38].

The physical mechanisms of interest for our study can be grouped into two classes: passive and active. The passive phenomenon is the fast force recovery which does not require the detachment of myosin cross-bridges from actin filaments and can operate without a specialized supply of ATP. It can be viewed as a collective folding-unfolding in the system of interacting bi-stable units and modeled by near equilibrium Langevin dynamics. The active force generation mechanism operates at much slower time scales, requires detachment from actin and is fueled by continuous ATP hydrolysis. The underlying processes take place far from equilibrium and are represented by stochastic models with broken time reversal symmetry implying non-potentiality, correlated noise, multiple reservoirs and other non-equilibrium mechanisms.

The physical modeling approaches reviewed in this paper support the biochemical perspective that phenomena involved in slow (active) and fast (passive) force generation are tightly intertwined. They reveal, however, that biochemical studies of the isolated proteins in solution, macroscopic physiological measurements of muscle fiber energetics and structural studies using electron microscopy, X-ray diffraction and spectroscopic methods do not provide by themselves all the necessary insights into the functioning of the organized contractile system. The importance of the microscopic physical modeling that goes beyond chemical kinetics is accentuated by our discussion of the mechanical stability of muscle response on the descending limb of the isometric tetanus (segment of the tension-elongation curve with negative stiffness) [17–19; 39].

An important general theme of this review is the cooperative mechanical response of muscle machinery which defies thermal fluctuations. To generate substantial force, individual contractile elements must act collectively and the mechanism of synchronization has been actively debated in recent years. We show that the factor responsible for the cooperativity is the inherent non-locality of the system ensured by a network of cross-linked elastic backbones. The cooperation is amplified because of the possibility to actively tune the internal stiffness of the system towards a critical state where correlation length diverges. The reviewed body of work clarifies the role of non-locality and criticality in securing the targeted mechanical response of muscle type systems in various physiological regimes. It also reveals that the "unusual" features of muscle mechanics, that one can associate with the idea of allosteric regulation, are generic in biological systems [40–43] and several non-muscle examples of such behavior are discussed in the concluding section of the paper.

### 1.1. Background

We start with recalling few minimally necessary anatomical and biochemical facts about muscle contraction.

Skeletal muscles are composed of bundles of non ramified parallel fibers. Each fiber is a multi-nuclei cell, from 100 µm to 30 cm long and 10 µm to 100 µm wide. It spans the whole length of the tissue. The cytoplasm of each muscle cell contains hundreds of 2 µm wide myofibrils immersed in a network of transverse tubules whose role is to deliver the molecules that fuel the contraction. When activated by the central nervous system the fibers apply tensile stress to the constraints. The main goal of muscle mechanics is to understand the working of the force generating mechanism which operates at sub-myofibril scale.

The salient feature of the skeletal muscle myofibrils is the presence of striations, a succession of dark an light bands visible under transmission electron microscope [16]. The 2 µm regions between two Z-disks, identified as sarcomeres in Fig. 1, are the main contractile units. In a typical striated muscle, the density of sarcomeres in the cross-section is of the order of $5 \times 10^{14}$ m$^{-2}$ As we see in this figure, each half-sarcomere contains smaller structures called myofilaments.

The thin filaments, which are 8 nm wide and 1 µm long, are composed of polymerized actin monomers. Their helix structure has a periodicity of about 38 nm, with each monomer having a 5 nm diameter. The thick filaments contain about 300 myosin II molecules per half-sarcomere, which corresponds to the density of $150 \times 10^{12}$ mm$^{-3}$. Each myosin II is a complex protein with 2 globular heads whose tails are assembled in a helix [44]. The tails of different myosins are packed together and constitute the backbone of the thick filament from which the heads, known as cross-bridges, project outward toward the surrounding actin filaments. The cross-bridges are organized in a 3 stranded helix with a periodicity of 43.5 nm and the axial distance between two adjacent double heads of about 14.5 nm [45].

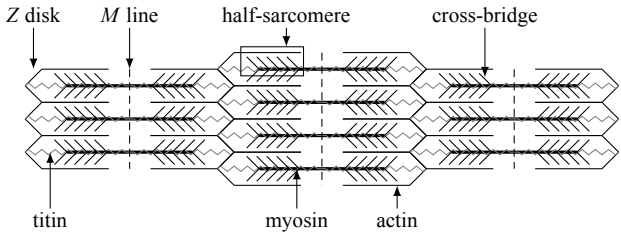

**Figure 1.** Schematic representation of a segment of myofibril showing the elementary force generating unit: the half-sarcomere. Z-disks are passive cross-linkers responsible for the crystalline structure of the muscle actin network; M-lines bundle myosin molecules into global active cross-linkers. Titin proteins connect the Z-disks inside each sarcomere.

Another important sarcomere protein, whose role in muscle contraction remains ambigous, is titin. This gigantic molecule is anchored on the Z-disks, spans the whole sarcomere structure and passively controls the overstretching; about its potentially active functions see Refs. [46–49].

A broadly accepted microscopic picture of muscle contraction was proposed by A.F Huxley and H.E. Huxley in the 1950's, see a historical review in Ref [50]. The development of electron myograph and X ray diffraction techniques at that time allowed the researcheres to observe the dynamics of the dark and light bands during fiber contraction [51–53]. The physical mechanism of force generation was first elucidated in [54], where contraction was explicitly linked to the relative sliding of the myofilaments and explained by a repeated, millisecond long attachement-pulling interaction between the thick and thin filaments; some conceptual alternatives are discussed in Refs. [55–57]

The sliding-filament hypothesis [53; 58] assumes that during contraction actin filaments move past myosin filaments while actively interacting with them through the myosin cross-bridges. Biochemical studies in solution showed that actomyosin interaction is powered by the hydrolysis of ATP into ADP and phosphate Pi [59]. The motor part of the myosin head acts as an enzyme which, on one side, increases the hydrolysis reaction rate and on the other side converts the released chemical energy into useful work. Each ATP molecule provides $100 \, zJ$ (zepto $= 10^{-21}$) which is equivalent to $\sim 25 \, k_B T$ at room temperature, where $k_b = 1.381 \times 10^{-23} \, J \, K^{-1}$ is the Boltzmann constant and $T$ is the absolute temperature in K. The whole system remains in permanent disequilibrium because the chemical potentials of the reactant (ATP) and the products of the hydrolysis reaction (ADP and Pi) are kept out of balance by a steadily operating exterior metabolic source of energy [16; 17; 60].

The stochastic interaction between individual myosin cross bridges and the adjacent actin filaments includes, in addition to cyclic attachment of myosin heads to actin binding sites, concurrent conformational change in the core of the myosin catalytic domain (of folding-unfolding type). A lever arm amplifies this structural transformation producing the power stroke, which is the crucial part of a mechanism allowing the attached cross bridges to generate macroscopic forces [16; 17].

A basic biochemical model of the myosin ATPase reaction in solution, linking together the attachment-detachment and the power stroke, is known as the Lymn–Taylor (LT) cycle [59]. It incorporates the most important chemical states, known as M-ATP, A-M-ADP-Pi, AM-ADP and AM, and associates them with particular mechanical configurations of the acto-myosin complex, see Fig. 2. The LT cycle consists of 4 steps [17; 35; 62; 63]:

(i) 1→2 Attachment. The myosin head (M) is initially detached from actin in a pre-power stroke configuration. ATP is in its hydrolyzed form ADP+Pi, which generates a high affinity to actin binding sites (A). The attachment takes place while the conformational mechanism is in pre-power stroke state.

(ii) 2→3 Power-stroke. Conformational change during which the myosin head executes a rotation around the binding site accompanied with a displacement increment of a few nm and a force generation of a few pN. During the power stroke, phosphate (Pi) is released.

(iii) 3→4 Detachment. Separation from actin filament occurs after the power stroke is completed while the myosin head remains in its post power stroke state. Detachment coincides with the release of the second hydrolysis product ADP which considerably destabilize the attached state. As the myosin head detaches, a fresh ATP molecule is recruited.

(iv) 4→1 Re-cocking (or repriming). ATP hydrolysis provides the energy necessary to recharge the power stroke mechanism.

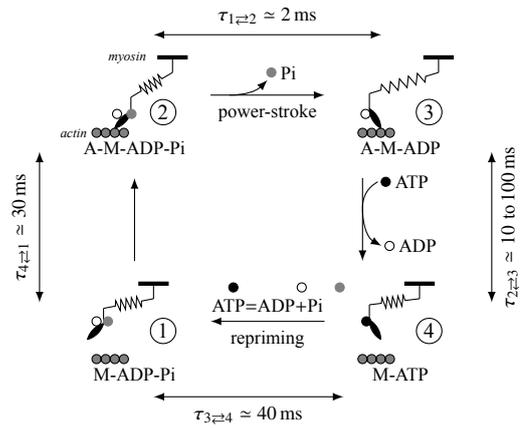

**Figure 2.** Representation of the Lymn–Taylor cycle, where each mechanical state (1 → 4) is associated with a chemical state (M-ADP-Pi, A-M-ADP-Pi, A-M-ADP and M-ATP). During one cycle, the myosin motor executes one power-stroke and splits one ATP molecule.

While this basic cycle has been complicated progressively to match an increasing body of experimental data [64–68], the minimal LT description is believed to be irreducible [69]. However, its association with microscopic structural details and relation to specific micro-mechanical interactions remain a subject of debate [70–72]. Another complication is that the influence of mechanical loading on the transition rates, that is practically impossible to simulate in experiments on isolated proteins, remains unconstrained by the purely biochemical models.

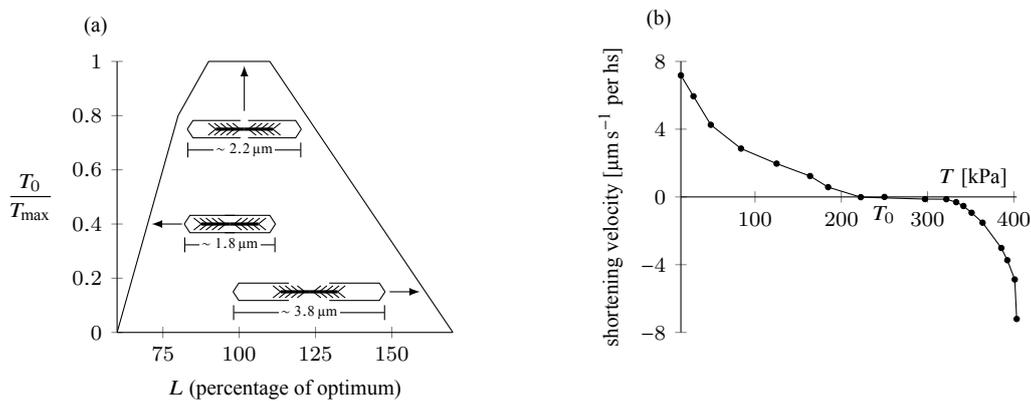

**Figure 3.** Isometric contraction (a) and isotonic shortening (b) experiments. (a) Isometric force $T_0$ as function of the sarcomere length linked to the amount of filament overlap. (b) Force-velocity relation obtained during isotonic shortening. Data in (b) are taken from Ref. [61].

An important feature of the LT cycle, which appears to be loading independent, is the association of vastly different timescales to individual biochemical steps, see Fig. 2. For instance, the power stroke, taking place at ∼1ms time scale, is the fastest step. It is believed to be independent of ATP activity which takes place at the orders of magnitude slower time scale, 30-100 ms [67; 73]. The rate limiting step of the whole cycle is the release of ADP with a characteristic time of ∼ 100 ms, which matches the rate of tension rise in an isometric tetanus.

### 1.2. Mechanical response

*1.2.1. Isometric force and isotonic shortening velocity.* Typical experimental setup for measuring the mechanical response of a muscle fibers involves a motor and a force transducer between which the muscle fiber is mounted. The fiber is maintained in an appropriate physiological solution and is electro stimulated. When the distance between the extremities of the fibers is kept constant (length clamp or hard device loading), the fully activated (tetanized) fiber generates an active force called the isometric tension $T_0$ which depends on the sarcomere length $L$ [77; 78].

The measured "tension-elongation" curve $T_0(L)$, shown in Fig. 3(a), reflects the degree of filament overlap in each half sarcomere. At small sarcomere lengths ($L \sim 1.8$ μm), the isometric tension level increases linearly as the detrimental overlap (frustration) diminishes. Around $L = 2.1$ μm, the tension reaches a plateau $T_{\max}$, the physiological regime, where all available myosin cross-bridges have a possibility to bind actin filament. The descending limb corresponds to regimes where the optimal filament overlap progressively reduces (see more about this regime in Section 5).

One of the main experiments addressing the mechanical behavior of skeletal muscles under applied forces (load clamp or soft loading device) was conducted by A.V. Hill [79], who introduced the notion of "force-velocity" relation. First the muscle fiber was stimulated under isometric conditions producing a force $T_0$. Then the control device was switched to the load clamp mode and a load step was applied to the fiber which shortened (or elongated) in response to the new force level. After a transient [80] the system reached a steady state where the shortening velocity could be measured. A different protocol producing essentially the same result was used in Ref. [81] where a ramp shortening (or stretch) was applied to a fiber in length clamp mode and the tension measured at a particular stage of the time response. Note that in contrast to the case of passive friction, the active force-velocity relation for tetanized muscle enters the quadrant where the dissipation is negative, see Fig. 3(b).

*1.2.2. Fast isometric and isotonic transients.* The mechanical responses characterized by the tension-elongation relation and the force-velocity relation are associated with timescales of the order of 100 ms. To shed light on the processes at the millisecond time scale, fast load clamp experiments were performed in Refs. [82–84]. Length clamp experiments were first conducted in Ref. [74], where a single fiber was mounted between a force transducer and a loudspeaker motor able to deliver length steps completed in 100 μs. More specifically, after the isometric tension was reached, a length step $\delta L$ (measured in nanometer per half sarcomere, nm hs$^{-1}$) was applied to the fiber, with a feedback from a striation follower device that allowed to control the step size per sarcomere, see Fig. 4(a). Such experimental protocols have since become standard in the field [76; 85–88].

The observed response could be decomposed into 4 phases:

$(0 \to 1)$ from 0 to about 100 μs (phase 1). The tension (respectively sarcomere length) is altered simultaneously with the length step (respectively force step) and reaches a level $T_1$ (respectively $L_1$) at the end of the step. The values $T_1$ and $L_1$ depend linearly on the loading (see Fig. 5, circles), and characterize the instant elastic response of the fiber. Various $T_1$ and $L_1$ measurements in different conditions allow one to link the instantaneous elasticity with different structural elements of the sarcomere, in particular to isolate the elasticity of the cross bridges from the elasticity of passive structures such as the myofilaments [89–91].

$(1 \to 2)$ from about 100 μs to about 3 ms (phase 2). In length clamp experiments, the tension is quickly recovered up to a plateau level $T_2$ close but below the original level $T_0$; see Fig. 4(a) and open squares in Fig. 5. Such quick recovery is too fast to engage the attachment-detachment processes and can be explained by the synchronized power stroke involving the attached heads [74]. For small step amplitudes $\delta L$, the tension $T_2$ is practically equal to the original tension $T_0$, see the plateau on the $T_2$ vs. elongation relation in Fig. 5. In load clamp experiment, the fiber shortens or elongates towards the level $L_2$, see filled squares in Fig. 5. Note that on Fig. 5, the measured $L_2$ points overlap with the $T_2$ points except that the plateau appears to be missing. In load clamp the value of $L_2$ at loads close to $T_0$ has been difficult to measure because of the presence of oscillations [92]. At larger steps, the tension $T_2$ start to depend linearly on the length step because the power stroke capacity of the attached heads has been saturated.

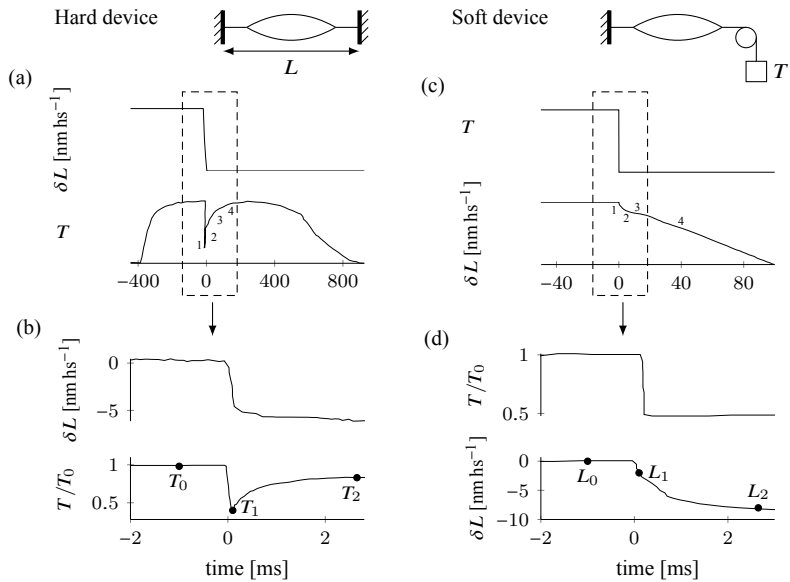

**Figure 4.** Fast transients in mechanical experiments on single muscle fibers in length clamp [hard device, (a) and (b)]; and in force clamp [soft device, (c) and (d)]. Typical experimental responses are shown separately on a slow timescale [(a) and (c)] and on a fast time scale [(b) and (d)]. In (a) and (c) the numbers indicate the distinctive steps of the transient responses: the elastic response (1), the processes associated with passive power stroke (2) and the ATP driven approach to steady state (3-4). Data are adopted from Refs. [74–76].

($2 \to 3 \to 4$) In force clamp transients after $\sim 3$ ms the tension rises slowly from the plateau to its original value $T_0$, see Fig. 4(a). This phase corresponds to the cyclic attachment and detachment of the heads see Fig. 2, which starts with the detachment of the heads that where initially attached in isometric conditions (phase 3). In load clamp transients phase 4 is clearly identified by a shortening at a constant velocity, see Fig. 4(c), which, being plotted against the force, reproduces the Hill's force-velocity relation, see Fig. 3(b).

First attempts to rationalize the fast stages of these experiments [74] have led to the insight that we deal here with mechanical snap-springs performing a transition between two configurations. The role of the external loading reduces to biasing mechanically one of the two states. The idea of bistability in the structure of myosin heads has been later fully supported by crystallographic studies [96–98].

Based on the experimental results shown in Fig. 5 one may come to a conclusion that the transient responses of muscle fibers to fast loading in hard (length clamp) and soft (load clamp) devices are identical. However, a careful analysis of Fig. 5 shows that the data for the load clamp protocol are missing in the area adjacent to the state of isometric contractions (around $T_0$). Moreover, the two protocols are clearly characterized by different kinetics.

Recall that the rate of fast force recovery can be interpreted as the inverse of the time scale separating the end of phase 1 and the end of phase 2. The experimental results obtained in soft and hard device can be compared if we present the recovery rate as a function of the final elongation of the system. In this way, one can compare kinetics in the two ensembles using the same initial and final states; see dashed lines in Fig. 5. A detailed quantitative comparison, shown in Fig. 6, reveals considerably slower response when the system follows the soft device protocol (filled symbols). The dependence of the relaxation rate on the type of loading was first noticed in Ref. [99] and then confirmed by the direct measurements in Ref. [100]. These discrepancies will be addressed in Section 2.

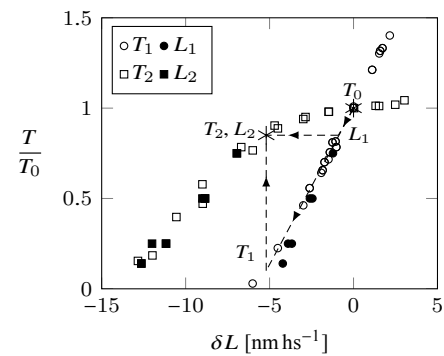

**Figure 5.** Tension-elongation relation reflecting the state of the system at the end of phase 1 (circles) and phase 2 (squares) in both length clamp (open symbols) and force clamp (filled symbols). Data are taken from Refs. [80; 85; 87; 93–95].

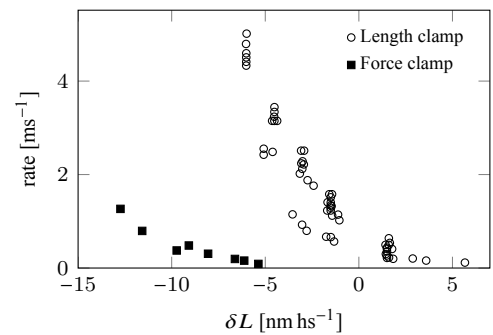

**Figure 6.** Drastically different kinetics in phase 2 of the fast load recovery in length clamp (circles) and force clamp (squares) experiments. Data are from Refs. [74; 80; 85–87; 93].

We complement this brief overview of the experimental results with an observation that a seemingly natural, purely passive interpretation of the power stroke is in apparent disagreement with the fact that the power stroke is an active force generating step in the Lymn–Taylor cross bridge cycle. The challenge of resolving this paradox served as a motivation for several theoretical developments reviewed in this paper.

### 1.3. Modeling approaches

*1.3.1. Chemomechanical models.* The idea to combine mechanics and chemistry in the modeling of muscle contraction was proposed by A.F. Huxley [54]. The original model was focused exclusively on the attachment-detachment process and the events related to the slow time scale (hundreds of milliseconds). The attachment-detachment process was interpreted as an out-of-equilibrium reaction biased by a drift with a given velocity [67; 73]. The generated force was linked to the occupancy of continuously distributed chemical states and the attempt was made to justify the observed force-velocity relations [see Fig. 3(b)] using appropriately chosen kinetic constants. This approach was brought to full generality by T.L. Hill and collaborators [101–105]. More recently, the chemo-mechanical modelling was extended to account for energetics, to include the power-stroke activity and to study the influence of collective effects [67; 86; 106–114].

In the general chemo-mechanical approach muscle contraction is perceived as a set of reactions among a variety of chemical states [67; 68; 86; 115; 116]. The mechanical feedback is achieved through the dependence of the kinetic constants on the total force exerted by the system on the loading device. The chemical states form a network which describes on one side, various stages of the enzymatic reaction, and on the other side, different mechanical configurations of the system. While some of crystallographic states have been successfully identified with particular sites of the chemical network (attached and detached [54], strongly and weakly attached [67], pre and post power stroke [74], associated with the first or second myosin head [94], etc.), the chemo-mechanical models remain largely phenomenological as the functions characterizing the dependence of the rate constants on the state of the force generating springs are typically chosen to match the observations instead of being derived from a microscopic model.

In other words, due to the presence of mechanical elements, the standard discrete chemical states are replaced by continuously parameterized configurational "manifolds". Even after the local conditions of detailed balance are fulfilled, this leads to the functional freedom in assigning the transition rates. This freedom originates from the lack of information about the actual energy barriers separating individual chemical states and the uncertainty was used as a tool to fit experimental data. This has led to the development of a comprehensive phenomenological description of muscle contraction that is almost fully compatible with available measurements, see, for instance, Ref. [68] and the references therein. The use of phenomenological expressions, however, gives only limited insight into the micro-mechanical functioning of the force generating mechanism, leaves some lagoons in the understanding, as in the case of ensemble dependent kinetics, and ultimately has a restricted predictive power.

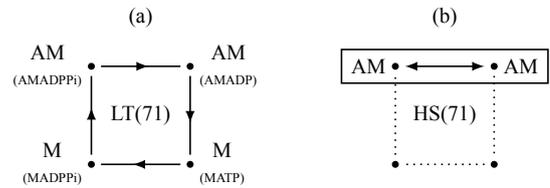

**Figure 7.** Biochemical vs purely mechanistic description of the power stroke in skeletal muscles: (a) The Lymn–Taylor four-state cycle, LT(71) and (b) the Huxley-Simmons two-state cycle, HS (71). Adapted from Ref. [121].

*1.3.2. Power-stroke modes* To model fast force recovery A.F. Huxley and R.M. Simmons (HS) [74] proposed to describe it as a chemical reaction between the folded and unfolded configurations of the attached cross-bridges with the reaction rates linked to the structure of the underlying energy landscape. Almost identical descriptions of mechanically driven conformational changes were proposed, apparently independently, in the studies of cell adhesion [117; 118], and in the context of hair cell gating [119; 120]. For all these systems the HS model can be viewed as a fundamental mean-field prototype [121].

While the scenario proposed by HS is in agreement with the fact that the power stroke is the fastest step in the Lymn–Taylor (LT) enzymatic cycle [16; 59], there remained a formal disagreement with the existing biochemical picture, see Fig. 7. Thus, HS assumed that the mechanism of the fast force recovery is passive and can be reduced to a mechanically induced conformational change. In contrast, the LT cycle for actomyosin complexes is based on the assumption that the power stroke can be reversed only actively through the completion of the biochemical pathway including ADP release, myosin unbinding, binding of uncleaved ATP, splitting of ATP into ADP and $P_i$, and then rebinding of myosin to actin [59; 68], see Fig. 2. While HS postulated that the power stroke can be reversed by mechanical means, most of the biochemical literature is based on the assumption that the power-stroke recocking cannot be accomplished without the presence of ATP. In particular, physiological fluctuations involving power stroke are almost exclusively interpreted in the context of *active* behavior [122–128]. Instead the purely mechanistic approach of HS, presuming that the power-stroke-related leg of the LT cycle can be decoupled from the rest of the biochemical pathway, was pursued in Refs [116; 129], but did not manage to reach the mainstream.

*1.3.3. Brownian ratchet models.* In contrast to chemo-mechanical models, the early theory of Brownian motors followed largely a mechanically explicit path [130–138]. In this approach, the motion of myosin II was represented by a biased diffusion of a particle on a periodic asymmetric landscape driven by a colored noise. The white component of the noise reflects the presence of a heat reservoir while the correlated component mimics the non-equilibrium chemical environment. Later, such purely mechanical approach was parallelled by the development of the equivalent chemistry-centered discrete models of Brownian ratchets, see for instance, Refs. [38; 139–142].

First direct applications of the Brownian ratchet models to muscle contraction can be found in Refs. [143–145], where the focus was on the attachment-detachment process at the expense of the phenomena at the short time scales (power stroke). In other words, the early models had a tendency to collapse the four state Lymn–Taylor cycle onto a two states cycle by absorbing the configurational changes associated with the transitions M-ATP $\to$ M-ADP-Pi and A-M-ADP-Pi $\to$ A-M-ADP into more general transitions M-ATP $\to$ AM-ADP and AM-ADP $\to$ M-ATP. Following Ref. [54], the complexity of the structure of the myosin head was reduced to a single degree of freedom representing a stretch of a series elastic spring. This simplification offered considerable analytical transparency and opened the way towards the study of stochastic thermodynamics and efficiency of motor systems, e.g. Refs. [140; 146; 147].

Later, considerable efforts were dedicated to the development of synthetic descriptions, containing both ratchet and power stroke elements [112; 113; 143; 144; 148–150]. In particular, numerous attempts have been made to unify the attachement-detachment-centered models with the power stroke-centered ones in a generalized chemo-mechanical framework [60; 67; 68; 86; 87; 105; 114; 116; 144; 151–154]. The ensuing models have reached the level of sophistication allowing their authors to deal with collective effects, including the analysis of traveling waves and coherent oscillations [60; 110; 114; 143; 155–159]. In particular, myosin-myosin coupling was studied in models of interacting motors [113; 152] and emergent phenomena characterized by large scale entrainment signatures were identified in Refs. [36; 110; 114; 122; 123; 148].

The importance of these discoveries is corroborated by the fact that macroscopic fluctuations in the groups of myosins have been also observed experimentally. In particular, considerable coordination between individual elements was detected in close to stall conditions giving rise to synchronized oscillations which could be measured even at the scale of the whole myofibril [26; 82; 92; 149; 160–162]. The synchronization also revealed itself through macro-scale spatial inhomogeneities reported near stall force condition [163–166].

In ratchet models the cooperative behavior was explained without direct reference to the power stroke by the fact that the mechanical state of one motor influences the kinetics of other motors. The long-range elastic interactions were linked to the presence of filamental backbones which are known to be elastically compliant [167; 168]. The fact, that similar cooperative behavior of myosin cross-bridges has been also detected experimentally at short time scales, during fast force recovery [92], suggests that at least some level of synchronization should be already within reach of the power-stroke-centered models disregarding motor activity and focusing exclusively on passive mechanical behavior. Elucidating the mechanism of such passive synchronization will be one of our main goals of Section 2.

*1.4. Organization of the paper*

In this review, we focus exclusively on models emphasizing the mechanical side of the force generation processes. The mechanical models affirm that in some situations the micro-scale stochastic dynamics of the force generating units can be adequately represented by chemical reactions. However, they also point to cases when one ends up unnecessarily constrained by the chemo-mechanical point of view.

The physical theories, emphasized in this review, are in tune with the approach pioneered by Huxley and Simmons in their study of fast force recovery and with the general approach of the theory of molecular motors. The elementary contractile mechanisms are modeled by systems of stochastic differential equations describing random walk in complex energy landscapes. These landscapes serve as a representation of both the structure and the interactions in the system, in particular, they embody various local and nonlocal mechanical feedbacks.

In contrast to fully microscopic molecular dynamical reconstructions of multi-particle dynamics, the reviewed mechano-centered models operate with few collective degrees of freedom. The loading is transmitted directly by applied forces while different types of noises serve as a representation of non-mechanical external driving mechanisms that contain both equilibrium and non-equilibrium components. Due to the inherent stochasticity of such mesoscopic systems [140], the emphasis is shifted from the averaged behavior, favored by chemo-mechanical approaches, to the study of the full probability distributions.

In Section 2 we show that even in the absence of metabolic fuel, long-range interactions, communicated by passive cross-linkers, can ensure a highly nontrivial cooperative behavior of interacting muscle cross-bridges. This implies ensemble dependence, metastability and criticality which all serve to warrant efficient collective stroke in the presence of thermal fluctuations. We argue that in the near critical regimes the barriers are not high enough for the Kramers approximation to be valid [169; 170] which challenges chemistry-centered approaches. Another important contribution of the physical theory is in the emphasis on fluctuations as an important source of structural information. A particularly interesting conclusion of this section is the realization that a particular number of cross-bridges in realistic half-sarcomeres may be a signature of an (evolutionary) fine tuning of the mechanical response to criticality.

In Section 3 we address the effects of correlated noise on force generation in isometric conditions. We focus on the possibility of the emergence of new noise-induced energy wells and stabilization of the states that are unstable in strictly equilibrium conditions. The implied transition from negative to positive rigidity can be linked to time correlations in the out-of-equilibrium driving and the reviewed work shows that subtle differences in the active noise may compromise the emergence of such "non-equilibrium" free energy wells. These results suggest that ATP hydrolysis may be involved in tuning the muscle system to near-criticality which appears to be a plausible description of the physiological state of isometric contraction.

In Section 4 we introduce mechanical models bringing together the attachment-detachment and the power stroke. To make a clear distinction between these models and the conventional models of Brownian ratchets we operate in a framework when the actin track is nonpolar and the bistable element is unbiased. The symmetry breaking is achieved exclusively through the coupling of the two subsystems. Quite remarkably, a simple mechanical model of this type formulated in terms of continuous Langevin dynamics can reproduce all four discrete states of the minimal LT cycle. In particular, it demonstrates that contraction can be propelled directly through a conformational change, which implies that the power stroke may serve as the leading mechanism not only at short but also at long time scales.

Finally, in Section 5 we address the behavior of the contractile system on the descending limb of the isometric tetanus, a segment of the force length relation with a negative stiffness. Despite potential mechanical instability, the isometric tetanus in these regimes is usually associated with a quasi-affine deformation. The mechanics-centered approach allows one to interpret these results in terms of energy landscape whose ruggedness is responsible for the experimentally observed history dependence and hysteresis near the descending limb. In this approach both the ground states and the marginally stable states emerge as fine mixtures of short and long half-sarcomeres and the negative overall slope of the tetanus is shown to coexists with a positive instantaneous stiffness. A version of the mechanical model, accounting for surrounding tissues, produces an intriguing prediction that the energetically optimal variation of the degree of nonuniformity with stretch must exhibits a devil's staircase-type behavior.

The review part ends with Section 7 where we go over some non-muscle applications of the proposed mechanical models In this Section 7 we formulate conclusions and discuss directions of future research.

## 2. Passive force generation

In this Section, we limit ourselves to models of passive force generation.

First of all we need to identify an elementary unit whose force producing function is irreducible. The second issue concerns the structure of the interactions between such units. The goal here is to determine whether the consideration of an isolated force-producing element is meaningful in view of the presence of various feedback loops. The pertinence of this question is corroborated by the presence of hierarchies that undermine the independence of individual units.

The schematic topological structure of the force generating network in skeletal muscles is shown in Fig. 8. Here we see that behind the apparent series architecture that one can expect to dominate in crystals, there is a system of intricate parallel connections accomplished by passive cross-linkers. Such elastic elements play the role of backbones linking elements at smaller scales. The emerging hierarchy is dominated by long-range interactions which make the "muscle crystal" rather different from the conventional inert solids.

The analysis of Fig. 8 suggests that the simplest nontrivial structural element of the network is a half-sarcomere that can be represented as a bundle of finite number of cross-bridges. The analysis presented below shows that such model cannot be simplified further because for instance the mechanical response of individual cross-bridges is not compatible by itself with observations.

The minimal model of this type was proposed by Huxley and Simmons (HS) who described myosin cross-bridges as hard spin elements connected to linear springs loaded in parallel [74]. In this Section, we show that the stochastic version of the HS model is capable of reproducing qualitatively the mechanical response of a muscle submitted to fast external loading in both length clamp (hard device) and force clamp (soft device) settings (see Fig. 6). We also address the question whether the simplest series connection of HS elements is compatible with the idea of an affine response of the whole muscle fiber.

Needless to say that the oversimplified model of HS does not address the full topological complexity of the cross-bridge organization presented in Fig. 8. Furthermore, the 3D steric effects that appear to be crucially important for the description of spontaneous oscillatory contractions [148; 162; 164; 166; 171–173], and the effects of regulatory proteins responsible for steric blocking [174–178], are completely outside the HS framework.

### 2.1. Hard spin model

Consider now in detail the minimal model [74; 99; 121; 179; 180] which interprets the pre- and post-power-stroke conformations of the myosin heads as discrete (chemical) states. Since these states can be viewed as two configurations of a "digital" switch such model belongs to the hard spin category.

The potential energy of an individual spin unit can be written in the form

$$u_{\text{HS}}(x) = \begin{cases} v_0 & \text{if } x = 0, \\ 0 & \text{if } x = -a. \end{cases} \quad (2.1)$$

where the variable $x$ takes two values, $0$ and $-a$, describing the unfolded and the folded conformations, respectively. By $a$ we denoted the "reference" size of the conformational change interpreted as the distance between the two energy wells. With the unfolded state we associate an energy level $v_0$ while the folded configuration is considered as a zero energy state, see Fig. 9(a). In addition to a spin unit with energy (2.1) we assume that each cross-bridge contains a linear spring with stiffness $\kappa_0$ in series with the bi-stable unit; see Fig. 9(b).

The attached cross-bridges are connecting myosin and actin filaments which play the role of elastic backbones. Their function is to provide mechanical feedback and coordinate the mechanical state of the individual cross-bridges [167; 168]. There is evidence [89; 95] that a lump description of the combined elasticity of actin and myosin filaments by a single spring is rather adequate, see also Refs. [89; 100; 181–183]). Hence we represent a generic half sarcomere as a cluster of $N$ parallel HS elements and assume that this parallel bundle is connected in series to a linear spring of stiffness $kb$.

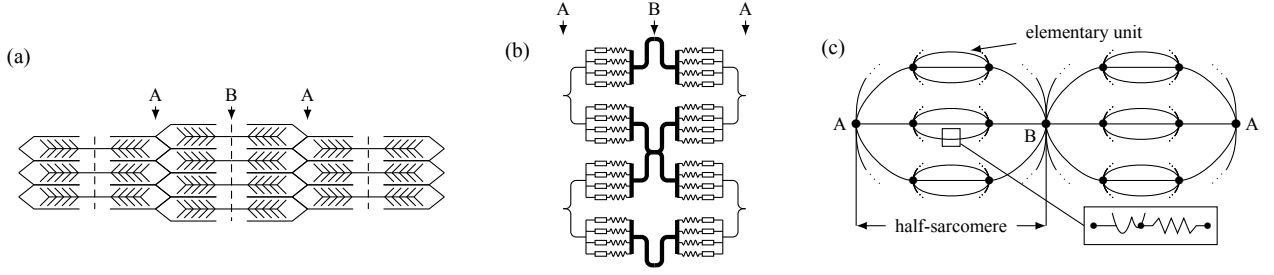

**Figure 8.** Structure of a myofibril. (a) Anatomic organization of half sarcomeres linked by $Z$ disks (A) and $M$ lines (B). (b) Schematic representation of the network of half sarcomeres; (c) Topological structure of the same network emphasizing the dominance of long-range interactions.

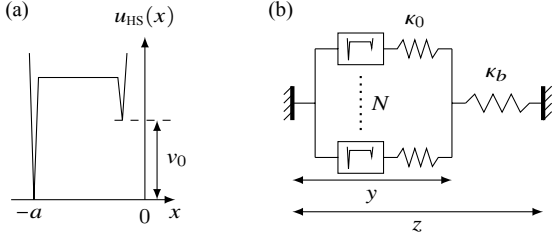

**Figure 9.** Hard spin model of a parallel bundle of bistable cross-bridges connected to a common elastic backbone. (a) Energy landscape of an individual power-stroke element; (b) $N$ cross-bridges connected to an elastic backbone with stiffness $\kappa_b$. Adapted from Ref. [99].

We chose $a$ as the characteristic length of the system, $\kappa_0 a$ as the characteristic force, and $\kappa_0 a^2$ as the characteristic energy. The resulting dimensionless energy of the whole system (per cross bridge) at fixed total elongation $z$ takes the form

$$v(\mathbf{x}; z) = \frac{1}{N} \sum_{i=1}^{N} \left[ (1 + x_i) v_0 + \frac{1}{2}(y - x_i)^2 + \frac{\lambda_b}{2}(z - y)^2 \right], \quad (2.2)$$

where $\lambda_b = \kappa_b/(N\kappa_0)$, $y$ represents the elongation of the cluster of parallel cross bridges and $x_i = \{0, -1\}$, see Fig. 9(b). Here, for simplicity, we did not modify the notations as we switched to non-dimensional quantities.

It is important to note that here we intentionally depart from the notations introduced in Section 1.2. For instance, the length of the half sarcomere was there denoted by $L$, which is now $z$. Furthermore, the tension which was previously $T$ will be now denoted by $\sigma$ while we keep the notation $T$ for the ambient temperature.

*2.1.1. Soft and hard devices.* It is instructive to consider first the two limit cases, $\lambda_b = \infty$ and $\lambda_b = 0$.

*Zero temperature behavior.* If $\lambda_b = \infty$, the backbone is infinitely rigid and the array of cross-bridges is loaded in a hard device with $y$ being the control parameter. Due to the permutational invariance of the energy

$$v(\mathbf{x}; y) = \frac{1}{N} \sum_{i=1}^{N} \left[ (1 + x_i) v_0 + \frac{1}{2}(y - x_i)^2 \right], \quad (2.3)$$

each equilibrium state is fully characterized by a discrete order parameter representing the fraction of cross-bridges in the folded (post power stroke) state

$$p = -\frac{1}{N} \sum_{i=1}^{N} x_i.$$

At zero temperature all equilibrium configurations with a given $p$ correspond to local minima of the energy (2.3), see Ref. [179]. These metastable states can be viewed as simple mixtures the two states, one fully folded with $p = 1$, and the energy $(1/2)(y+1)^2$, and the other one fully unfolded with $p = 0$, and the energy $(1/2)y^2 + v_0$. The energy of the mixture reads

$$\hat{v}(p; y) = p\frac{1}{2}(y+1)^2 + (1-p)\left[\frac{1}{2}y^2 + v_0\right]. \quad (2.4)$$

The absence of a *mixing* energy is a manifestation of the fact that the two populations of cross-bridges do not interact.

The energies of the metastable states parameterized by $p$ are shown in Fig. 10 (c–e). Introducing the reference elongation $y_0 = v_0 - 1/2$, one can show that the global minimum of the energy corresponds either to folded state with $p = 1$, or to unfolded state with $p = 0$. At the transition point $y = y_0$, all metastable states have the same energy, which means that the global switching can be performed at zero energy cost, see Fig. 10(d).

The tension-elongation relations along metastable branches parameterized by $p$ can be presented as $\hat{\sigma}(p; z) = \frac{\partial}{\partial z}\hat{v}(p; y) = y + p$, where $\sigma$ denotes the tension (per cross-bridge). At fixed $p$, we obtain equidistant parallel lines, see Fig. 10 [(a) and (b)]. At the crossing (folding) point $y = y_0$, the system following the global minimum exhibits a singular negative stiffness. Artificial metamaterial showing negative stiffness has been recently engineered by drawing on the Braess paradox for decentralized globally connected networks [13; 184; 185]. Biological examples of systems with non-convex energy and negative stiffness are provided by RNA and DNA hairpins and hair bundles in auditory cells [120; 186–188].

In the other limit $\lambda_b \to 0$, the backbone becomes infinitely soft ($z - y \to \infty$) and if $\lambda_b(z - y) \to \sigma$ the system behaves as if it was loaded in a soft device, where now the tension $\sigma$ is the control parameter. The relevant energy can be written in the form

$$w(\mathbf{x}, y; \sigma) = v(\mathbf{x}, y) - \sigma z$$
$$= \frac{1}{N} \sum_{i=1}^{N} \left[ (1 + x_i) v_0 + \frac{1}{2}(y - x_i)^2 - \sigma y \right], \quad (2.5)$$

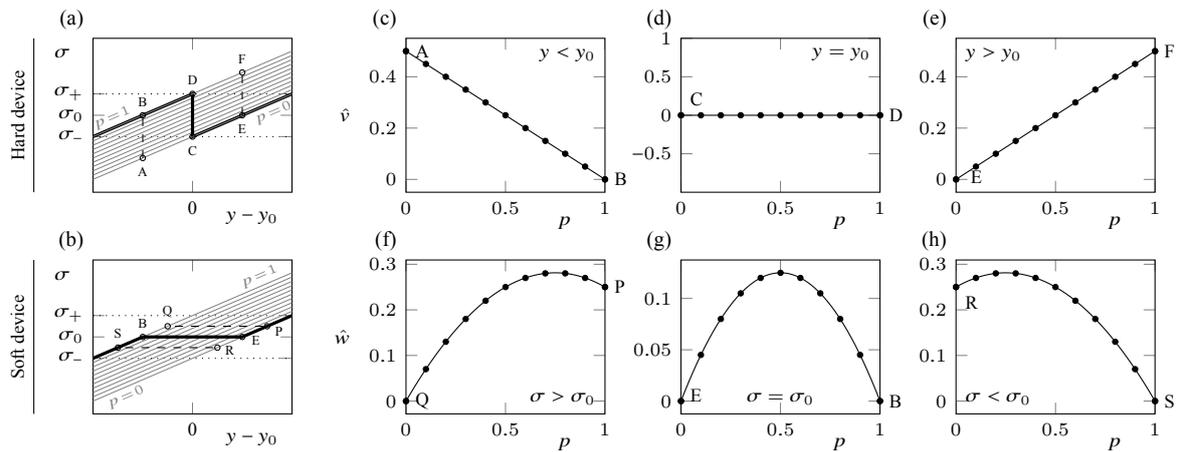

**Figure 10.** Behavior of a HS model with $N = 10$ at zero temperature. [(a) and (b)] Tension-elongation relations corresponding to the metastable states (gray) and along the global minimum path (thick lines), in hard (a) and soft (b) devices. (c–e) [respectively (f–h)] Energy levels of the metastable states corresponding to $p = 0, 0.1, \ldots, 1$, at different elongations $y$ (respectively tensions $\sigma$). Corresponding transitions (E→B, P→Q, ...) are shown in (a) and (b). Adapted from Ref. [179].

The order parameter $p$ parametrizes again the branches of local minimizers of the energy (2.5), see Ref. [179]. At a given value of $p$, the energy of a metastable state reads

$$\hat{w}(p;\sigma) = -\frac{1}{2}\sigma^2 + p\sigma + \frac{1}{2}p(1-p) + (1-p)v_0. \quad (2.6)$$

In contrast to the case of a hard device [see Eq. (2.4)], here there is a nontrivial coupling term $p(1-p)$ describing the energy of a regular solution. The presence of this term is a signature of a mean-field interaction among individual cross-bridges.

The tension-elongation relations describing the set of metastable states can be now written in the form $\hat{z}(p;\sigma) = -\frac{\partial}{\partial \sigma}\hat{w}(p;\sigma) = \sigma - p$. The global minimum of the energy is again attained either at $p = 1$ or $p = 0$, with a sharp transition at $\sigma = \sigma_0 = v_0$, which leads to a plateau on the tension-elongation curve, see Fig. 10 (b). Note that even in the continuum limit the stable "material" response of this system in hard and soft devices differ and this ensemble nonequivalence is a manifestation of the presence of long-range interactions. To illustrate this point further, we consider the energetic cost of mixing in the two loading devices at the conditions of the switch between pure states, see Fig. 10 [(d) and (g)]. In the hard device [see (d)] the energy dependence on $p$ in this state is flat suggesting that there is no barrier, while in the soft device [see (g)] the energy is concave which means that there is a barrier.

To develop intuition about the observed inequivalence, it is instructive to take a closer look at the minimal system with $N = 2$, see Fig. 11. Here for simplicity we assumed that $v_0 = 0$ implying $\sigma_0 = 0$ and $y_0 = -1/2$. The two pure configurations are labeled as $A$ ($p = 0$) and $C$ ($p = 1$) at $\sigma = \sigma_0$ and as $D$ ($p = 0$) and $B$ ($p = 1$) at $y = y_0$. In a hard device, where the two elements do not interact, the transition from state $D$ to state $B$ at a given $y = y_0$ goes through the configuration $B+D$, which has the same energy as configurations $D$ and $B$: the cross-bridges in folded and unfolded states are geometrically compatible and their mixing requires no additional energy. Instead, in a soft device, where individual elements interact, a transition from state $A$ to state $C$ taking place at a given $\sigma = 0$ requires passing through the transition state $A+C$ which has a nonzero pre-stress. Pure states in this mixture state have different values of $y$, and therefore the energy of the mixed configuration $A + C$, which is stressed, is larger than the energies of the pure unstressed states $A$ and $C$. We also observe that in a soft device the transition between the pure states is cooperative requiring essential interaction of individual elements while in a hard device it takes place independently in each element.

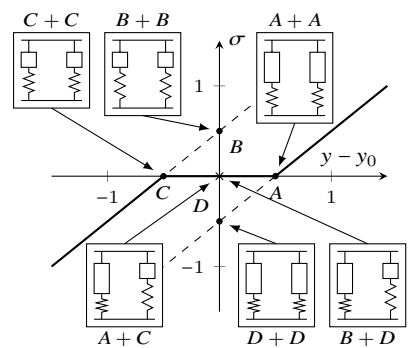

**Figure 11.** Behavior of two cross-bridges. Thick line: global minimum in a soft device ($\sigma_0 = 0$). Dashed lines, metastable states $p = 0$, and $p = 1$. The intermediate stress-free configuration is obtained either by mixing the two geometrically compatible states $B$ and $D$ in a hard device, which results in a $B + D$ structure without additional internal stress, or by mixing the two geometrically incompatible states $A$ and $C$ in a soft device, which results in a $A + C$ structure with internal residual stress. Adapted from Ref. [179].

*Finite temperature behavior.* We now turn to finite temperature to check the robustness of the observations made in the previous section.

Consider first the hard device case $\lambda_b = \infty$, considered chemo-mechanically in the seminal paper of HS [74], see Ref. [121] for the statistical interpretation. With the variable $y$ serving now as the control parameter, the equilibrium probability density for a micro-state $x$ with $N$ elements takes the form $\rho(x; y, \beta) = Z(y,\beta)^{-1} \exp[-\beta v(x;y)]$ where $\beta = (\kappa_0 a^2)/(k_B T)$ with $T$ being the absolute temperature and $k_B$ the Boltzmann constant. The partition function is $Z(y, \beta) = \sum_{\boldsymbol{x} \in \{0,-1\}^N} \exp\left[-\beta N v(\boldsymbol{x}; y)\right] = [Z_1(y, \beta)]^N$. Here $Z_1$ represents the partition function of a single element given by

$$Z_1(y, \beta) = \exp\left[-\frac{\beta}{2}(y+1)^2\right] + \exp\left[-\beta\left(\frac{y^2}{2} + v_0\right)\right]. \quad (2.7)$$

Therefore one can write $\rho(\boldsymbol{x}; y, \beta) = \prod_{i=1}^{N} \rho_1(x_i; y, \beta)$, where we have introduced the equilibrium probability distribution for a single element

$$\rho_1(x; y, \beta) = Z_1(y, \beta)^{-1} \exp\left[-\beta v(x; y)\right], \quad (2.8)$$

with $v(x; y)$—the energy of a single element.

The lack of cooperativity in this case is clear if one considers the marginal probability density at fixed $p$

$$\rho(p; y, \beta) = \binom{N}{Np} [\rho_1(-1; y, \beta)]^{Np} [\rho_1(0; y, \beta)]^{N(1-p)}$$
$$= Z(y, \beta)^{-1} \exp[-\beta N f(p; y, \beta)],$$

where $f(p; y, \beta) = \hat{v}(y, p) - (1/\beta) s(p)$ is the marginal free energy, $\hat{v}$ is given by Eq. (2.4) and $s(p) = \frac{1}{N} \log \binom{N}{Np}$, is the ideal entropy, see Fig. 12. In the thermodynamic limit $N \to \infty$ we obtain explicit expression $f_\infty(p; y, \beta) = \hat{v}(p; y) - (1/\beta) s_\infty(p)$, where $s_\infty(p) = -[p \log(p) + (1-p) \log(1-p)]$. The function $f_\infty(p)$ is always convex since $\frac{\partial^2}{\partial p^2} f_\infty(p; y, \beta) = [\beta p(1-p)]^{-1} > 0$, and therefore the marginal free energy always has a single minimum $p_*(y, \beta)$ corresponding to a microscopic mixture of de-synchronized elements, see Fig. 12(b).

These results show that the equilibrium (average) properties of a cluster of HS elements in a hard device can be fully recovered if we know the properties of a single element—the problem studied in [74]. In particular, the equilibrium free energy $\tilde{f}(z, \beta) = f(p_*; \sigma, \beta)$, where $p_*$ is the minimum of the marginal free energy $f$ [see Fig. 12(c)] can be written in the HS form

$$\tilde{f}(y, \beta) = -\frac{1}{\beta N} \log[Z(y, \beta)] = \frac{1}{2} y^2 + v_0 + \frac{y - y_0}{2}$$
$$- \frac{1}{\beta} \ln\left\{2 \cosh\left[\frac{\beta}{2}(y - y_0)\right]\right\}, \quad (2.9)$$

which is also an expression of the free energy in the simplest paramagnetic Ising model [189]. Its dependence on elongation is illustrated in Fig. 13(a). We observe that for $\beta \leq 4$ (supercritical temperatures), the free energy is convex while for $\beta > 4$ (sub-critical temperatures), it is non-convex. The emergence of an unusual "pseudo-critical" temperature $\beta = \beta_c = 4$ in this paramagnetic system is a result of the presence of the quadratic energy associated with the "applied field" $y$, see Eq. (2.9).

The ensuing equilibrium tension-elongation relation (per cross-bridge) is identical to the expression obtained in Ref. [74],

$$\langle \sigma \rangle (y, \beta) = \frac{\partial f}{\partial y} = \sigma_0 + y - y_0 - \frac{1}{2} \tanh\left[\frac{\beta}{2}(y - y_0)\right]. \quad (2.10)$$

As a result of the nonconvexity of the free energy, the dependence of the tension $\langle \sigma \rangle$ on $y$ can be non-monotone, see Fig. 13(b). Indeed, the equilibrium stiffness

$$\kappa(y, \beta) = \partial \langle \sigma \rangle (y, \beta) / \partial y$$
$$= 1 - (\beta/4) \left\{\text{sech}\left[\beta(y - y_0)/2\right]\right\}^2, \quad (2.11)$$

is a sign-indefinite sum of two terms: $\kappa_B = 1$, representing the Born elastic susceptibility associated with affine deformation and the fluctuation part $\kappa_F = (\beta/4) \left\{\text{sech}\left[\beta(y - y_0)/2\right]\right\}^2$ describing fluctuations induced non-affine contributions [175; 190].

In connection with these results we observe that the difference between the quasi-static stiffness of myosin II measured by single molecule techniques, and its instantaneous stiffness obtained from mechanical tests on myofibrils, may be due to the fluctuational term $\kappa_F$, see Refs. [91; 191; 192]. Note also that the fluctuation-related term does not disappear in the zero temperature limit (producing a delta function type contribution to the affine response at $y = y_0$), which is a manifestation of a (singular) glassy behavior [193; 194].

It is interesting that while fitting their experimental data HS used exactly the critical value $\beta = 4$, corresponding to zero stiffness in the state of isometric contraction. Negative stiffness, resulting from non-additivity of the system, prevails at subcritical temperatures; in this range a shortening of an element leads to tension increase which can be interpreted as a meta-material behavior [13; 99; 195].

In the soft device case $\lambda_b = 0$, the probability density associated with a microstate $\boldsymbol{x}$ is given by $\rho(\boldsymbol{x}, y; \sigma, \beta) = Z(\sigma, \beta)^{-1} \exp\left[-\beta N w(\boldsymbol{x}, y; \sigma)\right]$ where the partition function is now $Z(\sigma, \beta) = \int dy \sum_{\boldsymbol{x} \in \{0,1\}^N} \exp\left\{-\beta N \left[v(\boldsymbol{x}; y) - \sigma y\right]\right\}$.

By integrating out the internal variable $x_i$, we obtain the marginal probability density depending on the two order parameters, $y$ and $p$,

$$\rho(p, y; \sigma, \beta) = Z(\sigma, \beta)^{-1} \exp\left[-\beta N g(p; y; \sigma, \beta)\right]. \quad (2.12)$$

Here we introduced the marginal free energy

$$g(p, y; \sigma, \beta) = f(p, y, \beta) - \sigma y$$
$$= \hat{v}(p, y) - \sigma y - (1/\beta) s(p), \quad (2.13)$$

which is convex at large temperatures and non-convex (with two metastable wells) at low temperatures, see Fig. 14, signaling the presence of a genuine critical point.

By integrating the distribution (2.12) over $p$ we obtain the marginal distribution $\rho(y; \sigma, \beta) = Z^{-1} \exp\left[-\beta N g(y; \sigma, \beta)\right]$ where $g(y; \sigma, \beta) = \tilde{f}(y; \beta) - \sigma y$, with $\tilde{f}$ being the equilibrium free energy of the system in a hard device, see Eq. 2.9. This free energy has more than one stable state as long as the equation $\tilde{f}'(y) - \sigma = 0$ has more than one solution. Since $\tilde{f}'$ is precisely the average tension elongation-relation in the hard device case, we find that the critical temperature is exactly $\beta_c = 4$. The same result could be also obtained directly as a condition of the positive definiteness of the Hessian for the free energy (2.13) (in the thermodynamic limit).

The physical origin of the predicted second order phase transition becomes clear if instead of $p$ we now eliminate $y$ and introduce the marginal free energy at fixed $p$. In the (more transparent) thermodynamic limit we can write

$$g_\infty(p; \sigma, \beta) = \hat{w}(p, \sigma) - \beta^{-1} s_\infty(p), \quad (2.14)$$

where $\hat{w} = -(1/2)\sigma^2 + p(\sigma - \sigma_0) + (1/2)p(1-p) + v_0$, is the zero temperature energy of the metastable states parametrized by $p$, see Eq. 2.6 and Fig. 10. Since the entropy $s_\infty(p)$ is convex with a maximum at $p = 1/2$, the convexity of the free energy depends on the competition between the term $p(1-p)$ reflecting purely mechanical interaction and the term $s_\infty(p)/\beta$, with the the later dominating at low $\beta$.

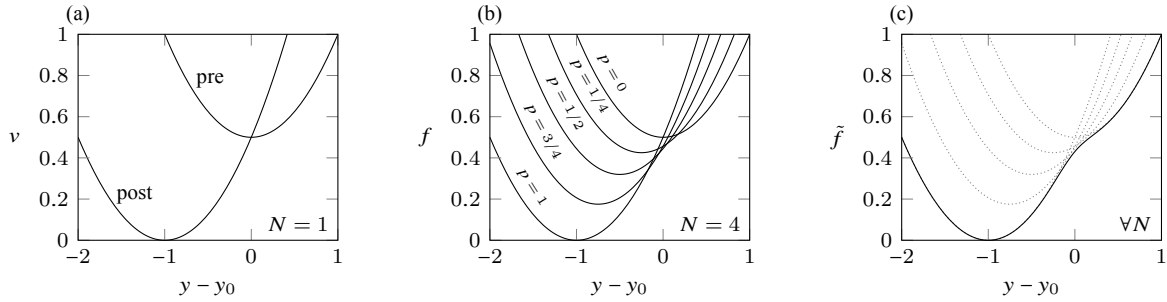

**Figure 12.** Hill-type energy landscapes in a hard device, for $N = 1$ (a) and $N = 4$ (b). (c) Equilibrium free-energy profile $\tilde{f}$ (solid line), which is independent of $N$ together with the metastable states for $N = 4$ (dotted lines). Here $v_0 = 1/2$ and $\beta = 10$. Adapted from Ref. [121].

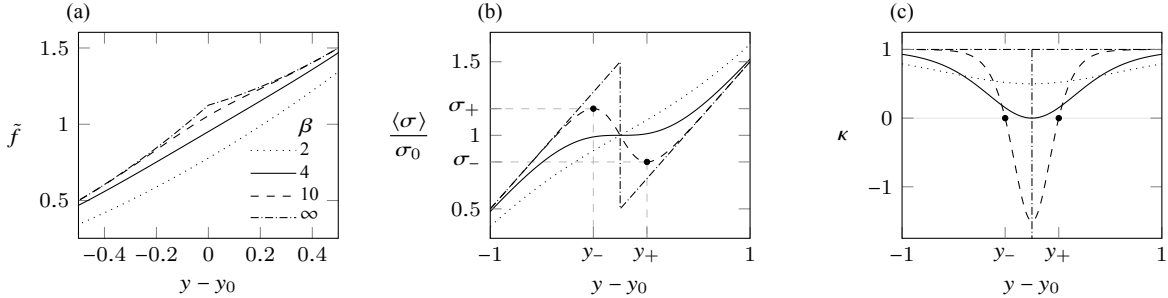

**Figure 13.** Equilibrium properties of the HS model in a hard device for different values of temperature. (a) Helmholtz free energy; (b) tension-elongation relations; (c) stiffness. In the limit $\beta \to \infty$ (dot-dashed line), corresponding to zero temperature, the stiffness $\kappa$ diverges at $y = y_0$. Adapted from Ref. [121].

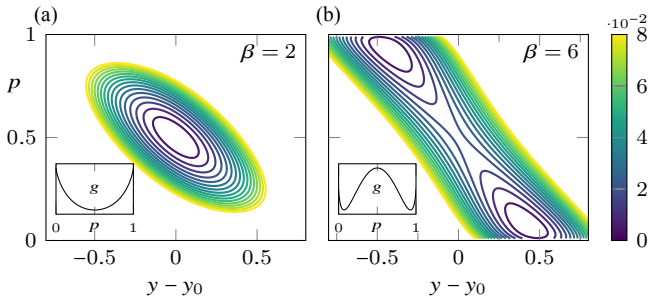

**Figure 14.** Marginal free energy $g$ at $\sigma = \sigma_0$ in the thermodynamic limit $N \to \infty$ for $\beta = 2$ (a) and $\beta = 6$ (b). The inserts show the energy profile of $g$ after elimination of $y$. Contours associated to energy higher than 0.08 are not shown. Here $v_0 = 1$.

The Gibbs free energy $g_\infty(\sigma, \beta)$ and the corresponding force-elongation relations are illustrated in Fig. 15. In (a), the energies of the critical points of the free energy (2.14) are represented as function of the loading and the temperature, with several isothermal sections of the energy landscape are shown in (b). For each critical point $\hat{p}$, the elongation $\hat{y} = \sigma - \hat{p}$ is shown in Fig. 15(c).

At $\sigma = \sigma_0 = v_0$, the free energy $g_\infty$ becomes symmetric with respect to $p = 1/2$ and therefore we have $\langle p \rangle (\sigma_0, \beta) = 1/2$, independently of the value of $\beta$. The structure of the second order phase transition is further illustrated in Fig. 16(a).

Both mechanical and thermal properties of the system can be obtained from the probability density (2.12). By eliminating $y$ and taking the thermodynamic limit $N \to \infty$ we obtain $\rho_\infty(p; \sigma, \beta) = Z^{-1} \exp\left[-\beta N g_\infty(p; \sigma, \beta)\right]$ with $Z(\sigma, \beta) = \sum_p \exp[-\beta N g_\infty(p; \sigma, \beta)]$. The average mechanical behavior of the system is now controlled by the global minimizer $p_*(\sigma, \beta)$ of the marginal free energy $g_\infty$, for instance, $\tilde{g}(\sigma, \beta) = g_\infty(p_*, \sigma, \beta)$ and $\langle p \rangle (\sigma, \beta) = p_*(\sigma, \beta)$. The average elongation $\langle y \rangle (\sigma, \beta) = \sigma - p_*(\sigma, \beta)$ is illustrated in Fig. 16 (c), for the case $\beta = 5$. The jump at $\sigma = \sigma_0$ corresponds to the switch of the global minimum from C to A, see Fig. 16[(a) and (c)].

In Fig. 16[(d)–(f)] we also illustrate typical stochastic behavior of the order parameter $p$ at fixed tension $\sigma = \sigma_0$ (ensuring that $\langle p \rangle = 1/2$). Observe that in the ordered (low temperature, ferromagnetic) phase [see (f)], the thermal equilibrium is realized through the formation of *temporal microstructure*, a domain structure *in time*, which implies intermittent jumps between ordered metastable (long living) configurations. Such transition are systematically observed during the unzipping of biomolecules, see, for instance, Ref. [196].

In Fig. 17 we show the equilibrium susceptibility $\chi(\sigma, \beta) = -\frac{\partial}{\partial \sigma} \langle p \rangle (\sigma, \beta) = N\beta \langle [p - \langle p \rangle (\sigma, \beta)]^2 \rangle \geq 0$, which diverges at $\beta = \beta_c$ and $\sigma = \sigma_0$. We can also compute the equilibrium stiffness $\kappa(\sigma, \beta)^{-1} = \frac{1}{N} \frac{\partial}{\partial \sigma} \langle y \rangle (\sigma, \beta) = \beta \langle [y - \langle y \rangle (\sigma, \beta)]^2 \rangle \geq 0$, where $\langle y \rangle (\sigma, \beta) = \sigma - \langle p \rangle (\sigma, \beta)$, and see that it is always positive in the soft device. This is another manifestation of the fact that the soft and hard device ensembles are not equivalent.

At the critical point ($\beta = 4$, $\sigma = \sigma_0$), the marginal energy of the system has a degenerate minimum corresponding to the configuration with $p = 1/2$; see Fig. 15[(c) dashed line]. Near the critical point, we have the asymptotics $p \sim 1/2 \pm (\sqrt{3}/4)[\beta - 4]^{1/2}$, for $\sigma = \sigma_0$, and $p \sim 1/2 - \text{sign}[\sigma - \sigma_0][(3/4)|\sigma - \sigma_0|]^{1/3}$, for $\beta = 4$, showing that the critical exponents take the classical mean field values [189]. Similarly we obtain $\langle y \rangle - y_0 = \pm(\sqrt{3}/4)[\beta - 4]^{1/2}$, for $\sigma = \sigma_0$, and $\langle y \rangle - y_0 = \text{sign}[\sigma - \sigma_0][(3/4)|\sigma - \sigma_0|]^{1/3}$, for $\beta = 4$. In critical conditions, where the stiffness is equal to 0, the system becomes anomalously reactive; for instance, being exposed to small positive (negative) force increment it instantaneously unfolds (folds).

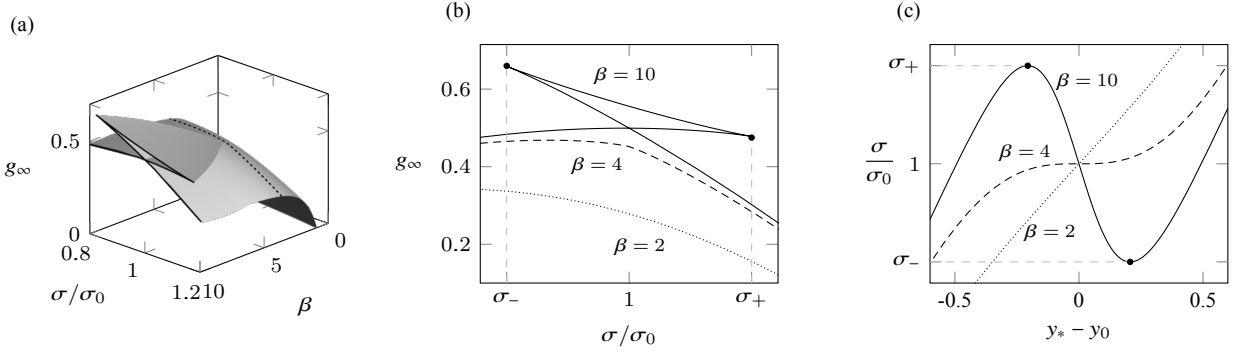

**Figure 15.** Mechanical behavior along metastable branches. (a) Free energy of the metastable states. For $\beta > 4$ (see dotted line), three energy levels coexist at the same tension. (b) Free energy at three different temperatures. (c) Tension-elongation curves.

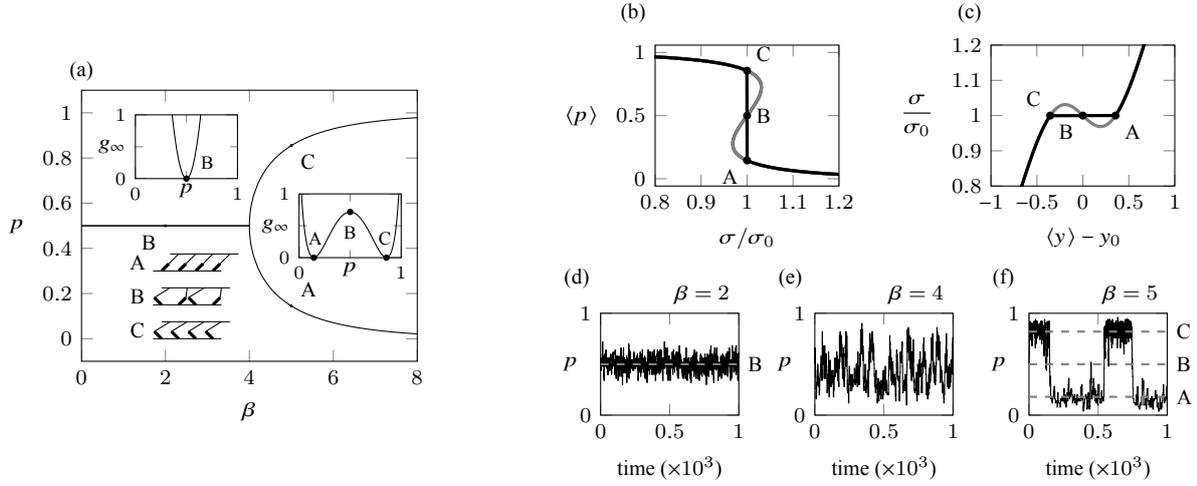

**Figure 16.** Phase transition at $\sigma = \sigma_0$, and its effect on the stochastic dynamics. (a) Bifurcation diagram at $\sigma = \sigma_0$. Lines show minima of the Gibbs free energy $g_\infty$. [(b) and (c)] Tension-elongation relations corresponding to the metastable states (gray) and in the global minimum path (black). [(d)-(f)] Collective dynamics with $N = 100$ in a soft device under constant force at different temperatures. Here the loading is such that $\langle p \rangle = 1/2$ for all values of $\beta$. Fig 16 (a) is adapted from Ref. [99].

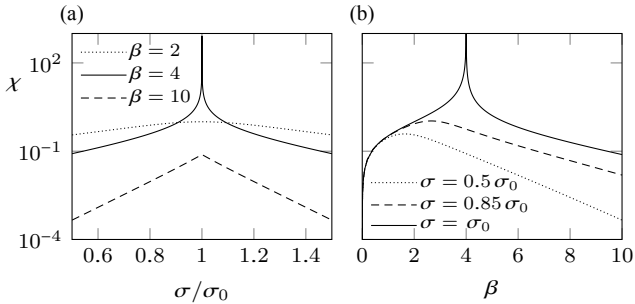

**Figure 17.** Susceptibility $\chi(\sigma, \beta)$ shown as a function of the loading parameter (a) and of the temperature (b).

In Fig. 18 we summarize the mechanical behavior of the system in hard [(a) and (b)] and soft devices [(c) and (d)]. In a hard device, the system develops negative stiffness below the critical temperature while remaining de-synchronized and fluctuating at fast time scale. Instead, in the soft device the stiffness is always non-negative. However, below the critical temperature the tension elongation relation develops a plateau which corresponds to cooperative (macroscopic) fluctuations between two highly synchronized metastable states. In the soft device ensemble, the pseudo-critical point of the hard device ensemble becomes a real critical point with diverging susceptibility and classical mean field critical exponents. For the detailed study of the thermal properties in soft and hard devices, see Refs. [121; 180].

*2.1.2. Mixed device.* Consider now the general case when $\lambda_b$ is finite. In the muscle context this parameter can be interpreted as the lump description of myofilaments elasticity [89; 91; 95], in cell adhesion it can be identified either with the stiffness of the extracellular medium or with the stiffness of the intracellular stress fiber [118; 197; 198], and for protein folding in optical tweezers, it can be viewed as the elasticity of the optical trap or the DNA handles [186; 188; 199–204].

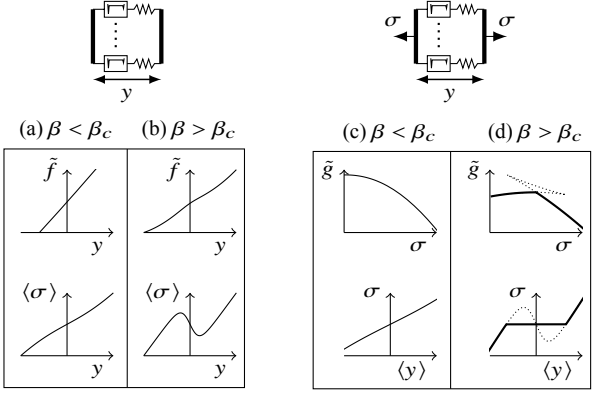

**Figure 18.** Different regimes for the HS model in the two limit cases of hard [(a) and (b)] and soft devices [(c) and (d)]. In a hard device, the pseudo critical temperature $\beta_c^{-1} = 1/4$ separates a regime where the tension elongation is monotone (a) from the region where the system develops negative stiffness (b). In soft device, this pseudo critical point becomes a real critical point above which ($\beta > \beta_c$) the system becomes bistable (d).

The presence of an additional series spring introduces a new macroscopic degree of freedom because the elongation of the bundle of parallel cross-bridges $y$ can now differ from the total elongation of the system $z$, see Fig. 9. At zero temperature, the metastable states are again fully characterized by the order parameter $p$, representing the fraction of cross-bridges in the folded (post-power-stroke) configuration. At equilibrium, the elongation of the bundle is given by $\hat{y} = (\lambda_b z - p)/(1 + \lambda_b)$, so that the energy of a metastable state is now $\hat{v}_b(p; z) = \hat{v}(p; \hat{y}) + (\lambda_b/2)(z - \hat{y})^2$, which can be rewritten as

$$\hat{v}_b(p; z) = \frac{\lambda_b}{2(1 + \lambda_b)} \left[ p(z+1)^2 + (1-p)z^2 \right] + (1-p)v_0 + \frac{p(1-p)}{2(1 + \lambda_b)}. \quad (2.15)$$

Notice the presence of the coupling term $\sim p(1-p)$, characterizing the mean field interaction between cross-bridges. One can see that this term vanishes in the limit $\lambda_b \to \infty$. Again, when $\lambda_b \to 0$ and $z - y \to \infty$, while $\lambda_b(z - y) \to \sigma$, we recover the soft device potential modulo an irrelevant constant.

The global minimum of the energy (2.15) corresponds to one of the fully synchronized configurations ($p = 0$ or $p = 1$). These two configurations are separated at the transition point $z = z_0 = (1 + \lambda_b)v_0/\lambda_b - 1/2$, by an energy barrier whose height now depends on the value of $\lambda_b$, see Ref. [179] for more details.

At finite temperature, the marginal free energy at fixed $p$ and $y$ can be written in the form

$$f_m(p, y; z, \beta) = f(p; y, \beta) + \frac{\lambda_b}{2}(z - y)^2, \quad (2.16)$$

where $f$ is the marginal free energy for the system in a hard device (at fixed $y$). Averaging over $y$ brings about the interaction among cross-bridges exactly as in the case of a soft device. The only difference with the soft device case is that the interaction strength now depends on the new dimensionless parameter $\lambda_b$.

The convexity properties of the energy (2.16) can be studied by computing the Hessian,

$$H(p, y; z, \beta) = \begin{pmatrix} 1 + \lambda_b & 1 \\ 1 & [\beta p(1-p)]^{-1} \end{pmatrix} \quad (2.17)$$

which is positive definite if $\beta < \beta_c$ where the critical temperature is now $\beta_c^* = 4(1 + \lambda_b)$. The latter relation also defines the critical line $\lambda_b = \lambda_c(\beta) = \beta/4 - 1$, separating disordered phase ($\lambda_b > \lambda_c$), where the marginal free energy has a single minimum, from the ordered phase ($\lambda_b < \lambda_c$), where the system can be bi-stable.

As in the soft device case, elimination of the internal variable $p$ allows one to write the partition function in a mixed device as $Z = \int \exp\{-\beta N [f_m(y; z, \beta)]\} \, dy$. Here $f_m$ denotes the marginal free energy at fixed $y$ and $z$

$$f_m(y; z, \beta) = \tilde{f}(y; \beta) + (\lambda_b/2)(z - y)^2 \quad (2.18)$$

and $\tilde{f}$ is the equilibrium free energy at fixed $y$, given by Eq. (2.9). We can now obtain the equilibrium free energy $\tilde{f}_m = -(1/\beta) \log [Z(z, \beta)]$ and compute its successive derivatives. In particular the tension-elongation relation $\langle \sigma \rangle (\langle y \rangle)$ and the equilibrium stiffness $\kappa_m$ can be written the form

$$\langle \sigma \rangle = \lambda_b [z - \langle y \rangle],$$
$$\kappa_m = \lambda_b \{1 - \beta N \lambda_b [\langle y^2 \rangle - \langle y \rangle^2]\}.$$

As in the soft device case, we have in the thermodynamic limit, $\langle y \rangle (z, \beta) = y_*(z, \beta)$, where $y_*$ is the global minimum of the marginal free energy (2.18). We can also write $\kappa_m = \frac{\kappa(y_*, \beta) \lambda_b}{\kappa(y_*, \beta) + \lambda_b}$, where $\kappa$ is the thermal equilibrium stiffness of the system at fixed $y$, see Eq. (2.11). Since $\lambda_b > 0$, we find that the stiffness of the system becomes negative when $\kappa$ becomes negative, which takes place at low temperatures when $\beta > 4$.

Our results in the mixed device case are summarized in Fig. 19(a) where we show the phase diagram of the system in the $(\lambda_b, \beta^{-1})$ plane. The hard and soft device limits, which we have already analyzed, correspond to points (a)–(d). At finite $\lambda_b$ there are three "phases": (i) In phase I, corresponding to $\beta < 4$, the marginal free energy (2.18) is convex and the equilibrium tension elongation relation is monotone; (ii) In phase II $[4 < \beta < 4(1 + \lambda_b)$, see (e)] the free energy is still convex but the tension-elongation becomes non monotone; (iii) In phase III $[\beta > 4(1 + \lambda_b)]$, the marginal free energy (2.18) is non convex and the equilibrium response contains a jump, see (f) in the right panel of Fig. 19.

*2.1.3. Kinetics.* Consider bi-stable elements described by microscopic variables $x_i$ whose dynamics can be represented as a series of jumps between the two states. The probabilities of the direct and reverse transitions in the time interval $dt$ can be written as

$$\mathbb{P}(x_i(t + dt) = -1 | x_i(t) = 0) = k_+(y, \beta) dt,$$
$$\mathbb{P}(x_i(t + dt) = 0 | x_i(t) = -1) = k_-(y, \beta) dt. \quad (2.19)$$

Here $k_+(y, \beta)$ [resp. $k_-(y, \beta)$] is the transition rate for the jump from the unfolded state (resp. folded state) to the folded state (resp. unfolded state). The presence of the jumps is a shortcoming of the hard spin model of Huxley and

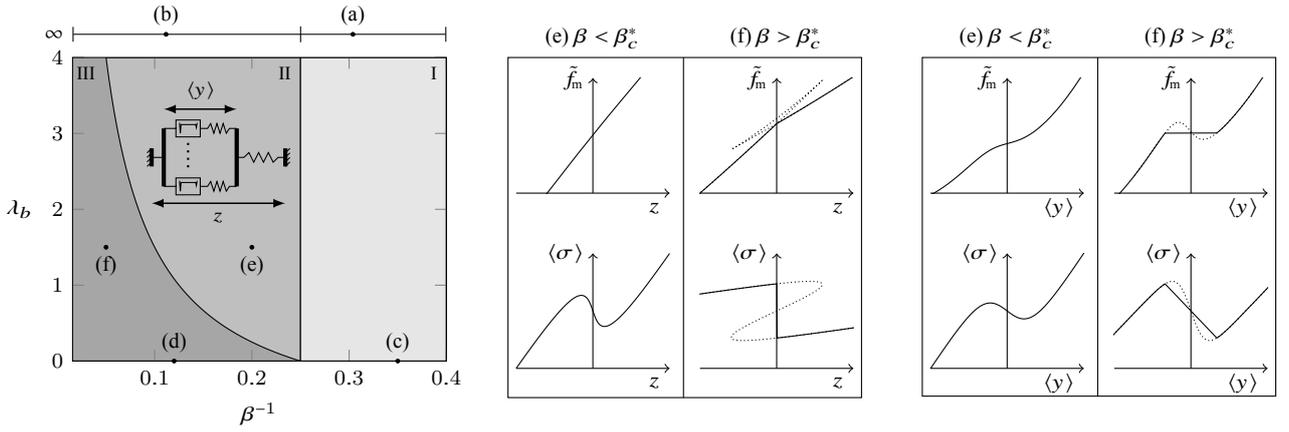

**Figure 19.** Phase diagram in the mixed device. The hard and soft device cases, already presented in Fig. 18, correspond to the limits (a)–(d). In the mixed device, the system exhibits three phases, labeled I, II and III in the left panel. The right panels show typical dependence of the energy and the force on the loading parameter $z$ and on the average internal elongation $\langle y \rangle$ in the subcritical (phase II, e) and in the supercritical (phase III, f) regimes. In phase I, the response of the system is monotone; it is analogous to the behavior obtained in a hard device for $\beta < 4$, see Fig. 18(b). In phase II, the system exhibits negative stiffness but no collective switching except for the soft device limit $\lambda_b \to 0$, see Fig. 18(d). In phase III (supercritical regime), the system shows an interval of macroscopic bistability (see dotted lines) leading to abrupt transitions in the equilibrium response (solid line).

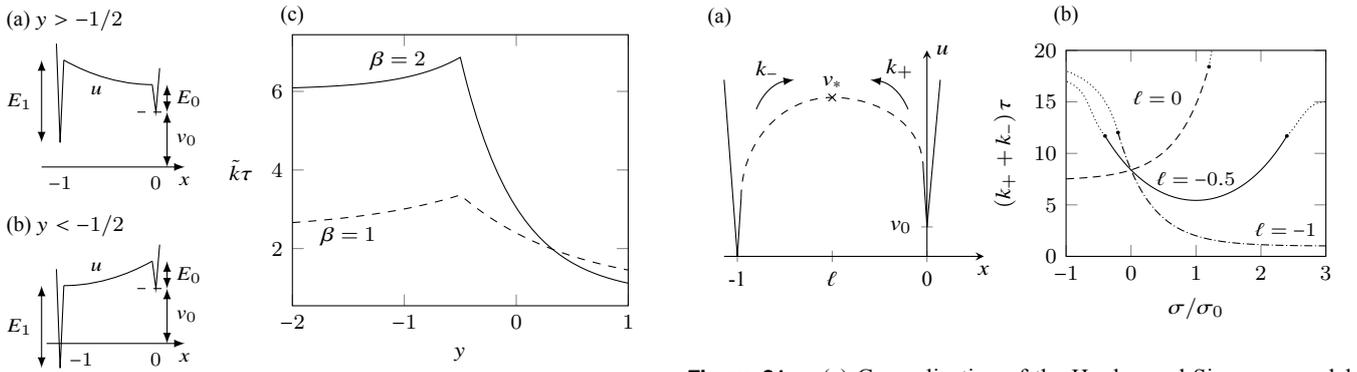

**Figure 20.** Energy barriers in the HS model. [(a) and (b)] Two functioning regimes. The regime (b) was not considered by Huxley and Simmons. (c), Relaxation rate as function of the total elongation $y$. The characteristic timescale is $\tau = \exp[\beta E_1]$. Adapted from Ref. [121].

**Figure 21.** (a) Generalization of the Huxley and Simmons model of the energy barriers based on the idea of the transition state $v_*$ corresponding to the conformation $\ell$. (b) Equilibration rate between the states as function of the loading parameter at different values of $\ell$. The original HS model corresponds to the case $\ell = -1$. In (b) $v_0 = 1$, $v_* = 1.2$ and $\beta = 2$. Dotted lines in (b) is a schematic representation of diffusion (versus reaction) dominated processes. Adapted from [180]

Simmons [74] and in the model with non-degenerate elastic bistable elements (soft spins) they are replaced by a continuous Langevin dynamics [99; 205], see Section 2.2.4.

To compute the transition rates $k_\pm(y,\beta)$ without knowing the energy landscape separating the two spin states, we first follow [74] who simply combined the elastic energy of the linear spring with the idea of the flat microscopic energy landscape between the wells, see Fig. 20(a,b) for the notations. Assuming further that the resulting barriers $E_0$ and $E_1 = E_0 + v_0$ are large comparing to $k_B T$, we can use the Kramers approximation and write the transition rates in the form

$$k_+(y,\beta) = k_- \exp[-\beta(y - y_0)],$$
$$k_-(y,\beta) = \exp[-\beta E_1] = \text{const}, \quad (2.20)$$

where $k_-$ determines the timescale of the dynamic response: $\tau = 1/k_- = \exp[\beta E_1]$. The latter is fully controlled by a single parameter $E_1$ whose value was chosen by HS to match the observations.

Note that Eq. (2.20) is only valid if $y > -1/2$ [see Fig. 20(a)], which ensures that the energy barrier for the transition from pre- to post- power stroke is actually affected by the load. In the range $y < -1/2$, omitted by HS, the forward rate becomes constant, see Fig. 20(a).

The fact that only one transition rate in the HS approach depends on the load makes the kinetic model non-symmetric: the overall equilibration rate between the two states $r = k_+ + k_-$ monotonously decreases with stretching. For a long time this seemed to be in accordance with experiments [74; 85; 87; 206], however, a recent reinterpretation of the experimental results in Ref. [207] suggested that the recovery rate may eventually increase with the amplitude of stretching. This finding can be made compatible with the HS framework if we assume that both energy barriers, for the power stroke and for the reverse power stroke, are load dependent, see Fig. 21, and Ref. [180] for more details. This turns out to be a built-in property of the soft spin model considered in Section 2.2.

In the hard spin model with $N$ elements, a single stochastic trajectory can be viewed as a random walk characterized by the transition probabilities

$$\mathbb{P}\left[p^{t+dt} = p^t + 1/N\right] = \phi_+(p^t, t)dt,$$
$$\mathbb{P}\left[p^{t+dt} = p^t - 1/N\right] = \phi_-(p^t, t)dt,$$
$$\mathbb{P}\left[p^{t+dt} = p^t\right] = 1 - \left[\phi_+(p^t, t) + \phi_-(p^t, t)\right] dt \quad (2.21)$$

where the rate $\phi_+$ (resp. $\phi_-$) describes the probability for one of the unfolded (resp. folded) elements to fold (resp. unfold) within the time-window $dt$. While in the case of a hard device we could simply write $\phi_+(t) = N(1-p^t)k_+(y, \beta)$, and $\phi_-(t) = Np^t k_-$, in both soft and mixed devices, $y$ becomes an internal variable whose evolution become dependent on $p$, making the corresponding dynamics non-linear.

The isothermal stochastic dynamics of the system specified by the transition rates (2.20) is most naturally described in terms of the probability density $\rho(p, t)$. It satisfies the master equation,

$$\frac{\partial}{\partial t}\rho(p, t) = \phi_+ \left(1 - p + 1/N, t\right) \rho\left(p - 1/N, t\right)$$
$$+ \phi_- \left(p + 1/N; t\right) \rho\left(p + 1/N, t\right)$$
$$- \left[\phi_+(1 - p; t) + \phi_-(p; t)\right] \rho(p, t), \quad (2.22)$$

where $\phi_+$ and $\phi_-$ are the transition rates introduced in Eq. (2.21). This equation generalizes the HS mean-field kinetic equation dealing with the evolution of the first moment $\langle p \rangle (t) = \sum p\rho(p, t)$, namely

$$\frac{\partial}{\partial t} \langle p \rangle (t) = \langle \phi_+(1 - p, t) \rangle - \langle \phi_-(p, t) \rangle. \quad (2.23)$$

In the case of a hard device, studied by HS, the linear dependence of $\phi_\pm$ on $p$ allows one to compute the averages on the right hand side of (2.23) explicitly. The result is the first order reaction equation of HS

$$\frac{\partial}{\partial t} \langle p \rangle = k_+(y) (1 - \langle p \rangle) - k_-(y) \langle p \rangle. \quad (2.24)$$

In (2.24) the transition probabilities (2.19) depend only on the control parameter $y$ and the trajectories of individual elements are independent. Hence, at a given $y$ each macro-configuration can be viewed as a realization of $N$ equivalent Bernoulli processes with the probability of success $\rho_1(t) = \rho_1(-1; y(t), \beta) = \langle p(y(t), \beta) \rangle$ represented by a solution of the HS reaction equation (2.24). Therefore the probability density $\rho(p, t) = \mathbb{P}(p^t = p)$ is a binomial distribution with parameters $N$ and $\langle p(t) \rangle$:

$$\rho(p, t) = \binom{N}{Np} [\langle p(t) \rangle]^{Np} [1 - \langle p(t) \rangle]^{N-Np}. \quad (2.25)$$

The entire distribution is then enslaved to the dynamics of the order parameter $\langle p \rangle (t)$ captured by the original HS model. It is then straightforward to show that in the long time limit the distribution (2.25) converges to the Boltzmann distribution (2.8).

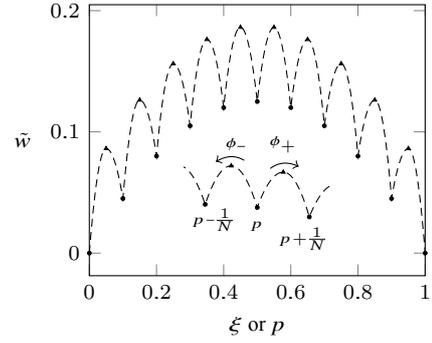

**Figure 22.** Energy landscape characterizing the sequential folding process of $N = 10$ bistable elements in a soft device with $\sigma = \sigma_0$. Parameters are $v_0 = 1$, $v_* = 1.2$, and $\ell = -0.5$. Adapted from Ref. [180]

In the soft and mixed devices the cross-bridges interact and the kinetic picture is more complex. To simplify the setting, we assume that the relaxation time associated with the internal variable $y$ is negligible comparing to other time scales. This implies that the variable $y$ can be considered as equilibrated, meaning in turn that $y = \hat{y}(p, \sigma) = \sigma - p$ in a soft device and $y = \hat{y}(p, z) = (1 + \lambda_b)^{-1}(\lambda_b z - p)$, in a mixed device. Below, we briefly discuss the soft device case, which already captures the effect of the mechanical coupling in the kinetics of the system. Details of this analysis can be found in Ref. [180].

To characterize the transition rates in a cluster of $N > 1$ elements under fixed external force, we introduce the energy $\tilde{w}(p, p_*)$ corresponding to a configuration where $p$ elements are folded ($x_i = -1$) and $p_*$ elements are at the transition state ($x_i = \ell$), see Fig. 21. The energy landscape separating two configurations $p$ and $q$ can be represented in terms of the "reaction coordinate" $\xi = p - x(q - p)$, see Fig. 22. The transition rates between neighboring metastable states can be computed explicitly using our generalized HS model (see Fig. 21),

$$\tau\phi_+(p; \sigma, \beta) = N(1 - p) \exp\left[-\beta\Delta\tilde{w}_+(p; \sigma)\right],$$
$$\tau\phi_-(p; \sigma, \beta) = Np \exp\left[-\beta\Delta\tilde{w}_-(p; \sigma)\right], \quad (2.26)$$

where $\Delta\tilde{w}_\pm$ are the energy barriers separating neighboring states,

$$\Delta\tilde{w}_+(p; \sigma) = -\ell(\sigma - p) - \sigma_0 + (1 + \frac{3}{N})\frac{\ell^2}{2}$$
$$\Delta\tilde{w}_-(p; \sigma) = -(\ell + 1)(\sigma - p) + (1 + \frac{3}{N})\frac{\ell^2}{2} - \frac{1 + N + 2\ell}{2N}.$$

In (2.26) $1/\tau = \alpha \exp[-\beta v_*]$, with $\alpha =$ const, determining the overall timescale of the response. The mechanical coupling appearing in the exponent of (2.26) makes the dynamics non-linear.

To understand the peculiarities of the time dependent response of the parallel bundle of $N$ cross-bridges brought about by the above nonlinearity, it is instructive to first examine the expression for the mean first passage time $\tau(p \to p')$ characterizing transitions between two metastable states with $Np$ and $Np'$ ($p < p'$) folded elements.

Following Ref. [208] (and omitting the dependence on $\sigma$ and $\beta$), we can write

$$\tau(p \to p') = \sum_{Nk=Np}^{Np'} \left[\rho(k) \phi_+(k)\right]^{-1} \sum_{Ni=0}^{Nk} \rho(i), \quad (2.27)$$

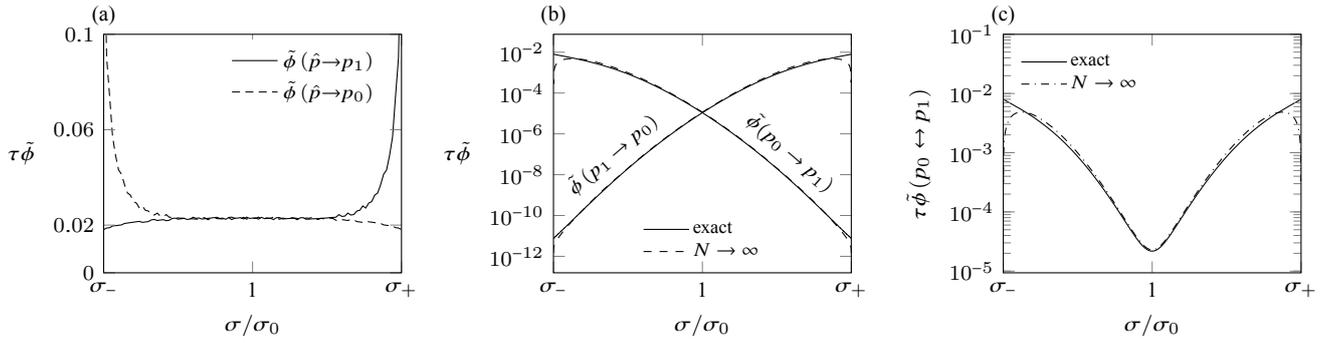

**Figure 23.** Intra- and inter-bassin relaxation rates in a soft device. (a) Relaxation towards to the metastable state in the case of a reflecting barrier at $p = \hat{p}$ (intra-bassin relaxation). [(b) and (c)] Transition between the two macroscopic configurations $p_0(\sigma)$ and $p_1(\sigma)$ (inter-bassin relaxation). (b) Forward [$\tilde{\phi}(p_0 \to p_1)$] and reverse [$\tilde{\phi}(p_1 \to p_0)$] rates. (c) Equilibration rate $\tilde{\phi}(p_0 \leftrightarrow p_1) = \tilde{\phi}(p_0 \to p_1) + \tilde{\phi}(p_1 \to p_0)$. Solid line, computation based on Eq. (2.27); dot-dashed line, thermodynamic limit approximation, see Eq. (2.29). The parameters are $N = 200$, $\beta = 5$ and $\ell = -0.5$. Adapted from Ref. [180].

where $\rho$ is the marginal equilibrium distribution at fixed $p$ and $\phi_+$ is the forward rate. In the case $\beta > \beta_c$ for the interval of loading $[\sigma_-, \sigma_+]$, the marginal free energy $g_\infty$ [see (2.14)] has two minima which we denote $p = p_1$ and $p = p_0$, with $p_0 < p_1$. The minima are separated by a maximum located at $p = \hat{p}$. We can distinguish two process: (i) The intra-bassin relaxation, which corresponds to reaching the metastable states ($p = 0$ or $p = 1$) starting from the top of the energy barrier $\hat{p}$ and (ii) The inter-bassin relaxation, which deals with transitions between macro-states.

For the intra-bassin relaxation, the first passage time can be computed using Eq. (2.27), see Ref. [180]. The resulting rates $\tilde{\phi}(\hat{p} \to p_{0,1}) \equiv 1/[\tau(\hat{p} \to p_{0,1})]$ are practically independent of the load and scale with $1/N$, see Fig. 23(a).

Regarding the transition between the two macrostates, we note that Eq. (2.27) can be simplified if $N$ is sufficiently large. In this case, the sums in Eq. (2.27) can be transformed into integrals

$$\tau(p_0 \to p_1) = N^2 \int_{p_0}^{p_1} [\rho_\infty(u)\phi_+(u)]^{-1} \left[ \int_0^u \rho_\infty(v)\,dv \right] du, \quad (2.28)$$

where $\rho_\infty \sim \exp[-\beta N g_\infty]$ is the marginal distribution in the thermodynamic limit. The inner integral in Eq. (2.28) can be computed using Laplace method. Noticing that the function $g_\infty$ has a single minimum in the interval $[0, u > p_0]$ located at $p_0$, we can write

$$\tau(p_0 \to p_1) = \left[ \frac{2\pi N}{\beta |g_\infty''(p_0)|} \right]^{\frac{1}{2}} \int_{p_0}^{p_1} [\rho_\infty(u)\,\phi_+(u)]^{-1} \rho_\infty(p_0)\,du.$$

In the remaining integral, the inverse density $(1/\rho_\infty)$ is sharply peaked at $p = \hat{p}$ so again using Laplace method we obtain

$$\tau(p_0 \to p_1) = 2\pi (N/\beta) \phi_+(\hat{p})^{-1} \left| g_\infty''(p_0) g_\infty''(\hat{p}) \right|^{-\frac{1}{2}} \\ \times \exp\left\{ \beta N [g_\infty(\hat{p}) - g_\infty(p_0)] \right\}. \quad (2.29)$$

We see that the first passage time is of the order of $\exp[N \Delta g_\infty]$, see Eq. (2.29), where $\Delta g_\infty$ is the height of the energy barrier separating the two metastable states. In the thermodynamic limit, this energy barrier grows exponentially with $N$, which freezes collective inter-basin dynamics and generates metastability, see Fig. 23[(b) and (c)] and Ref. [180]. The above analysis can be generalized for the case of a mixed device by replacing the soft device marginal free energy $g$ by its mixed device analog.

The kinetic behavior of the system in the general case is illustrated in Fig. 24. The individual trajectories generated by the stochastic Eq. 2.21 are shown for $N = 100$. The system is subjected to a slow stretching in hard [(a) and (b)], soft [(c) and (d)] and mixed [(e) and (f)] devices. These numerical experiments mimic various loading protocols used for unzipping tests in biological macro-molecules [186; 199; 203; 209].

Observe that individual trajectories at finite $N$ show a succession of jumps corresponding to collective folding-unfolding events. At large temperatures, see Fig. 24[(a), (c) and (e)], the transition between the folded and the unfolded state is smooth and is associated with a continuous drift of a unimodal density distribution, see inserts in Fig. 24. In the hard device such behavior persists even at low temperatures, see (b), which correlates with the fact that the marginal free energy in this case is always convex. Below the critical temperature [(d) and (f)], the mechanical response becomes hysteretic. The hysteresis is due to the presence of the macroscopic wells in the marginal free energy which is also evident from the bimodal distribution of the cross-bridges shown in the inserts. A study of the influence of the loading rate on the mechanical response of the system can be found in Ref. [180].

To illustrate the fast force recovery phenomenon, consider the response of the system to an instantaneous load increment. We compare the behaviors predicted by the master equation (2.22) and by the mean-field chemical kinetic equation of HS. In Fig. 25 we see the anticipated slowing down induced by the collective effects at high loads and low temperatures [see (b) and (c), solid lines]. The corresponding probability distributions at different times are illustrated in Fig. 25[(e) and (f)]. The chemical kinetics approximation is accurate at large temperatures [see thin lines in Fig. 25(a)] but fails to reproduce the exact two scale dynamics at low temperatures, event though the final equilibrium states are captured correctly.

The difference between the chemo-mechanical description of HS and the stochastic simulation targeting the full probability distribution is due to the fact that in the equation describing the mean-field kinetics the transition rates are computed based on the average values of the order parameter. At large temperatures, where the distribution is uni-modal, the average values faithfully describe the most probable states and therefore the mean-field kinetic theory captures the timescale of the response adequately; see Fig. 25 (a). Instead, at low temperatures, when the distribution is bi-modal, the averaged values correspond to the states that are poorly populated; see Fig. 25 (b) where $\langle p \rangle_{\text{in}} = 1/2$. The value of the order parameter, which actually makes kinetics slow, describes a particular metastable configuration rather than the average state and therefore the mean-field kinetic equation fails to reproduce the real dynamics; see Fig. 25[(b) and (c)].

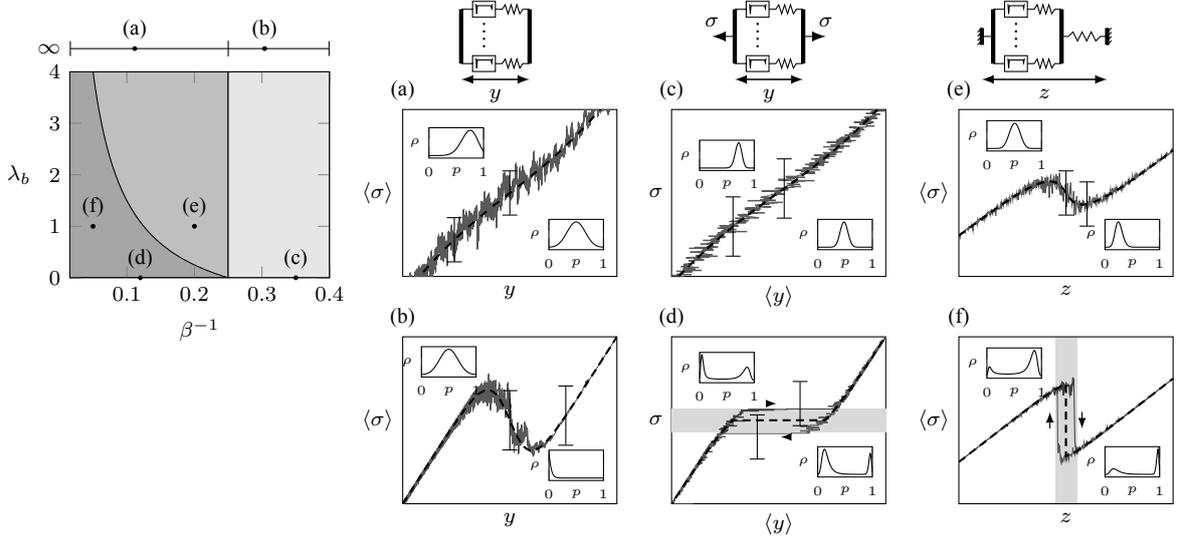

**Figure 24.** Quasi-static response to ramp loading in different points of the phase diagram. [(a) and (b)] Hard device, see Ref. [121]; [(c) and (d)] soft device, see Ref. [180] and [(e) and (f)], mixed device. In each point, stochastic trajectories obtained from Eq. 2.21 (solid lines) are superimposed on the thermal equilibrium response (dashed lines). The inserts show selected snapshots of the probability distribution solving the master equation (2.22).

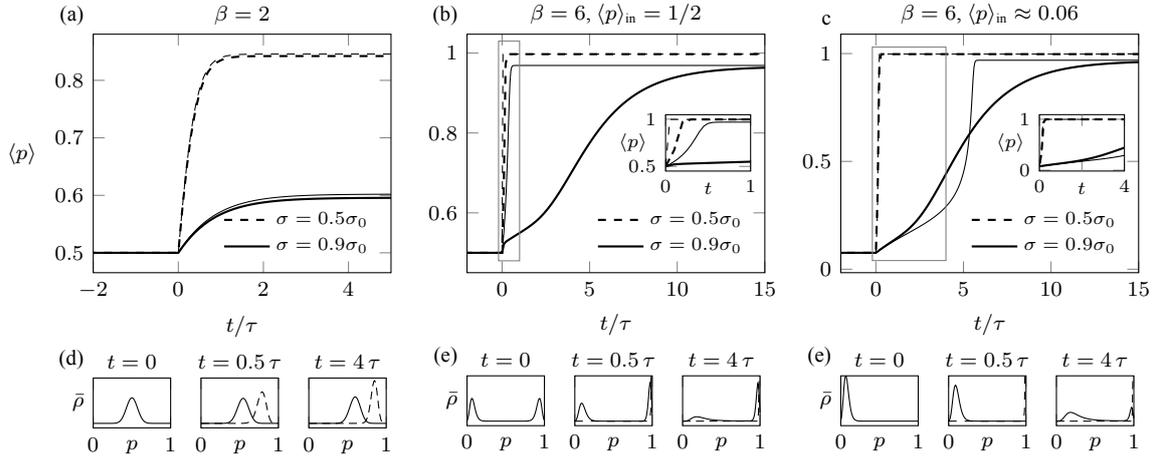

**Figure 25.** Relaxation of the average conformation in response to fast force drops at different temperatures and initial conditions $\langle p \rangle_{\text{in}}$. Thick lines, solutions of the master equation (2.22); thin lines, solutions of the mean-field HS equation. In (b), the initial condition corresponds to thermal equilibrium: bimodal distribution and $\langle p \rangle_{\text{in}} = 1/2$. In (c), the initial condition corresponds to the unfolded metastable state: unimodal distribution and $\langle p \rangle_{\text{in}} \approx 0.06$. Snapshots at different times show the probability density profiles.

## 2.2. Soft spin model

The hard spin model states that the slope of the $T_1$ curve, describing instantaneous stiffness of the fiber, and the slope of the $T_2$ curve are equal, which differs from what is observed experimentally, see Fig. 5. The soft spin model [99; 205] was developed to overcome this problem and to provide a purely mechanical continuous description of the phenomenon of fast force recovery. To this end, the discrete degrees of freedom were replaced by the continuous variables ($x_i$); the latter can be interpreted as projected angles formed by the segment S1 of the myosin head with the actin filament. Most importantly, the introduction of continuous variables has eliminated the necessity of using multiple intermediate configurations for the head domain [67; 68; 86].

The simplest way to account for the bistability in the configuration of the myosin head is to associate a bi-quadratic double-well energy $u_{ss}(x_i)$ with each variable $x_i$, see Fig. 26(a); interestingly, a comparison with the reconstructed potentials for unfolding biological macro-molecules shows that a biquadratic approximation may be quantitavely adequate [186]. A nondegenerate spinodal region can be easily incorporated into this model, however, in this case we lose the desirable analytical transparency. It is sufficient for our purposes to keep the other ingredients of the hard spin model intact; the original variant of the soft spin model model (see Ref. [210]) corresponded to the limit $\kappa_b/(N\kappa_0) \to \infty$.

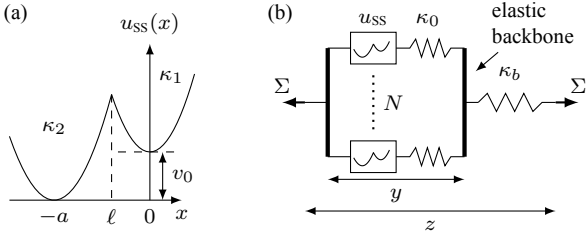

**Figure 26.** Soft spin (snap-spring) model of a parallel cluster of cross-bridges. (a) Dimensional energy landscape of a bistable cross-bridge. (b) Structure of a parallel bundle containing $N$ cross-bridges. Adapted from Ref. [99].

In the soft spin model the total energy of the cross bridge can be written in the form

$$v(x, y) = u_{ss}(x) + (\kappa_0/2)(y - x)^2, \qquad (2.30)$$

where

$$u_{ss}(x) = \begin{cases} \frac{1}{2}\kappa_1(x)^2 + v_0 & \text{if } x > \ell, \\ \frac{1}{2}\kappa_2(x+a)^2 & \text{if } x \leq \ell. \end{cases} \qquad (2.31)$$

The parameter $\ell$ describes the point of intersection of the 2 parabolas in the interval $[-a, 0]$, and therefore $v_0 = (\kappa_2/2)(\ell + a)^2 - (\kappa_1/2)\ell^2$, is the energy difference between the pre-power-stroke and the post-power-stroke configurations. It will be convenient to use normalized parameters to characterize the asymmetry of the energy wells: $\lambda_2 = \kappa_2/(1 + \kappa_2)$ and $\lambda_1 = \kappa_1/(1 + \kappa_1)$.

The dimensionless total internal energy per element of a cluster now reads

$$v(\boldsymbol{x}, y; z) = \frac{1}{N} \sum_{i=1}^{N} \left[ u_{ss}(x_i) + \frac{1}{2}(y - x_i)^2 + \frac{\lambda_b}{2}(z - y)^2 \right], \qquad (2.32)$$

where $\lambda_b = \kappa_b/(N\kappa_0)$. Here $z$ is the control parameter. In a soft device case, the energy takes the form

$$w(\boldsymbol{x}, y; \sigma) = \frac{1}{N} \sum_{i=1}^{N} \left[ u_{ss}(x_i) + \frac{1}{2}(y - x_i)^2 - \sigma y \right]. \qquad (2.33)$$

where $\sigma$ is the applied tension per cross-bridge, see Ref. [179] for the details.

*2.2.1. Zero temperature behavior.* By minimizing out the internal variable $y$ and introducing again the fraction of cross cross bridges in the post power stroke state, $p = \frac{1}{N} \sum \alpha_i$, where $\alpha_i = 1$ if $x_i > \ell$ and 0 otherwise, we find that the global minimum of the energy again corresponds to one of the homogeneous states $p = 0, 1$ with a sharp transition at $z = z_0$.

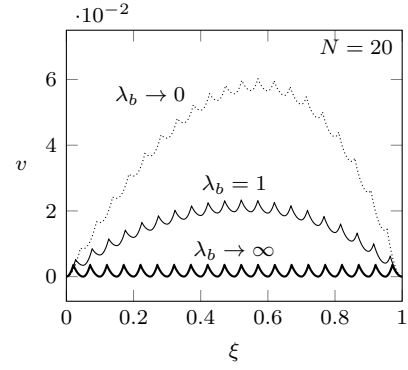

**Figure 27.** Energy landscape along the global minimum path for the soft-spin model in a hard device at different values of the coupling parameter $\lambda_b$ with $N = 20$. Adapted from Ref. [179]. The asymmetry in the potential is the results of choosing $\lambda_2 \neq \lambda_1$. Parameters are, $\lambda_2 = 0.4, \lambda_1 = 0.7, \ell = -0.3$.

We can also take advantage of the fact the soft spin model deals with continuous variables $x_i$ and define a continuous reaction path connecting metastable states with different number of folded units. Each folding event is characterized by a micro energy barrier that can be now computed explicitly. The typical structure of the resulting energy landscape is illustrated in Fig. 27 as a function of a continuous "reaction coordinate" $\xi$ interpolating between the different values of $p$, for different values of the coupling parameter $\lambda_b$, see Ref. [179] for the details. In Fig. 28 we illustrate the zero temperature behavior of the soft-spin model with a realistic set of parameters, see Tab. 1 below.

*2.2.2. Finite temperature behavior.* When $z$ is the control parameter (mixed device), the equilibrium probability distribution for the remaining mechanical degrees of freedom can be written in the form $\rho(\boldsymbol{x}, y; z, \beta) = Z^{-1}(z, \beta) \exp\left[-\beta N v(\boldsymbol{x}, y; z)\right]$, where $\beta = (\kappa_0 a^2)/(k_B T)$ and $Z(z, \beta) = \int \exp\left[-\beta N v(\boldsymbol{x}, y; z)\right] d\boldsymbol{x} dy$. In the soft device ensemble, $z$ becomes a variable and the equilibrium distribution takes the form,

$$\rho(\boldsymbol{x}, y, z; \sigma, \beta) = Z^{-1}(\sigma, \beta) \exp\left[-\beta N w(\boldsymbol{x}, y, z; \sigma)\right], \quad (2.34)$$

with $Z(\sigma, \beta) = \int \exp\left[-\beta N w(\boldsymbol{x}, y, z, \sigma)\right] d\boldsymbol{x} dy dz$.

When $z$ is fixed, the internal state of the system can be again characterized by the two mesoscopic paramters $y$ and $p$. By integrating (2.34) over $\boldsymbol{x}$ and $y$ we can define the marginal density $\rho(p; z, \beta) = Z^{-1} \exp\left[-\beta N f(p; z, \beta)\right]$. Here $f$ is the marginal free energy at fixed $(p, z)$ which is illustrated in Fig. 29.

As we see, the system undergoes an order-disorder phase transition which is controlled by the temperature and by the elasticity of the backbone. If the double well potential is symmetric ($\lambda_1 = \lambda_2$), this transition is of second order as in the hard spin model. A typical bifurcation diagrams for the case of slightly nonsymmetric energy wells are shown in Fig. 30. The main feature of the model without symmetry is that the second order phase transition becomes a first order phase transition.

A phase diagram obtained with realistic parameters (justified later in the paper) is shown in Fig. 31[(a) and (b)]. Recall-ing that $\lambda_b = \kappa_b/(N\kappa_0)$, we use $(T, \lambda_b)$ [see Fig. 31(a)] and $(T, N)$ [see Fig. 31(b)] planes to represent the same configuration of phases. In Phases I and II, the marginal free energy $f$ has a single minimum while in Phase III it may have three critical points, two corresponding to metastable states and one to an unstable state. The equilibrium response can be obtained by computing the partition function $Z$ numerically. In the thermodynamic limit, we can employ the same methods as in the previous section and identify the equilibrium mechanical properties of the system with the global minimum of the marginal free energy $f$. In Fig. 31[(c)–(e)], we illustrate the equilibrium mechanical response of the system : similar phase diagrams have been also obtained for other systems with long-range interactions [211].

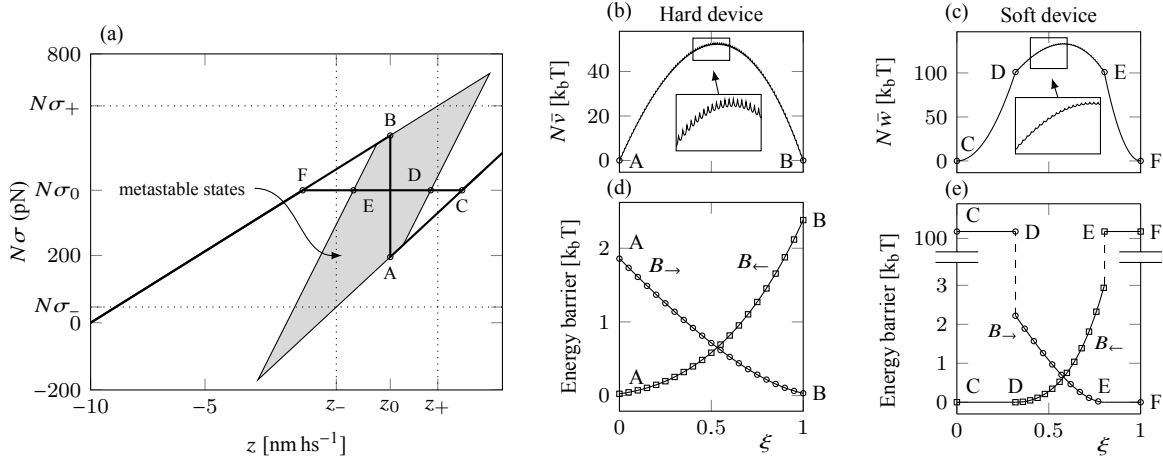

**Figure 28.** Soft spin model at zero temperature with parameters adjusted to fit experimental data, see Tab. 1 in Section 2.2.3. (a) Tension-elongation relations in the metastable states (gray area) and along the global minimum path (solid lines). [(b) and (c)] Energy landscape corresponding to successive transitions between the homogeneous states (A→B and C→F), respectively. [(d) and (e)] Size of the energy barriers corresponding to the individual folding ($B_\rightarrow$) and ($B_\leftarrow$) at finite $N$ (open symbols) and in the thermodynamic limit (solid lines). Adapted from Ref. [179].

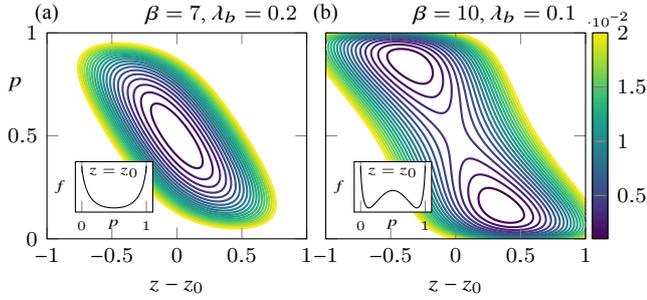

**Figure 29.** Non equilibrium free energy at fixed $(p, z)$ for the case of symmetric power-stroke element with different parameters $(\beta, \lambda_b)$ at $z = z_0$. Other parameters are $\lambda_1 = \lambda_2 = 0.5$, $\ell = -0.5$ and $z_0 = -1/2$, which here ensures that $\langle p \rangle (z_0) = 1/2$.

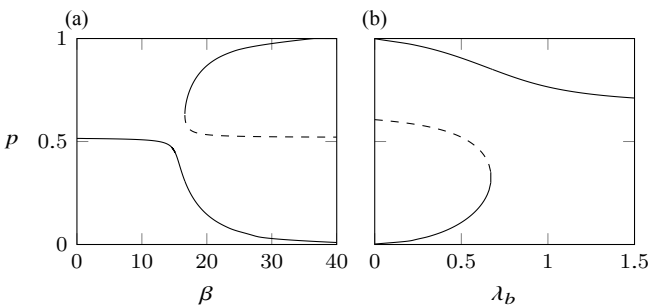

**Figure 30.** Bifurcation diagram with non-symmetric wells. Solid (dashed) lines correspond to local minima (maxima) of the free energy. Parameters are, $\lambda_2 = 0.47$, $\lambda_1 = 0.53$, $\ell = -0.5$, $\lambda_b = 0.5$ and $\beta = 20$. Here $z$ is such that $\langle p \rangle = 1/2$ at $\beta = 20$ and $\lambda_b = 0.5$.

While the soft spin model is analytically much less transparent than the hard spin model, we can still show analytically that the system develops negative stiffness at

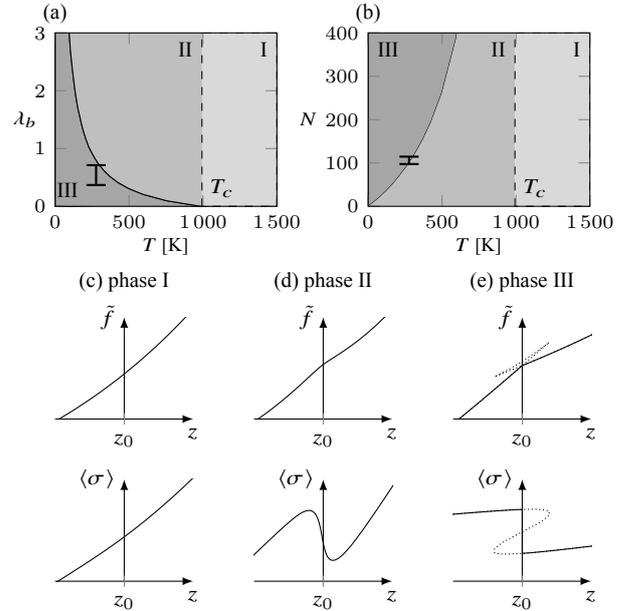

**Figure 31.** Phase diagram of the soft spin model. [(a) and (b)] Boundaries between the phases I, II and III in the $(T, \lambda_b)$ space and in the $(T, N)$ space, respectively. [(c)-(e)] Typical free energy $\tilde{f}$, tension-elongation relation $\langle \sigma \rangle (z)$ and marginal free energy in each phase. The parameters are listed in Tab.1, see Section 2.2.3.

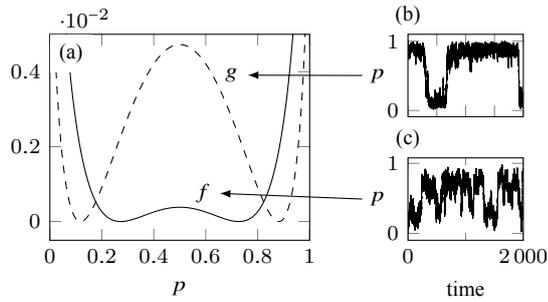

**Figure 32.** Non-equilibrium energy landscapes: $f$ (solid lines) and $g$ (dashed lines) at $z = z_0$. Trajectories on the right are obtained from stochastic simulations. Minima are arbitrarily set to 0. Parameters are: $\beta = 10$, $\lambda_1 = \lambda_2 = 1/2$, $v_0 = 0$, $\lambda_b = 0.1$ (symmetric system).

sufficiently low temperatures. Indeed, we can write

$$\tilde{f}'' = \langle \sigma \rangle' = \lambda_b \left[ 1 - \beta N \lambda_b \left\langle (y - \langle y \rangle)^2 \right\rangle \right],$$

where $\tilde{f}$ is the equilibrium free energy of the system in a mixed device. This expression is sign indefinite and by the same reasoning as in the hard spin case, on can show that the critical line separating Phase I and Phase II is represented in Fig. 31 by a vertical line $T = T_c$. In phase I ($T > T_c$) the equilibrium free energy is convex and the resulting tension-elongation is monotone. In Phase II ($T < T_c$) the equilibrium free energy is non-convex and the tension-elongation relation exhibits an interval with negative stiffness. In phase III the energy is non-convex within a finite interval around $z = z_0$, see dotted line in Fig. 31(e). As a result the system has to oscillate between two metastable states to remain in the global minimum of the free energy [solid line in Fig. 31(e)]. The ensuing equilibrium tension-elongation curve is characterized by a jump located at $z = z_0$.

Observe that the critical line separating Phase II and Phase III in Fig. 31 (b) represents the minimum number of cross-bridges necessary to obtain a cooperative behavior at a given value of the temperature. We see that for temperatures around 300 K, the critical value of $N$ is about 100 which corresponds approximately to the number of cross-bridges involved in isometric contraction in each half-sarcomere, see Section 2.2.3. This observation suggests that muscle fibers may be tuned to work close to the critical state [99]. A definitive statement of this type, however, cannot be made at this point in view of the considerable error bars in the data presented in Table 1.

In a soft device, a similar analysis can be performed in terms of the marginal Gibbs free energy $g(p; \sigma, \beta)$. A comparison of the free energies of a symmetric system in the hard and the soft device ensembles is presented in Fig. 32, where the parameters are such that the system in the hard device is in phase III, see Fig. 31.

We observe that both free energies are bi-stable in this range of parameters, however the energy barrier separating the two wells in the hard device case is about three times smaller than in the case of a soft device. Since the macroscopic energy barrier separating the two states is proportional to $N$, the characteristic time of a transition increases exponentially with $N$ as in the hard spin model, see Section 2.1.3. Therefore the kinetics of the power-stroke will be exponentially slower in a soft device than in a hard device as it is observed in experiment, see more about this in the next section. Note also that the macroscopic oscillations are more coherent in a soft device than in a hard device.

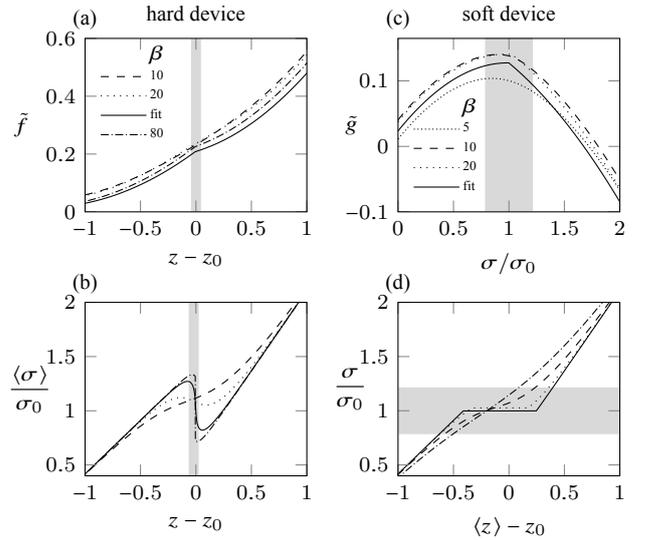

**Figure 33.** Soft spin model in hard [(a) and (b)] and soft [(c) and (d)] devices. [(a) and (c)] Free energies; [(b) and (d)] tension elongation relations. The solid lines correspond to the parameters listed in Tab.1 and the gray regions indicate the corresponding domains of bistability. The tension and elongation are normalized to their values at the transition point where $\langle p \rangle = 1/2$.

By differentiating the equilibrium Gibbs free energy $\tilde{g}(\sigma, \beta) = -1/(\beta N) \log [Z(\sigma, \beta)]$ with respect to $\sigma$, we obtain the tension-elongation relation, which in a soft device is always monotone since

$$\tilde{g}'' = -\left[ 1 + \beta N \left\langle (z - \langle z \rangle)^2 \right\rangle \right] < 0.$$

This shows once again that soft and hard device ensembles are non-equivalent, in particular, that only the system in a hard device can exhibit negative susceptibility.

In Fig. 33, we illustrate the behavior of the equilibrium free energies $\tilde{f}$ and $\tilde{g}$ in thermodynamic limit [(a) and (b)] together with the corresponding tension-elongation relations [(b) and (d)], see Ref. [212] for the details. The tension and elongation are normalized by their values at the transition point where $\langle p \rangle = 1/2$ while the value of $\beta$ is taken from experiments (solid line). The bi-stability (metastability) takes place in the gray region and we see that this region is much wider in the soft device than in the hard device, which corroborates that the energy barrier is higher in a soft device.

*2.2.3. Matching experiments.* The next step is to match the model with experimental data. The difficulty of the parameter identification lies in the fact that the experimental results vary depending on the species, and here we limit our analysis to the data obtained from *rana temporaria* in Refs. [73; 80; 87; 89]. Typical values of the parameters of the non-dimensional model obtained from these data are listed in Table 1.

The first parameter $a$ is obtained from structural analysis of myosin II [73; 96–98]. It has been shown that that its tertiary structure can be found in two conformations forming an angle of ~70°. This corresponds to an axial displacement of the lever

| | Dimensional | | | Non-dimensional |
|---|---|---|---|---|
| $a$ | $10 \pm 1$ | nm | | |
| $\kappa_0$ | $2.7 \pm 0.9$ | pN nm$^{-1}$ | $N$ | $100 \pm 30$ |
| $T$ | $277.15$ | K | $\beta$ | $80 \pm 30$ |
| $\kappa_b$ | $150 \pm 10$ | pN nm$^{-1}$ | $\lambda_b$ | $0.56 \pm 0.25$ |
| $\kappa_1$ | $3 \pm 1$ | pN nm$^{-1}$ | $\lambda_1$ | $0.5 \pm 0.1$ |
| $\kappa_2$ | $1.05 \pm 0.75$ | pN nm$^{-1}$ | $\lambda_2$ | $0.25 \pm 0.15$ |
| $v_0$ | $50 \pm 10$ | zJ | $v_0$ | $0.15 \pm 0.30$ |

**Table 1.** Realistic values (with estimated error bars) for the parameters of the snap-spring model ( 1 zJ $= 10^{-21}$ J).

arm end of ~10 nm. We therefore fix the characteristic length in our model at $a = (10 \pm 1)$ nm.

The absolute temperature $T$ is set to 277.15 K which correspond to 4 °C. This is the temperature at which most experiments on frog muscles are performed [206].

Several experimental studies aimed at measuring the stiffness of the myosin head and of the myofilaments (our backbone). One technique consists in applying rapid (100 μs) length steps to a tetanized fiber to obtain its overall stiffness $\kappa_{\text{tot}}$, which corresponds to the elastic backbone in series with $N$ cross-bridges: $\kappa_{\text{tot}} = (N\kappa_0 \kappa_b)/(N\kappa_0 + \kappa_b)$. The stiffness associated with the double well potential ($\kappa_{1,2}$) is not included into this formula because the time of the purely elastic response is shorter than the time of the conformational change. This implies an assumption that the conformational degree of freedom is "frozen" during the purely elastic response. Such assumption is supported by experiments reported in Ref. [213], where shortening steps were applied at different stages of the fast force recovery, which means during the power-stroke. The results show that the overall stiffness is the same in the recovery process and in the isometric conditions.

If we change the chemical environment inside the fiber by removing the cell membrane ("skinning") it is possible to perform the length steps under different calcium concentrations. We recall that calcium ions bind to the tropomyosin complex to allow the attachment of myosin heads to actin. Therefore, by changing the calcium environment, one can change the number of attached motors ($N$) and thus their contribution to the total stiffness while the contribution of the filaments remains the same [87; 89; 214]. Another solution is to apply rapid oscillations during the activation phase when force rises [100; 215]. These different techniques give $\kappa_b = (150 \pm 10)$ pN nm$^{-1}$, a value which is compatible with independent X-ray measurements [76; 91; 93; 100; 167; 168; 181; 216].

To determine the stiffness of a single element, elastic measurements have been performed on fibers in *rigor mortis* where all the 294 cross-bridges of a half-sarcomere are attached, see Ref. [89]. Under the assumption that the filament elasticity is the same in rigor and in the state of tetanus, one can deduce the stiffness of a single cross-bridge. The value extracted from experiment is $\kappa_0 = (2.7 \pm 0.9)$ pN nm$^{-1}$ [80; 91; 100]. Given that we know the values of $\kappa$ and $a$, we can estimate the non-dimensional inverse temperature, $\beta = (\kappa_0 a^2)/(k_B T) = 71 \pm 26$.

Once $\kappa_b$ and $\kappa_0$ are known, the number of cross-bridges attached in the state of isometric contraction can be obtained directly from the formula $\kappa_{\text{tot}} = (N\kappa_0 \kappa_b)/(N\kappa_0 + \kappa_b)$. Experimental data indicate that $N = 106 \pm 11$ [80; 87; 100].

We can then deduce the value of our coupling parameter, $\lambda_b = \kappa_b/(N\kappa_0) = 0.54 \pm 0.19$.

As we have seen, the phase diagram shown in Fig. 31(b), suggests a way to understand why $N \approx 100$. Larger values of $N$ are beneficial from the perspective of the total force developed by the system. However, reaching deep inside phase III means highly coherent response, which gets progressively more sluggish as $N$ increases. In this sense being around the would be a compromise between a high force and a high responsiveness. It follows from the developed theory that for the normal temperature the corresponding value of $N$ would be exactly around 100; for an attempt of a similar evolutionary justification for the size of titin molecule [217]. There are, of course, other functional advantages of a near-criticality associated, for instance, with diverging correlation length and the possibility of fast coherent response.

At the end of the second phase of the fast force recovery (see Section 1.2.2), the system reaches an equilibrium state characterized by the tension $T_2$ in a hard device or by the shortening $L_2$ in a soft device. The values of these parameters are naturally linked with the equilibrium tension $\langle \sigma \rangle$ in a hard device and equilibrium length $\langle z \rangle$ in a soft device. In particular, the theory predicts that in the large deformation (shortening or stretching) regimes, the tension-elongation relation must be linear, see Fig. 33. The linear stiffness in these regimes corresponds to the series arrangement of $N$ elastic elements, each one with stiffness equal to either $\kappa_1$ or $\kappa_2$ and with a series spring characterized by the stiffness $\kappa_b$. Using the classical dimensional notations—$(T, L)$ instead of the non dimensional $(\sigma, z)$—the tension elongation relation at large shortening takes the form

$$T_2(L) = \frac{\frac{\kappa_0 \kappa_2}{\kappa_0 + \kappa_2} \kappa_b}{\frac{\kappa_0 \kappa_2}{\kappa_0 + \kappa_2} + \kappa_b}(L + a)$$

In experiment, the tension $T_2$ drops to zero when a step $L_2 \simeq -14$ nm hs$^{-1}$ (nanometer per half-sarcomere) is applied to the initial configuration $L_0$. Therefore $L_0 = -a - L_2$. Since $a = 11$ nm, we obtain $L_0 = 3.2$ nm. Using a linear fit of the experimental curve shown in Fig. 5 (shortening) we finally obtain $\kappa_2 \simeq 1$ pN nm$^{-1}$.

The value of $\kappa_1$ is more difficult to determine since there are only few papers dealing with stretching [94; 218]. Based on the few available measurements, we conclude that the stiffness in stretching is ~1.5 larger than in shortening which gives $\kappa_1 \simeq 3.6$ pN nm$^{-1}$. A recent analysis of the fast force recovery confirms this estimate [207].

The last parameter to determine is the intrinsic bias of the double well potential, $v_0$, which controls the active tension in the isometric state. The tetanus of a single sarcomere in physiological conditions is of the oder of 500 pN [80; 100]. If we adjust $v_0$ to ensure that the equilibrium tension matches this value, we obtain $v_0 \simeq 50$ zJ. This energetic bias can also be interpreted as the maximum amount of mechanical work that the cross-bridge can produce during one stroke. Since the amount of metabolic energy resulting from the hydrolysis of one ATP molecule is of the order of 100 zJ we obtain a maximum efficiency around 50 % which agrees with the value currently favoured in the literature [18; 219].

*2.2.4. Kinetics.* After the values of the nondimensional parameters are identified, one can simulate numerically the kinetics of fast force recovery by exposing the mechanical system to a Langevin thermostat. For simplicity, we assume that the macroscopic variables $y$ and $z$ are fast and are always mechanically equilibrated. Such quasi-adiabatic approximation is not essential but it will allow us to operate with a single relaxation time-scale associated with the microscopic variables $x_i$. Denoting by $\eta$ the corresponding drag coefficient we construct the characteristic timescale $\tau = \eta/\kappa_0$, which will be adjusted to fit the overall rate of fast force recovery.

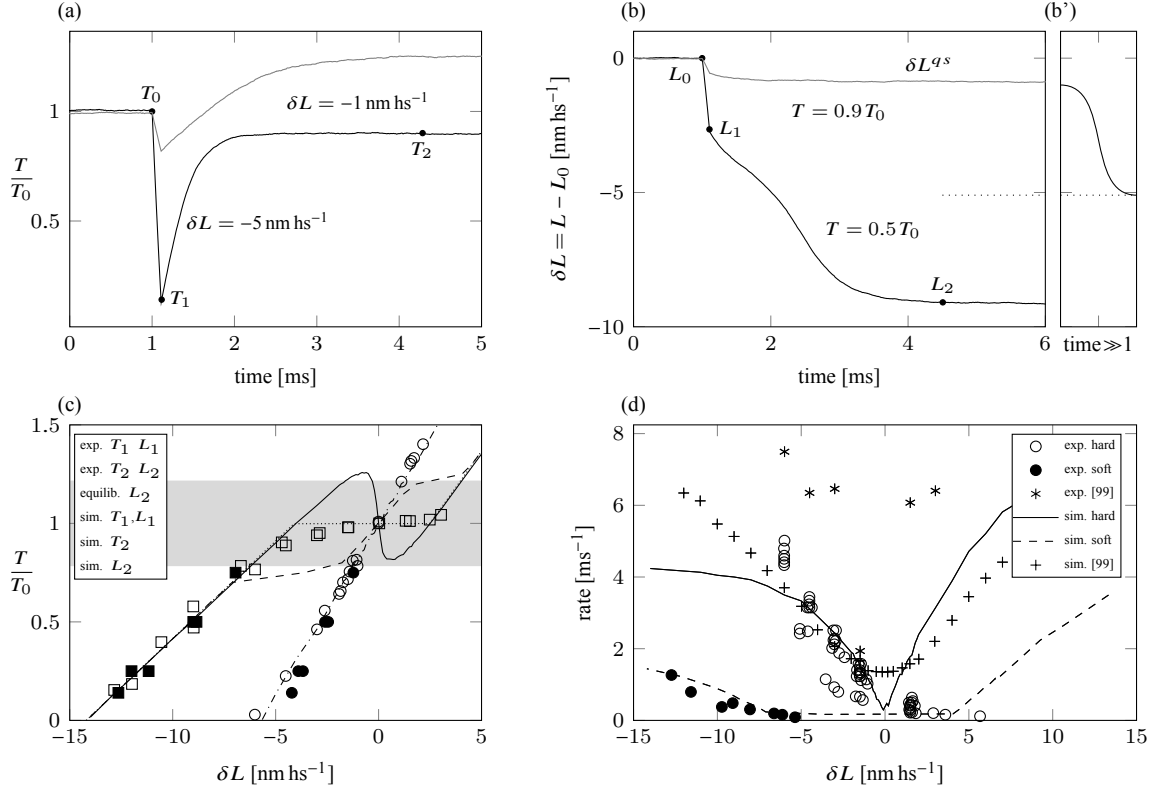

**Figure 34.** Soft-spin model compared with experimental data from Fig. 5 and 6. [(a) and (b)] Average trajectories were obtained from stochastic simulations, after the system was exposed to various load steps in hard (a) and soft (b) devices. (b') Schematic representation of the regime shown in (b) for large times illustrating eventual equilibration (dotted line). (c) Tension-elongation relation obtained from the numerical simulations (sim.) compared with experimental data (symbols, exp.); dotted line, thermal equilibrium in a soft device. (d) Comparison of the rates of recovery: crosses show the result of the chemomechanical model from Ref [207]; asterisks show the "fast component" of the recovery rate (see explanations of such fast-slow decomposition in Ref [207]). Figure adapted from Ref. [99]. Here parameters are: $\kappa_2 = 0.41\,\mathrm{pN\,nm^{-1}}$, $\kappa_1 = 1.21\,\mathrm{pN\,nm^{-1}}$, $\lambda_b = 0.72$, $\ell = -0.08\,\mathrm{nm}$, $N = 112$, $\beta = 52$ ($\kappa_0 = 2\,\mathrm{pN\,nm^{-1}}$, $a = 10\,\mathrm{nm}$, $T = 277.13\,\mathrm{K}$, $z_0 = 4.2\,\mathrm{nm\,hs^{-1}}$.

The response of the internal variables $x_i$ is governed by the non-dimensional system

$$\mathrm{d}x_i = b(x_i)\mathrm{d}t + \sqrt{2\beta^{-1}}\mathrm{d}B_i$$

where the drift is

$$b(\boldsymbol{x}, z) = -u'_{ss}(x_i) + (1+\lambda_b)^{-1}(\lambda_b z + \tfrac{1}{N}\sum x_i) - x_i,$$
$$b(\boldsymbol{x}, \sigma) = -u'_{ss}(x_i) + \sigma + N^{-1}\sum x_i - x_i$$

in a hard and a soft device, respectively. In both cases the diffusion term $\mathrm{d}B_i$ represents a standard Wiener processes.

In Fig. 34, we illustrate the results of stochastic simulations imitating fast force recovery, using the same notations as in actual experiments. The system, initially in thermal equilibrium at fixed $L_0$ (or $T_0$), was perturbed by applying fast ($\sim 100\,\mu\mathrm{s}$) length (load) steps with different amplitudes.

Typical ensemble-averaged trajectories are shown in Fig. 34[(a) and (b)] in the cases of hard and soft device, respectively. In a soft device (b) the system was not able to reach equilibrium within the realistic time scale when the applied load was sufficiently close to $T_0$, see, for instance, the curve $T = 0.9T_0$ in Fig. 34(b), where the expected equilibrium value is $L_2 = -5\,\mathrm{nm\,hs^{-1}}$. Instead, it remained trapped in a quasi-stationary (glassy) state because of the high energy barrier required to be crossed in the process of the collective power-stroke. The implied kinetic trapping, which fits the pattern of two-stage dynamics exhibited by systems with long-range interactions [211; 220; 221], may explain the failure to reach equilibrium in experiments reported in Refs. [92; 161; 222]. In the hard device case, the cooperation among the cross-bridges is much weaker and therefore the kinetics is much faster, which allows the system to reach equilibrium at experimental time scale.

A quantitative comparison of the obtained tension-elongation curves with experimental data [see Fig. 34(c)] shows that for large load steps the equilibrium tension fits the linear behavior observed in experiment as it can be expected from our calibration procedure. For near isometric tension in a soft device the model also predicts the correct interval of kinetic trapping, see the gray region in Fig. 34(c).

While the model suggests that negative stiffness should be a characteristic feature of the realistic response in a hard device

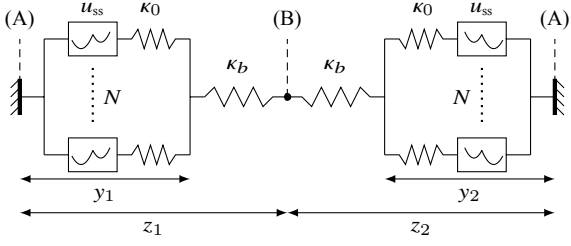

**Figure 35.** Model of a single sarcomere. A single sarcomere is located between two Z-disks (A). The M-line (B) separates the two half-sarcomeres. A single sarcomere contains two arrays of $N$ parallel cross-bridges connected by two linear springs

for a single half-sarcomere (see Fig. 31), such behavior has not been observed in experiments on whole myofibrils. Note, however, that in the model all cross bridges are considered to be identical and, in particular, it is assumed that they are attached with the same initial pre-strain. If there exists a considerable quenched disorder resulting from the randomness of the attachment/detachment positions, the effective force elongation curve will be flatter [151]. Another reason for the disappearence of the negative susceptibility may be that the actual spring stiffness inside a cross-bridge is smaller due to nonlinear elasticity [223]. One can also expect the unstable half-sarcomeres to be stabilized actively through processes involving ATP , see Refs. [158; 224] and our Section 3. The softening can be also explained by the collective dynamics of many half sarcomeres organized in series, see our Section 2.3.

The comparison of the rates of fast recovery obtained in our simulations with experimental data (see Fig. 34) shows that the soft-spin model reproduces the kinetic data in both hard and soft ensembles rather well. Note, in particular, that the rate of recovery in both shortening and stretching protocols increases with load. This is a direct consequence of the fact that the energy barriers for forward and the reverse transitions depend on the mechanical load. Instead, in the original formulation of the HS, and in most subsequent chemomechanical models, the reverse rate was kept constant and this effect was missing. In Ref. [207], the authors proposed to refine the HS model by introducing a load dependent barrier also for the reversed stroke, see the results of their modeling in Fig. 34.

### 2.3. Interacting half-sarcomeres

So far, attention has been focused on (passive) behavior of a single force generating unit, a half-sarcomere. We dealt with a zero dimensional, mean field model without spatial complexity. However, as we saw in Fig. 8(a), such elementary force generating units are arranged into a complex, spatially extended structure. Various types of cross-links in this structure can be roughly categorized as parallel or series connections.

A prevalent perspective in physiological literature is that interaction among force generating units is so strong that the mean field model of a single unit provides an adequate description of the whole myofibril. The underlying assumption is that the deformation, associated with muscle contractions, is globally affine.

To challenge this hypothesis, we consider in this Section the simplest arrangement of force generating units. We assume that the whole section of a muscle myofibril between the neighboring Z disk and M-line deforms in an affine way and treat such transversely extended unit as a (macro) half-sarcomere. The neighboring (macro) half-sarcomeres, however, will be allowed to deform in an non-affine way. The resulting model describes a chain of (macro) half-sarcomeres arranged in series and the question is whether the fast force recovery in such a chain takes place in an affine way [225].

Chain models of a muscle myofibril were considered in Refs. [2; 114; 226] where the nonaffinity of the deformation was established based on the numerical simulations of kinetics. Analytical studies of athermal chain models with bi-stable elements were conducted in Refs [227–230] where the non-affinity of the deformation (a non-Cauchy-Born behavior) was linked to phase coexistence. More recent studies of the finite temperature behavior can be found in Refs. [196; 231–234].

Here we present a simple analytical study of the equilibrium properties of a chain of half-sarcomeres which draws on Ref. [231] and allows one to understand the outcome of the numerical experiments conducted in Ref. [114].

*2.3.1. Two half-sarcomeres.* Consider first the most elementary series connection of two half-sarcomeres, each of them represented as a parallel bundle of $N$ cross-bridges. This system can be viewed as a schematic description of a single sarcomere, see Fig.35(b). To understand the mechanics of this system, we begin with the case where the temperature is equal to zero. The total (nondimensional) energy per cross bridge reads

$$v = \frac{1}{2}\left\{\frac{1}{N}\sum_{i=1}^{N}\left[u_{\text{ss}}(x_{1i}) + \frac{1}{2}(y_1 - x_{1i})^2 + \frac{\lambda_b}{2}(z_1 - y_1)^2\right]\right.$$
$$\left. + \frac{1}{N}\sum_{i=1}^{N}\left[u_{\text{ss}}(x_{2i}) + \frac{1}{2}(y_2 - x_{2i})^2 + \frac{\lambda_b}{2}(z_2 - y_2)^2\right]\right\}.$$
(2.35)

In a hard device case, when we impose the average elongation $\bar{z} = (1/2)(z_1 + z_2)$, none of the half-sarcomeres is loaded individually in either soft or hard device. In a soft device case, the applied tension $\sigma$, which we normalized by the number of cross bridges in a half-sarcomere, is the same in each half-sarcomere when the whole system is in equilibrium. The corresponding dimensionless energy per cross bridge is $w = v - \sigma\bar{z}$.

The equilibrium equations for the continuous variables $x_i$ are the same in hard and soft devices, and have up to 3 solutions,

$$\begin{cases} \hat{x}_{k1}(y_k) = (1 - \lambda_1)\,\hat{y}_k, & \text{if } x_{ki} \geq \ell, \\ \hat{x}_{k2}(y_k) = (1 - \lambda_2)\,\hat{y}_k - \lambda_1, & \text{if } x_{ki} < \ell, \\ \hat{x}_{k*} = \ell, \end{cases} \quad (2.36)$$

where again $\lambda_{1,2} = \kappa_{1,2}/(1 + \kappa_{1,2})$ and $\hat{y}_k$ denotes the equilibrium elongation of the half-sarcomere with index $k = 1, 2$.

We denote by $\boldsymbol{\xi} = \{\xi_1, \xi_2\}$, the micro-configuration of a sarcomere where the triplets $\xi_k = (p_k, r_k, q_k)$, with $p_k + q_k + r_k = 1$, characterize the fractions of cross bridges in half-sarcomere $k$ that occupy position $\hat{x}_{k1}$, $\hat{x}_{k*}$ (spinodal state) and

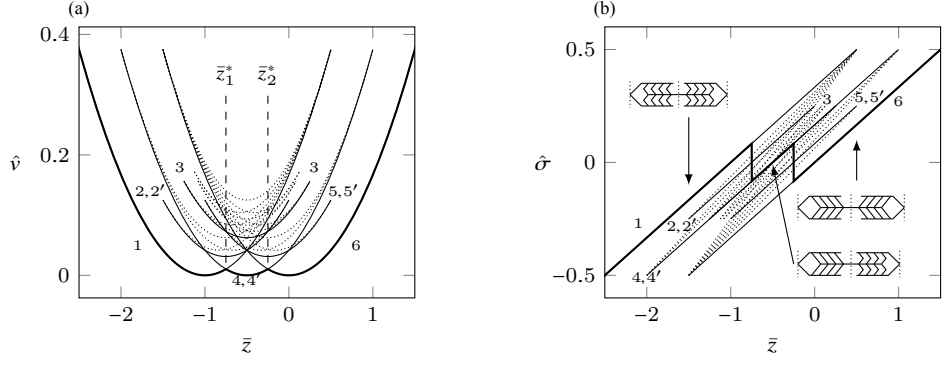

**Figure 36.** Mechanical equilibrium in a half-sarcomere chain with $N = 2$ and symmetric double well potential in a hard device. (a) Energy levels; (b) Tension-elongation relation. Solid lines, metastable states; dashed lines, unstable states; bold lines:, global minimum. Parameters: $\lambda_1 = \lambda_2 = 0.5$, $u_0 = 0$, $\ell = -0.5$, $\lambda_b = 1$.

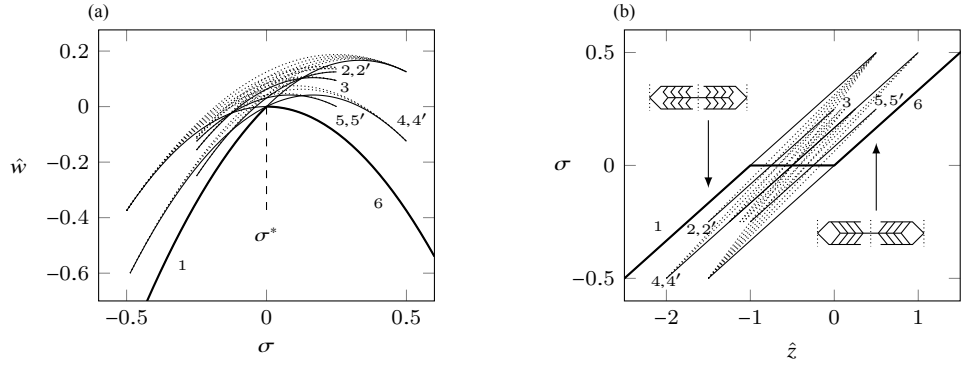

**Figure 37.** Mechanical equilibrium with $N = 2$ and symmetric double well potential in a soft device. (a) Energy levels; (b) Tension-elongation relation. Solid lines, metastable states; dashed lines, unstable states; bold lines, global minimum. Parameters are as in Fig.36.

$\hat{x}_{k2}$, respectively. For a given configuration $\xi_k$, the equilibrium value of $y_k$ is given by

$$\hat{y}_k(\xi_k, z_k) = \frac{\lambda_b \hat{z}_k + r_k \ell - p_k \lambda_2}{\lambda_b + \lambda_{xb}(\xi_k)},$$

where $\lambda_{xb}(\xi_k) = p_k\lambda_2 + q_k\lambda_1 + r_k$, is the stiffness of each half-sarcomere. The elongation of a half-sarcomere in equilibrium is $\hat{z}_k = \hat{y}_k + \sigma/\lambda_b$, where $\sigma$ is a function of $\bar{z}$ and $\boldsymbol{\xi}$ in the hard device case and a parameter in the soft device case.

To close the system of equations we need to add the equilibrium relation between the tension $\sigma$ and the total elongation $\bar{z} = (1/2)(\hat{y}_1 + \hat{y}_2) + \sigma/\lambda_b$. After simplifications, we obtain

$$\hat{\sigma}(\bar{z}, \boldsymbol{\xi}) = \lambda(\boldsymbol{\xi}) \left[ \bar{z} + \frac{1}{2}\left( \frac{p_1\lambda_2 - r_1\ell}{\lambda_{xb}(\xi_2)} + \frac{p_2\lambda_2 - r_2\ell}{\lambda_{xb}(\xi_2)} \right) \right], \quad (2.37)$$

$$\hat{z}(\sigma, \boldsymbol{\xi}) = \frac{\sigma}{\lambda(\boldsymbol{\xi})} - \frac{1}{2}\left( \frac{p_1\lambda_2 - r_1\ell}{\lambda_{xb}(\xi_1)} + \frac{p_2\lambda_2 - r_2\ell}{\lambda_{xb}(\xi_2)} \right) \quad (2.38)$$

in a hard and a soft devices, respectively, where $\lambda(\boldsymbol{\xi})^{-1} = \lambda_b^{-1} + (1/2)[\lambda_{xb}(\xi_1)^{-1} + \lambda_{xb}(\xi_2)^{-1}]$ is compliance of the whole sarcomere. The stability of a configuration $(\xi_1, \xi_2)$ can be checked by computing the Hessian of the total energy and one can show, that configurations containing cross-bridges in the spinodal state are unstable, see Refs. [212; 227] for detail.

We illustrate the metastable configurations in Fig.36 (hard device) and Fig.37 (soft device). For simplicity, we used a symmetric double well potential ($\lambda_1 = \lambda_2 = 0.5$, $\ell = -0.5$). Each metastable configuration is labeled by a number representing a micro-configuration in the form $\{(p_1, q_1), (p_2, q_2)\}$ where $p_k = 0, 1/2, 1$ (resp. $q_k = 0, 1/2, 1$) denotes the fraction of cross bridges in the post-power-stroke state (resp. pre-power-stroke) in half-sarcomere $k$. The correspondence between labels and configurations goes as follows: 1: $\{(1,0),(1,0)\}$ – 2 and 2': $\{(1,0),(\frac{1}{2},\frac{1}{2})\}$ and $\{(\frac{1}{2},\frac{1}{2}),(1,0)\}$ – 3: $\{(\frac{1}{2},\frac{1}{2}),(\frac{1}{2},\frac{1}{2})\}$ – 4 and 4': $\{(1,0),(0,1)\}$ and $\{(0,1),(1,0)\}$ – 5 and 5': $\{(\frac{1}{2},\frac{1}{2}),(0,1)\}$ and $\{(0,1),(\frac{1}{2},\frac{1}{2})\}$ – 6: $\{(0,1),(0,1)\}$. For instance the label 2': $\{(\frac{1}{2},\frac{1}{2}),(1,0)\}$ corresponds to a configuration where in the first half-sarcomere, half of the cross bridges are in post-power-stroke and another half are in pre-power-stroke; in the second half-sarcomere, all the cross bridges are in post-power-stroke. In the hard device case (see Fig.36) the system, following the global minimum path (bold line), evolves through non affine states 4 $\{(1,0),(0,1)\}$ and 4' $\{(0,1),(1,0)\}$, where one half-sarcomere is fully in pre-power-stroke, and the other one is fully in post-power-stroke. This path is marked by two transitions located at $\bar{z}_1^*$ and $\bar{z}_2^*$ see Fig.36(a).

The inserted sketches in Fig.36 (b) show a single sarcomere in the 3 configurations encountered along the global minimum path. Note that along the two affine branches, where the sarcomere is in affine state (1 and 6), the M-line (see the middle vertical dashed line) is in the middle of the structure. Instead, in the non-affine state (branch 4), the two half-sarcomeres are not equally stretched, and the M-line is not positioned in the center of the sarcomere. As a result of the (spontaneous) symmetry breaking, the M-line can be shifted in any of the two possible directions to form either configuration 4 or 4', see also Ref. [225]. In the soft device case [see Fig. 37], the system following the global minimum path never explores non-affine states. Instead both half-sarcomeres undergo a full unfolding transition at the same threshold tension $\sigma^*$.

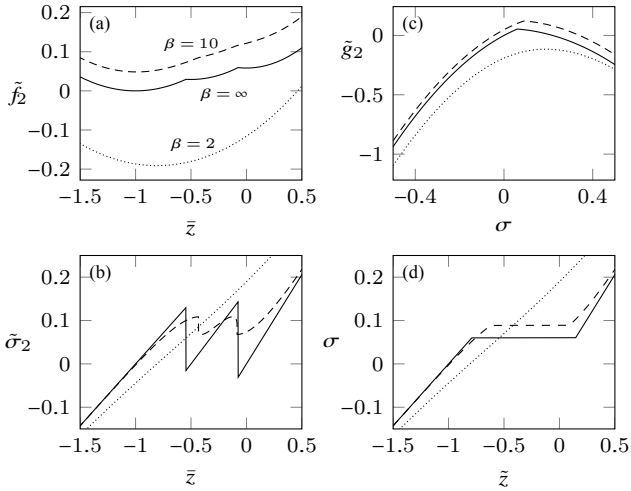

**Figure 38.** Equilibrium response of a single sarcomere in the thermodynamic limit. [(a) and (b)] Hard device; [(c) and (d)] soft device. [(a) and (c)] Gibbs and Helmholtz free energy; [(b) and (d)] corresponding tension-elongation . Parameters are, $\lambda_1 = 0.7$, $\lambda_2 = 0.4$, $\ell = -0.3$, $\lambda_b = 1$.

If the temperature is different from zero we need to compute the partition functions

$$Z_2(\bar{z}, \beta) = \int \exp\left[-2\beta N v(\bar{z}, \boldsymbol{x})\right] \delta(z_1 + z_2 - 2\bar{z}) \, d\boldsymbol{x} \quad (2.39)$$

$$Z_2(\sigma, \beta) = \int \exp\left[-2\beta N w(\sigma, \boldsymbol{x})\right] d\boldsymbol{x}, \quad (2.40)$$

in a hard and a soft device, respectively, where again $\beta = (\kappa a^2/(k_B T)$. The corresponding free energies are $\tilde{f}_2(\bar{z}, \beta) = -(1/\beta) \log[Z_2(\bar{z})]$ and $\tilde{g}_2(\sigma, \beta) = -(1/\beta) \log[Z_2(\sigma)]$.

The explicit expressions of these free energies can be obtained in the thermodynamic limit $N \to \infty$, but they are too long to be presented here, see Refs. [212; 231] for more details. We illustrate the results in Fig.38 where we show both, the energies and the tension-elongation isotherms.

We see that a sarcomere exhibits different behavior in the two loading conditions. In particular, the Gibbs free energy remains concave in the soft device case for all temperatures while the Helmholtz free energy becomes non-convex at low temperatures in the hard device case. Non-convexity of the Helmholtz free energy results in non-monotone tension-elongation relations with the developments of negative stiffness.

It is instructive to compare the obtained non-affine tension-elongation relations with the ones computed under the assumption that each half-sarcomere is an elementary constitutive element with a prescribed tension-elongation relation. We suppose that such a relation can be extracted from the response of a half-sarcomere in either soft or hard device

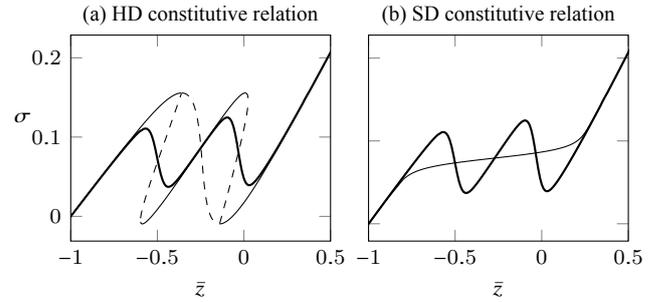

**Figure 39.** Tension-elongation relations for a sarcomere in a hard device. Thick lines: equilibrium tension-elongation relations based on the computation of the partition function (2.39). Thin lines: response of two half-sarcomere in series, each one endowed with the constitutive relation illustrated in Fig. 33(b). (a) Hard device constitutive law. (b) Soft device constitutive law, see Ref. [212] for more details. Parameters are: $\lambda_1 = 0.7$, $\lambda_2 = 0.4$, $\ell = -0.3$, $N = 10$, $\beta = 20$ and $\lambda_b = 1$.

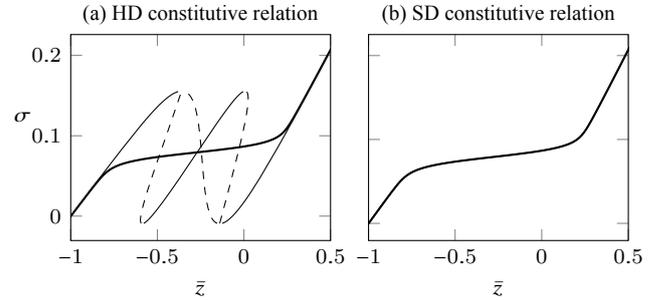

**Figure 40.** Tension-elongations for a sarcomere in a soft device. Thick lines: equilibrium tension-elongation relations based on the computation of the partition function (2.40). Thin lines: response of two half-sarcomere in series, each one endowed with the constitutive relation illustrated in Fig. 33(d). (a) Hard device constitutive law. (b) Soft device constitutive law, see Ref. [212] for more details. Parameters are as in Fig. 39.

which allows us to use expressions obtained earlier, see Fig. 33.

The hard device case is presented in Fig. 39. With thick lines we show the equilibrium tension-elongation relation while thin lines correspond to the behavior of the two phenomenologically modeled half-sarcomeres in series exhibiting each either soft or hard device constitutive behavior. Note that if the chosen constitutive relation corresponds to the hard device protocol [illustrated in Fig 33(b)], we obtain several equilibrium states for a given total elongation which is a result of the imposed constitutive constraints, see Fig. 39(a). The global minimum path predicted by the "constitutive model" shows discontinuous transitions between stable branches which resemble continuous transitions along the actual equilibrium path. If instead we use the soft device constitutive law for the description of individual half-sarcomeres [illustrated in Fig. 33(d)], the tension-elongation response becomes monotone and is therefore completely unrealistic, see Fig. 39(b). We reiterate that in both comparisons, the misfit is due to the fact that in a fully equlibrated sarcomere none of the half-sarcomeres is loaded in either soft or hard device. It would be interesting to show that a less schematic system of this type can reproduce non-affinities observed experimentally [235].

In Fig. 40 we present the result of a similar analysis for a sarcomere loaded in a soft device. In this case, if the "constitutive model" is based on the hard device tension-elongation relations [from Fig. 33(b)], we obtain the same (constrained) metastable states as in the previous case, see Fig. 39(a), thin lines. This means, in particular, that the response contains jumps while the actual equilibrium response is monotone, see Fig. 40(a). Instead, if we take the soft device tension-elongation relation as a "constitutive model", we obtain the correct overall behavior, see Fig. 40(b). This is expected since in the (global) soft device case both half-sarcomeres are effectively loaded in the same soft device and the overall response is affine.

The fact that the model generates different constitutive relations in soft and hard device, and that another, for instance mixed, loading conditions may be associated with yet other constitutive relations, makes the task of building a macroscopic continuum theory of skeletal muscles rather challenging. One conclusion may be that the approach based on local constitutive relation may not be adequate for such a medium, dominated by long range interactions, and one may have to search instead a nonlocal constitutive closure for the system of balance laws. Such closure would then involve integral equations involving the kernels which depend on both the size and the shape of the domain.

*2.3.2. A chain of half-sarcomeres.* Next, consider the behavior of a chain of $M$ half-sarcomeres connected in series. As before, each half-sarcomere is modeled as a parallel bundle of $N$ cross bridges.

We first study the mechanical response of this system at zero temperature. Introduce $x_{ki}$—the continuous degrees of freedom characterizing the state of the cross bridges in half-sarcomere $k$, $y_k$—the position of the backbone that connects all the cross bridges of the half-sarcomere $k$ and $z_k$—the total elongation of the half-sarcomere $k$. The total energy (per cross bridge) of the chain takes the form

$$v(\boldsymbol{x},\boldsymbol{y},\boldsymbol{z}) = \frac{1}{MN}\sum_{k=1}^{M}\left\{\sum_{i=1}^{N}\left[u_{\text{ss}}(x_{ki}) + \frac{1}{2}(y_k - x_{ki})^2\right.\right.$$
$$\left.\left. + \lambda_b \frac{1}{2}(z_k - y_k)^2\right]\right\}, \quad (2.41)$$

where $\boldsymbol{x} = \{x_{ki}\}$, $\boldsymbol{y} = \{y_k\}$ and $\boldsymbol{z} = \{z_k\}$. In the hard device, the total elongation of the chain is prescribed: $M\bar{z} = \sum_{k=1}^{M} z_k$, where $\bar{z}$ is the average imposed elongation (per half-sarcomere). In the soft device case, the tension $\bar{\sigma}$ is imposed and the energy of the system also includes the energy of the loading device $w = v - \bar{\sigma}\sum_{k=1}^{M} z_k$.

We again characterize the microscopic configuration of each half-sarcomere $k$ by the triplet $\xi_k = (p_k, q_k, r_k)$, denoting as before the fraction of cross bridges in each of the wells and in the spinodal point, with $p_k + q_k + r_k = 1$ for all $1 \leq k \leq M$. The vector $\boldsymbol{\xi} = (\xi_1, \ldots, \xi_M)$ then characterizes the configuration of the whole chain.

In view of the complexity of the ensuing energy landscape, here we characterize only a subclass of metastable configurations describing homogeneous (affine) states of individual half-sarcomeres. More precisely, we limit our attention to configurations with $q_k = 0$, $p_k = 1, 0$ and $r_k = 1, 0$ for all $1 \leq k \leq M$. In this case, a single half-sarcomere can be characterized by a spin variable $m_k = 1, 0$.

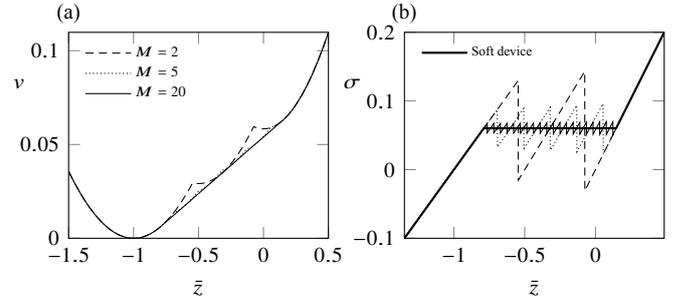

**Figure 41.** Global minimum of the (hard device) energy in the zero temperature limit ($\beta \to \infty$) for a sarcomere chain with different $M$: (a) - energies; (b) - tension-elongation relations. In (b) the solid line represents the tension-elongation relation in a soft device. Parameters are: $\lambda_1 = 0.7, \lambda_2 = 4, \ell = -0.3$.

The resulting equilibrium tension-elongation relations in hard and soft devices take the form

$$\hat{\sigma}(\bar{z},\boldsymbol{m}) = \left[\frac{1}{\lambda_b} + \frac{1}{M}\sum_{k=1}^{M}\frac{1}{m_k\lambda_2 + (1-m_k)\lambda_1}\right]^{-1}$$
$$\times \left[\bar{z} + \frac{1}{M}\sum_{k=1}^{M}\frac{m_k\lambda_1}{m_k\lambda_2 + (1-m_k)\lambda_1}\right], \quad (2.42)$$

$$\hat{z}(\bar{\sigma},\boldsymbol{m}) = \left[\frac{1}{\lambda_b} + \frac{1}{M}\sum_{k=1}^{M}\frac{1}{m_k\lambda_2 + (1-m_k)\lambda_1}\right]\bar{\sigma}$$
$$- \frac{1}{M}\sum_{k=1}^{M}\frac{m_k\lambda_2}{m_k\lambda_2 + (1-m_k)\lambda_1}, \quad (2.43)$$

where $\boldsymbol{m} = (m_1, \ldots, m_M)$.

In Fig. 41 we show the energy and the tension-elongation relation for the system following the global minimum path in a hard device. Observe that the tension-elongation relation contains a series of discontinuous transitions as the order parameter $M^{-1}\sum m_k$ increases monotonously from 0 to 1 and their number increases with $M$ while their size decreases. In the limit $M \to \infty$, the relaxed (minimum) energy is convex but not strictly convex, see the interval where the energy depends linearly on the elongation for the case $M = 20$ in Fig. 41(a), see also Refs. [227; 236]. The corresponding tension-elongation curves [see Fig. 41(b)] exhibit a series of transitions. In contrast to the case of a single half sarcomere, the limiting behavior of a chain is the same in the soft and hard devices (see the thick line). The obtained analytical results are in full agreement with the numerical simulations reported in Refs. [114; 164; 235; 237].

Fig. 42 illustrates the distribution of elongations of individual half-sarcomere in a hard device case as the system evolves along the global minimum path. One can see that when deformation becomes non-affine the population of half-sarcomere splits into 2 groups: one group is stretched at the level above average (top trace above diagonal) and the other one at the level below average (bottom trace below diagonal). The numbers beside the curves indicates the amount of half-sarcomeres in each group. In the soft device case, the system always remains in the affine state: all half-sarcomeres change conformation at the same moment and therefore the system stays on the diagonal (the dashed lines) in Fig. 42.

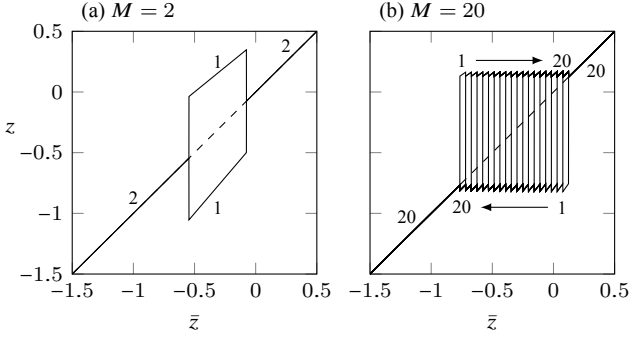
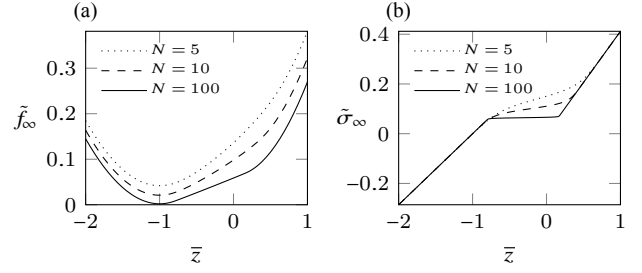

**Figure 42.** Elongation of half-sarcomeres along the global minimum path for $M = 2$ (a) and $M = 20$ (b) in a hard device. Upper branch, pre-power-stroke half-sarcomeres; lower branch, post-power-stroke half-sarcomeres. Numbers indicate how many half-sarcomere are in each branch at a given $\bar{z}$. Dashed lines, Soft device response. Parameters are: $\lambda_1 = 0.7$, $\lambda_2 = 0.4$, $\ell = -0.3$.

**Figure 43.** Influence of the parameter $N$ on the equilibrium response of an infinitely long chain ($M \to \infty$) in a hard device: (a) free energy; (b) tension-elongation relation. The asymmetry of the tension curve is a consequence of the asymmetry of the double well potential. Parameters are: $\lambda_1 = 0.7$, $\lambda_2 = 0.4$, $\ell = -0.3$, $\lambda_b = 1$, and $\beta = 10$.

Assume now that the temperature is different from zero. The partition function for the chain in a soft device can be obtained as the product of individual partition functions:

$$Z_M(\bar{\sigma}, \beta) = [Z_s(\bar{\sigma}, \beta)]^M$$
$$= \left[ \sqrt{\frac{2\pi}{N\beta\lambda_b}} \int \exp\left[-\beta N g(\bar{\sigma}, x, \beta)\right] dx \right]^M,$$

which reflects he fact that the half-sarcomeres in this setting are independent. In the hard device, the analysis is more involved because of the total length constraint. In this case we need to compute

$$Z_M(\bar{z}, \beta) = \int \exp\left[-\beta NM v(\bar{z}, x)\right] \delta\left[\frac{1}{M}\sum z_k - \bar{z}\right] dx \quad (2.44)$$

A semi-explicit asymptotic solution can be obtained for the hard device case in the limit $\beta \to \infty$ and $M \to \infty$. Note first, that the partition function depends only on the "average magnetization" $m$ – the fraction of half-sarcomeres in post-power-stroke conformation. At $MN \to \infty$ we obtain asymptotically (see Refs. [212; 231] for the details)

$$Z_M(\bar{z}, \beta) \approx C \frac{\phi(m^*) \exp\left[-\beta MN \Psi(m^*; \bar{z}, \beta)\right]}{\left[\beta MN \left.\partial_m^2 \Psi(m; \bar{z}, \beta)\right|_{m=m^*}\right]^{\frac{1}{2}}}, \quad (2.45)$$

where $C = \left(\frac{2\pi}{\beta}\right)^{\frac{(N+2)M-1}{2}} N^{\frac{1}{2}-M}$. Using the notations $\mu_{1,2} = (\lambda_{1,2}\lambda_b)/(\lambda_{1,2}+\lambda_b)$, we can now write the expression for the marginal free energy at fixed $m$ in the form

$$\Psi(m; \bar{z}, \beta) = \frac{1}{2}\left[\frac{m}{\mu_2} + \frac{1-m}{\mu_1}\right]^{-1}(\bar{z}+m)^2 + (1-m)v_0$$
$$- \frac{1}{2\beta}\left[m\log(1-\lambda_2) + (1-m)\log(1-\lambda_1)\right]$$
$$+ \frac{1}{\beta N}\left[m\log(m) + (1-m)\log(1-m)\right.$$
$$\left. + \frac{m}{2}\log(\lambda_2\lambda_b) + \frac{1-m}{2}\log(\lambda_1\lambda_b)\right], \quad (2.46)$$

where $\phi(m) = \left\{[m/\mu_2 + (1-m)/\mu_1][m(1-m)]\right\}^{-\frac{1}{2}}$. Here $m^*$ is the minimum of $\Psi$ in the interval $]0, 1[$. A direct computation of the second derivative of (2.46) with respect to $m$ shows that $\Psi$ is always convex. In other words, our assumption that individual half-sarcomeres respond in an affine way, implies that the system does not undergo a phase transition in agreement with what is expected for a 1D system with short range interactions.

Now we can compute the Helmholtz free energy and the equilibrium tension-elongation relation for a chain in a hard device

$$\tilde{f}_\infty(\bar{z}, \beta) = \Psi(m^*; \bar{z}, \beta), \quad (2.47)$$
$$\tilde{\sigma}_\infty(\bar{z}, \beta) = \left(\frac{m^*}{\mu_2} + \frac{1-m^*}{\mu_1}\right)^{-1}(\bar{z}+m^*). \quad (2.48)$$

In the case of a soft device, the Gibbs free energy and the corresponding tension-elongation relation are simply the re-scaled versions of the results obtained for a single half-sarcomere, see Section 2.2.

In Fig. 43 we illustrate a typical equilibrium behavior of a chain in a hard device. The increase of temperature enhances the convexity of the energy, as in the case of a single half-sarcomere, however, when the temperature decreases we no longer see the negative stiffness. Instead, when $N$ is sufficiently large, we see a tension-elongation plateau similar to what is observed in experiments on myofibrils, see Fig. 43(b).

The obtained results can be directly compared with experimental data. Consider, for instance, the response of a chain with $M = 20$ half-sarcomeres submitted to a rapid length step. The equilibrium model with realistic parameters predicts in this case a tension-elongation plateau close to the observed $T_2$ curve, see dashed line in 44(a). Our numerical experiments, however, could not reproduce the part of this plateau in the immediate vicinity of the state of isometric contractions. This may mean that even in the chain placed in a hard device, individual half-sarcomeres end up being loaded in a mixed device and can still experience kinetic trapping. Our stochastic simulations for a chain in a soft device reproduce the whole trapping domain around the state of isometric contractions, see Fig. 44(a).

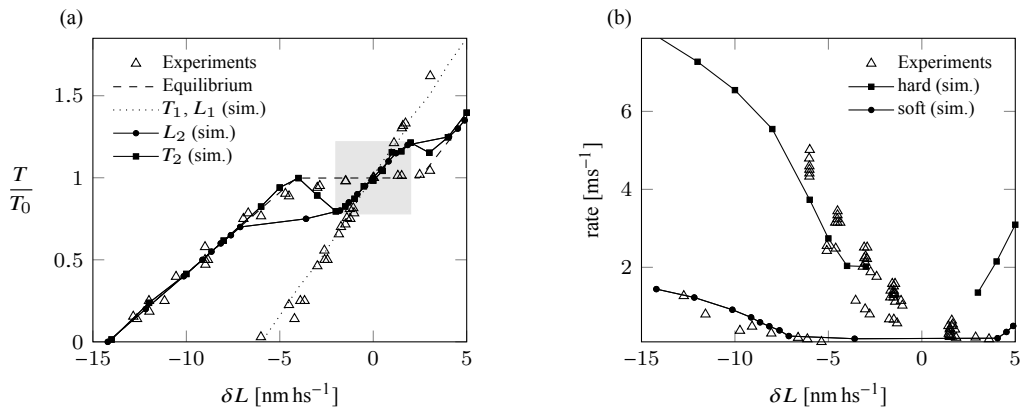

**Figure 44.** Quick recovery response of a chain with $M = 20$ half-sarcomeres. (a) Tension elongation relation obtained with $M = 20$ in a hard device (circles) and in a soft device (squares) compared with the same experiments as in Fig. 5 (triangles). (b) Corresponding rates in hard (circles) and soft (squares) devices compared with experimental data from Fig. 6 (triangles).

The computed rate of the quick recovery for the chain is shown in Fig. 44(b). We see that the model is able to capture quantitatively the difference between the two loading protocols. However, the hard device response of the chain (see squares) is more sluggish than in the case of a single half-sarcomere. Once again, we see an interval around the state of isometric contractions where our system cannot reach its equilibrium state at the experimental time scale. Note, however, that the rate of relaxation to equilibrium increases with both stretching and shortening, saturating for large applied steps as it was experimentally observed in Ref. [207].

## 3. Active rigidity

As we have seen in Section 2.2.3, individual half-sarcomeres with attached cross-bridges operate in an unstable (spinodal) or near critical regime, see Refs. [99; 159]. The analysis in Section 2.3 shows that it warrants strain inhomogeneities at the level of a myofibril, see also Refs. [227; 238]. However, the implied non-affinity has not been observed in experiment. Purely entropic stabilization is excluded in this case because the temperature alone is not sufficiently high to ensure positive stiffness of individual half-sarcomeres [114].

Here we discuss a possibility that the homogeneity of the myofibril configuration is due to active stabilization of individual half-sarcomeres [224]. We conjecture that metabolic resources are used to modify the mechanical susceptibility of the system and to stabilize configurations that would not have existed in the absence of ATP hydrolysis [239–241].

We present the simplest model showing that active rigidity can emerge through resonant non-thermal excitation of molecular degrees of freedom. The idea is to immitate the inverted Kapitza pendulum [242], aside from the fact that in biological systems the inertial stabilization has to be replaced by its overdamped analog. The goal is to show that a macroscopic mechanical stiffness can be controlled at the microscopic scale by a time correlated noise which in biological setting may serve as a mechanical representation of a nonequilibrium chemical reaction [243].

### 3.1. Mean field model

To justify the prototypical model with one degree of freedom, we motivate it using the modeling framework developed above.

Suppose that we model a half-sarcomere by a parallel array of $N$ cross-bridges attached to a single actin filament following Section 2.2. We represent again attached cross bridges as bistable elements in series with linear springs but now assume additionally that there is a nonequilibrium driving provided through stochastic rocking of the bi-stable elements. More specifically, we replace the potential $u_{ss}(x)$ for individual cross-bridges by $u_{ss}(x) - xf(t)$, where $f(t)$ is a correlated noise with zero average simulating out of equilibrium environment, see Ref. [244] for more details.

If such a half-sarcomere is subjected to a time dependent deterministic force $f_{ext}(t)$, the dynamics can be described by the following system of nondimensional Langevin equations

$$\dot{x}_i = -\partial_{x_i} W + \sqrt{2D}\xi(t),$$
$$\nu \dot{y} = -\partial_y W,$$
(3.1)

where $\xi(t)$ a white noise with the properties $\langle \xi(t) \rangle = 0$, and $\langle \xi(t_1)\xi(t_2) \rangle = \delta(t_2 - t_1)$. Here $D$ is a temperature-like parameter, the analog of the parameter $\beta^{-1}$ used in previous sections. The (backbone) variable $y$, coupled to $N$ fast soft-spin type variables $x_i$ through identical springs with stiffness $\kappa_0$, is assumed to be macroscopic, deterministic and slow due to the large value of the relative viscosity $\nu$. We write the potential energy in the form $W = \sum_{i=1}^{N} v(x_i, y, t) - f_{ext}y$, where $v(x, y, t)$ is the energy (2.30) with a time dependent tilt in $x$ and the function $f_{ext}(t)$ is assumed to be slowly varying. The goal now is to average out fast degrees of freedom $x_i$ and to formulate the effective dynamics in terms of a single slow variable $y$.

Note, that the equation for $y$ can be re-written as

$$\frac{\nu}{N}\dot{y} = \kappa_0 \left( \frac{1}{N} \sum_{i=1}^{N} x_i - y \right) + \frac{f_{\text{ext}}}{N}, \quad (3.2)$$

which reveals the mean field nature of the interaction between $y$ and $x_i$. If $N$ is large, we can replace $\frac{1}{N}\sum_{i=1}^{N} x_i$ by $\langle x \rangle$ using the fact that the variables $x_i$ are identically distributed and exchangeable [245]. Denoting $\nu_0 = \nu/N$ and $g_{\text{ext}} = f_{\text{ext}}/(\kappa_0 N)$ and assuming that these variables remain finite in the limit $N \to \infty$, we can rewrite the equation for $y$ in the form

$$\nu_0 \dot{y} = \kappa_0 [(\langle x \rangle - y) + g_{\text{ext}}(t)].$$

Assume now for determinacy that the function $f_{\text{ext}}(t)$ is periodic and choose its period $\tau_0$ in such a way that $\Gamma = \nu_0/\kappa_0 \gg \tau_0$. We then split the force $\kappa_0(\langle x \rangle - y)$ acting on $y$ into a slow component $\kappa_0 \psi(y) = \kappa_0(\overline{\langle x \rangle} - y)$ and a slow-fast component $\kappa_0 \phi(y,t) = \kappa_0(\langle x \rangle - \overline{\langle x \rangle})$ where $\overline{\langle x \rangle} = \lim_{t\to\infty}(1/t)\int_0^t \int_{-\infty}^{\infty} x\rho(x,t)\,dx\,dt$, and $\rho(x,t)$ is the probability distribution for the variable $x$. We obtain $\Gamma \dot{y} = \psi(y) + \phi(y,t) + g_{\text{ext}}$ and the next step is to average this equation over $\tau$.

To this end we introduce a decomposition $y(t) = z(t) + \zeta(t)$, where $z$ is the averaged (slow) motion and $\zeta$ is a perturbation with time scale $\tau$. Expanding our dynamic equation in $\zeta$, we obtain,

$$\Gamma(\dot{z} + \dot{\zeta}) = \psi(z) + \partial_z \psi(z)\zeta + \phi(z,t) + \partial_z \phi(z,t)\zeta + g_{\text{ext}}. \quad (3.3)$$

Since $g_{\text{ext}}(t) \simeq \tau_0^{-1} \int_t^{t+\tau_0} g_{\text{ext}}(u)\,du$, we obtain at fast time scale $\Gamma \dot{\zeta} = \phi(z,t)$, see Ref. [246] for the general theory of these type of expansions. Integrating this equation between $t_0$ and $t \leq t_0 + \tau_0$ at fixed $z$ we obtain $\zeta(t) - \zeta(t_0) = \Gamma^{-1} \int_{t_0}^t \phi(z(t_0), u)du$ and since $\phi$ is $\tau_0$ periodic with zero average, we can conclude that $\zeta(t)$ is also $\tau_0$ periodic with zero average. If we now formally average (3.3) over the fast time scale $\tau_0$, we obtain $\Gamma \dot{z} = \psi(z) + r + g_{\text{ext}}$, where $r = (\Gamma \tau_0)^{-1} \int_0^\tau \int_0^t \partial_z \phi(z,t)\phi(z,u)\,du dt$. Given that both $\phi(z,t)$ and $\partial_z \phi(z,t)$ are bounded, we can write $|r| \leq (\tau_0/\Gamma)c \ll 1$, where the "constant" $c$ depends on $z$ but not on $\tau_0$ and $\Gamma$. Therefore, if $N \gg 1$ and $\nu/(\kappa_0 N) \gg \tau_0$, the equation for the coarse grained variable

$$z(t) = \tau_0^{-1} \int_t^{t+\tau_0} y(u)\,du$$

can be written in terms of an effective potential

$$(\nu/N)\dot{z} = -\partial_z F + f_{\text{ext}}/N.$$

To find the effective potential we need to compute the primitive of the averaged tension $F(z) = \int^z \sigma(s)\,ds$, where $\sigma(y) = \kappa_0[y - \overline{\langle x \rangle}]$. The problem reduces to the study of the stochastic dynamics of a variable $x(t)$ described by a dimensionless Langevin equation

$$\dot{x} = -\partial_x w(x,y,t) + \sqrt{2D}\xi(t). \quad (3.4)$$

The potential $w(x,y,t) = w_p(x,t) + v_e(x,y)$ is the sum of two components: $w_p(x,t) = u_{ss}(x) - xf(t)$, mimicking an out of equilibrium environment and $v_e(x,y) = (\kappa_0/2)(x-y)^2$, describing the linear elastic coupling of the "probe" with a "measuring device" characterized by stiffness $\kappa_0$. We assume that the energy is supplied to the system through a time-correlateded rocking force $f(t)$ which is characterized by an amplitude $A$ and a time scale $\tau$. To have analytical results, we further assume that the potential $u_{ss}(x)$ is bi-quadratic, $u_{ss}(x) = (1/2)(|x| - 1/2)^2$. Similar framework has been used before in the studies of directional motion of molecular motors [247].

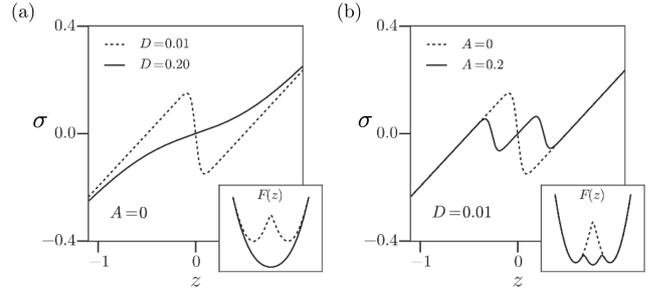

**Figure 45.** Tension elongation curves $\sigma(z)$ in the case of periodic driving (adiabatic limit). The equilibrium system ($A = 0$) is shown in (a) and and out-of-equilibrium system ($A \neq 0$) - in (b). The insets show the effective potential $F(z)$. Here $\kappa_0 = 0.6$. Adapted from Ref. [224].

The effective potential $F(z)$ can be viewed as a non-equilibrium analog of the free energy [248–251]. While in our case, the mean-field nature of the model ensures the potentiality of the averaged tension, in a more general setting, the averaged stochastic forces may lose their gradient structure and even the effective "equations of state" relating the averaged forces with the corresponding generalized coordinates may not be well defined [252–257].

### 3.2. Phase diagrams

Suppose first that the non-equilibrium driving is represented by a periodic (P), square shaped external force

$$f(t) = A(-1)^{n(t)} \text{ with } n(t) = \lfloor 2t/\tau \rfloor, \quad (3.5)$$

where the brackets denote the integer part. The Fokker-Planck equation for the time dependent probability distribution $\rho(x,t)$ reads

$$\partial_t \rho = \partial_x [\rho \, \partial_x w(x,t) + D \partial_x \rho]. \quad (3.6)$$

Explicit solution of (3.6) can be found in the adiabatic limit when the correlation time $\tau$ is much larger than the escape time for the bi-stable potential $u_{ss}$ [132; 258]. The idea is that the time average of the steady state probability can be computed from the mean of the stationary probabilities with constant driving force (either $f(t) = A$ or $f(t) = -A$).

The adiabatic approximation becomes exact in the special case of an equilibrium system with $A = 0$, when the stationary probability distribution can be written explicitly

$$\rho_0(x) = Z^{-1} \exp[-\tilde{v}(x)/D].$$

Here $Z = \int_{-\infty}^{\infty} \exp(-\tilde{v}(x)/D)dx$, and $\tilde{v}(x,z) = (1/2)(|x| - 1/2)^2 + (\kappa_0/2)(x-z)^2$. The tension elongation curve $\sigma(z)$ can then be computed analytically, since we know $\overline{\langle x \rangle} = \langle x \rangle = \int_{-\infty}^{\infty} x\rho_0(x)\,dx$. The resulting curve and the corresponding

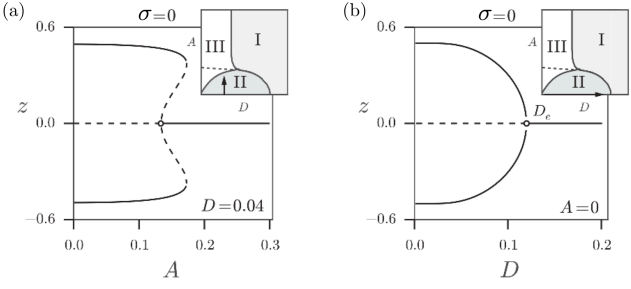
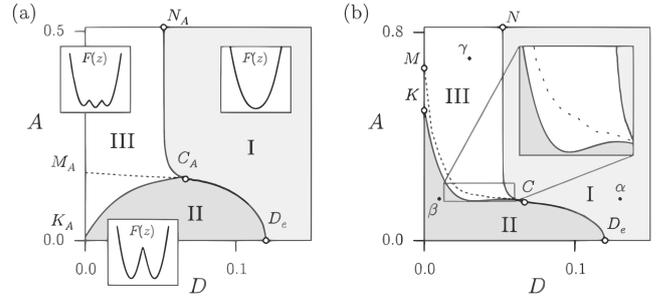

**Figure 46.** The parameter dependence of the roots of the equation $\sigma(z) = 0$ in the adiabatic limit: (a) fixed $D = 0.04$ and varying $A$, first order phase transition [line $C_A - M_A$ in Fig. 47 (a)]; (b) fixed $A = 0$ and varying $D$, second order phase transition [line $D_e - C_A$ in Fig. 47 (a)]. The dashed lines correspond to unstable branches. Here $\kappa = 0.6$. Adapted from Ref. [224].

**Figure 47.** Phase diagram in $(A, D)$ plane showing phases I,II and III: (a) - adiabatic limit, (b) - numerical solution at $\tau = 100$ (b). $C_A$ is the tri-critical point, $D_e$ is the point of a second order phase transion in the passive system. The "Maxwell line" for a first order phase transition in the active system is shown by dots. Here $\kappa_0 = 0.6$. Adapted from Ref. [224].

potential $F(z)$ are shown in Fig. 45(a). At zero temperature the equilibrium system with $A = 0$ exhibits negative stiffness at $z = 0$ where the effective potential $F(z)$ has a maximum (spinodal state). As temperature increases we observe a standard entropic stabilization of the configuration $z = 0$, see Fig. 45(a).

By solving equation $\partial_z \sigma|_{z=0} = 0$, we find an explicit expression for the critical temperature $D_e = r/[8(1 + \kappa_0)]$ where $r$ is a root of a transcendental equation $1 + \sqrt{r/\pi} e^{-1/r}/[1 + \mathrm{erf}(1/\sqrt{r})] = r/(2\kappa_0)$. The behavior of the roots of the equation $\sigma(z) = -\kappa_0(\langle x \rangle - z) = 0$ at $A = 0$ is shown in Fig. 46 (b) which illustrates a second order phase transition at $D = D_e$.

In the case of constant force $f \equiv A$ the stationary probability distribution is also known [259]

$$\rho_A(x) = Z^{-1} \exp\left[-(\tilde{v}(x) - Ax)/D\right],$$

where again $Z = \int_{-\infty}^{\infty} \exp(-\tilde{v}(x)/D)\,\mathrm{d}x$. In adiabatic approximation we can write the time averaged stationary distribution in the form, $\rho_{\mathrm{Ad}}(x) = \frac{1}{2}[\rho_A(x) + \rho_{-A}(x)]$, which gives $\overline{\langle x \rangle} = \frac{1}{2}\left[\langle x \rangle(A) + \langle x \rangle(-A)\right]$.

The force-elongation curves $\sigma(z)$ and the corresponding potentials $F(z)$ are shown in Fig. 45 (b). We see the main effect: as the degree of non-equilibrium, characterized by $A$, increases, not only the stiffness in the state $z = 0$, where the original double well potential $u_{\mathrm{ss}}$ had a maximum, changes from negative to positive, as in the case of entropic stabilization, but we also see that the effective potential $F(z)$ develops around this point a new energy well.

We interpret this phenomenon as the emergence of active rigidity because the new equilibrium state becomes possible only at a finite value of the driving parameter $A$, while the temperature $D$ can be arbitrarily small. The behavior of the roots of the equation $\sigma(z) = -\kappa_0(\overline{\langle x \rangle} - z) = 0$ at $A \neq 0$ is shown in Fig. 46(a) which now illustrates a first order phase transition.

The full steady state regime map (dynamic phase diagram) summarizing the results obtained in adiabatic approximation is presented in Fig. 47 (a). There, the "paramagnetic" phase I describes the regimes where the effective potential $F(z)$ is convex, the "ferromagnetic" phase II is a bi-stability domain where the potential $F(z)$ has a double well structure and, finally, the "Kapitza" phase III is where the function $F(z)$ has three convex sections separated by two concave (spinodal) regions. We interpret the boundary $C_A - D_e$ separating phases I and II as a line of (zero force) second order phase transitions and the dashed line $C_A - M_A$ as a Maxwell line for the (zero force) first order phase transition, see Fig. 46. Then $C_A$ can be interpreted as a tri-critical point.

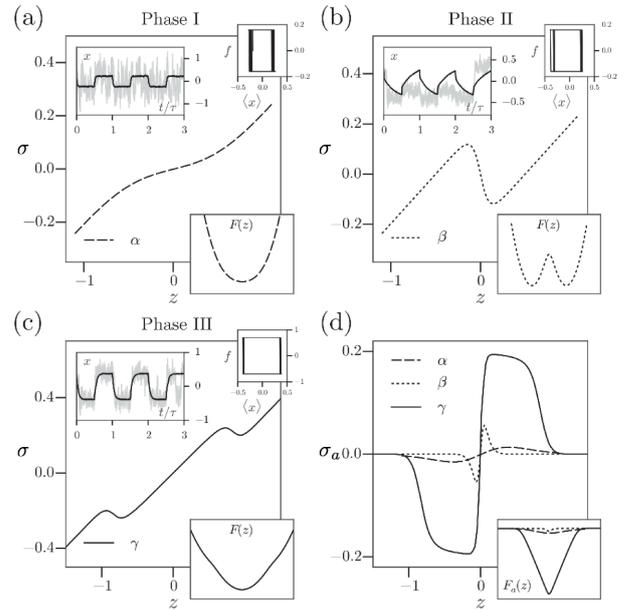

**Figure 48.** (a-c) Typical tension-length relations in phases I, II and III. Points $\alpha$, $\beta$ and $\gamma$ are the same as in Fig. 47 (b); (d) shows the active component of the force. Inserts show the behavior of stochastic trajectories in each of the phases at $z \simeq 0$ (gray lines) superimposed on their ensemble averages (black lines); the stationary hysteretic cycles, the structure of the effective potentials $F(z)$ and the active potential $F_a(z)$ defined as a primitive of the active force $\sigma_a(z)$. The parameters: $\kappa_0 = 0.6$, $\tau = 100$. Adapted from Ref. [224].

The adiabatic approximation fails at low temperatures (small $D$) where the escape time diverges and for these regimes the phase diagram has to be corrected numerically, see Fig. 47 (b). Direct numerical simulation based on Eq. 3.4 shows that the main features of the resulting diagram (tri-critical point, point $D_e$ and the vertical asymptote of the boundary separating phases I and III at large values of $A$) have been captured adequately by the adiabatic approximation. The new features of the non-adiabatic phase diagram is a dip of the boundary separating Phases II and III at some $D < D_e$ leading to an interesting re-entrant behavior (see Refs. [260; 261]). This is an effect of stochastic resonance which is beyond reach of the adiabatic approximation.

Force-elongation relations characterizing the mechanical response of the system at different points on the $(A, D)$ plane [see Fig. 47 (b)] are shown in Fig. 48 where the upper insets illustrate the typical stochastic trajectories and the associated cycles in $\{\langle x(t)\rangle, f(t)\}$ coordinates. We observe that while in phase I thermal fluctuations dominate periodic driving and undermine the two well structure of the potential, in phase III the jumps between the two energy wells are fully synchronized with the rocking force. In phase II the system shows intermediate behavior with uncorrelated jumps between the wells.

In Fig. 48(d) we illustrate the active component of the force $\sigma_a(z) = \sigma(z; A) - \sigma(z; 0)$ in phases I, II and III. A salient feature of Fig. 48(d) is that active force generation is significant only in the resonant (Kapitza) phase III. A biologically beneficial plateau (tetanus) is a manifestation of the triangular nature of a pseudo-well in the active landscape $F_a(z) = \int^z \sigma_a(s)ds$; note also that only slightly bigger $(f, \langle x\rangle)$ hysteresis cycle in phase III, reflecting a moderate increase of the extracted work, results in considerably larger active force. It is also of interest that the largest active rigidity is generated in the state $z = 0$ where the active force is equal to zero.

If we now estimate the non-dimensional parameters of the model by using the data on skeletal muscles, we obtain $A = 0.5, D = 0.01, \tau = 100$ [224]. This means that muscle myosins in stall conditions (physiological regime of isometric contractions), may be functioning in resonant phase III. The model can therefore provide an explanation of the observed stability of skeletal muscles in the negative stiffness regime [99]; similar stabilization mechanism may be also assisting the titin-based force generation at long sarcomere lengths [262].

The results presented in this Section for the case of periodic driving were shown in Ref. [224] to be qualitatively valid also for the case of dichotomous noise. However, the Ornstein-Uhlenbeck noise was unable to generate a nontrivial Kapitza phase.

To conclude, the prototypical model presented in this Section shows that by controlling the degree of non-equilibrium in the system, one can stabilize apparently unstable or marginally stable mechanical configurations and in this way modify the structure of the effective energy landscape (when it can be defined). The associated pseudo-energy wells with resonant nature may be crucially involved not only in muscle contraction but also in hair cell gating [119], integrin binding [263], folding/unfolding of proteins subjected to periodic forces [264] and other driven biological phenomena [265–268].

## 4. Active force generation

In this Section we address the slow time scale phase of force recovery which relies on attachment-detachment processes [79]. We review two types of models. In models of the first type the active driving comes from the interaction of the myosin head with actin filament, while the power stroke mechanism remains passive [269]. In models of the second type, the active driving resides in the power stroke machinery [244]. The latter model is fully compatible with the biochemical Lynm-Taylor cycle of muscle contractions.

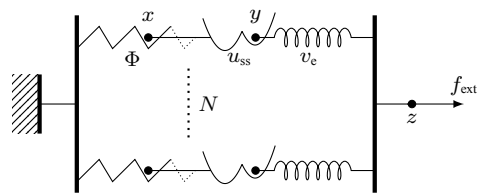

**Figure 49.** Schematic representation of a parallel bundle of cross-bridges that can attach and detach. Each cross bridge is modeled as a series connection of a ratchet $\Phi$, a bi-stable snap-spring $u_{ss}$, and linear elastic element $v_e$.

### 4.1. Contractions driven by the attachment-detachment process

A physiological perspective that the power-stroke is the driving force of active contraction was challenged by the discovery that myosin catalytic domain can operate as a Brownian ratchet, which means that it can move and produce contraction without any assistance from the power-stroke mechanism [136; 137; 142]. It is then conceivable that contraction is driven directly by the attachment-detachment machinery which can rectify the correlated noise and select a directionality following the polarity of actin filaments [60; 143].

To represent the minimal set of variables characterizing skeletal Myosin II in both attached and detached states— position of the motor domain, configuration of the lever domain and the stretch state of the series elastic element—we use three continuous coordinates [269]. To be maximally transparent we adopt the simplest representation of the attachement-detachment process provided by the rocking Brownian ratchet model [132; 247; 270; 271].

We interpret again a half-sarcomere as a HS type parallel bundle of $N$ cross bridges. We assume, however, that now each cross-bridge is a three-element chain containing a linear elastic spring, a bi-stable contractile element, and a molecular motor representing the ATP-regulated attachment-detachment process, see Fig. 49. The system is loaded either by a force $f_{ext}$ representing a cargo or is constrained by the prescribed displacement of the backbone.

The elastic energy of the linear spring can be written as $v_e(x) = \frac{1}{2}\kappa_0(z-y-\ell)^2$, where $\kappa_0$ is the elastic modulus and $\ell$ is the reference length. The energy $u_{ss}$ of the bi-stable mechanism is taken to be three-parabolic

$$u_{ss}(y-x) = \begin{cases} \frac{1}{2}\kappa_1(y-x)^2 & y-x \geq b_1 \\ -\frac{1}{2}\kappa_3(y-x-b)^2 + c & b_2 \leq y-x < b_1 \\ \frac{1}{2}\kappa_2(y-x-a)^2 + v_0 & y-x < b_2 \end{cases}$$
(4.1)

where $\kappa_{1,2}$ are the curvatures of the energy wells representing pre-power stroke and post-power stroke configurations, respectively and $a < 0$ is the characteristic size of the power stroke. The bias $v_0$ is again chosen to ensure that the two wells have the same energy in the state of isometric contraction. The en-ergy barrier is characterized by its position $b$, its height $c$ and its curvature $\kappa_3$. The values of parameters $b_1$ and $b_2$ are chosen to ensure the continuity of the energy function.

We model the myosin catalytic domain as the Brownian ratchet of Magnasco type [132]. More specifically, we view it as a particle moving in an asymmetric periodic potential while being subjected to a correlated noise. The periodic potential is assumed to be piece-wise linear in each period

$$\Phi(x) = \begin{cases} \frac{Q}{\lambda_1 L}(x - nL), & 0 < x - nL < \lambda_1 L \\ \frac{Q}{\lambda_2} - \frac{Q}{\lambda_2 L}(x - nL), & \lambda_1 L < x - nL < L \end{cases} \quad (4.2)$$

where $Q$ is the amplitude, $L$ is the the period, $\lambda_1 - \lambda_2$ is the measure of the asymmetry; $\lambda_1 + \lambda_2 = 1$. The variable $x$ marks the location of a particle in the periodic energy landscape: the head is attached if $x$ is close to one of the minima of $\Phi(x)$ and detached if it is close to one of the maxima.

The system of $N$ cross-bridges of this type connected in parallel is modeled by the system of Langevin equations [269]

$$\begin{cases} \nu_x \dot{x}_i = -\Phi'(x_i) + u'_{ss}(y_i - x_i) + f(t + t_i) + \sqrt{2\nu_x k_B T}\xi_x(t) \\ \nu_y \dot{y}_i = -u'_{ss}(y_i - x_i) - \kappa_0(y_i - z - \ell_i) + \sqrt{2\nu_y k_B T}\xi_y(t) \\ \nu_z \dot{z} = \sum_{i=1}^{N} \kappa_0(y_i - z - \ell_i) + f_{\text{ext}} + \sqrt{2\nu_z k_B T}\xi_z(t) \end{cases}$$

(4.3)

where $\nu_{x,y,z}$ denote the relative viscosities associated with the macroscopic variables, and $\xi$ is a standard white noise. The correlated component of the noise $f(t)$, imitating the activity of the ATP, is assumed to be periodic and piece-wise constant, see Eq. (3.5).

Since our focus here is on active force generation rather than on active oscillations, we de-synchronize the dynamics by introducing phase shifts $t_i$, assumed to be independent random variables uniformly distributed in the interval $[0, T]$; we also allow the pre-strains $\ell_i$ to be random and distribute them in the intervals $[iL - a/2, iL + a/2]$. Quenched disorder disfavors coherent oscillations observed under some special conditions (e.g. Ref. [166]). While we leave such collective effects outside our review, several comprehensive expositions are availbale in the literature, see Refs. [12; 36; 38; 112; 114; 143; 157; 158; 165; 166; 171; 171; 173; 272–282].

To illustrate the behavior of individual mechanical units we first fix the parameter $z = 0$, and write the total energy of a cross-bridge as a function of two remaining mechanical variables $y$ and $x$:

$$v(x, y) = \Phi(x) + u_{ss}(y - x) + v_e(-y) \quad (4.4)$$

The associated energy landscape is shown in Fig. 50, where the upper two local minima $A$ and $B$ indicate the pre-power stroke and the post-power stroke configurations of a motor attached in one position on actin potential, while the two lower local minima $A'$ and $B'$ correspond to the pre-power stroke and the post-power stroke configurations of a motor shifted to a neighboring attached position. We associate the detached state with an unstable position around the maximum separating the minima $(A, B)$ and $(A', B')$, see Ref. [269] for more details.

In Fig. 51 we show the results of numerical simulations of isotonic contractions at $f_{\text{ext}} = 0.5 T_0$, where $T_0$ is the stall tension. One can see that the catalytic domain of an individual head, described by the variable $x$, evolves through three different attachment sites [see Fig. 51(a)]. In Fig. 51(b) we show the time history of the variable $x - y$ characterizing the conformational state of a single myosin head during the cycle. The first vertical line shows the moment in which the power stroke $A \to B$ takes place. The second vertical line shows the motion from the active site $i$ on the actin filament to the next site $i' = i + 1$, corresponding to the transition $B \to A'$. This motion induces a change of state in the bi-stable element which brings the lever arm into the pre-power stroke position. Due to the advance of the variable $z$ during such isotonic contractions, see Fig. 51(d), the elastic element whose configuration can be read on Fig. 51(c), relaxes and the post-power stroke minimum $B'$ becomes preferable. The third vertical line shows the moment in which the new power stroke $A' \to B'$ takes place.

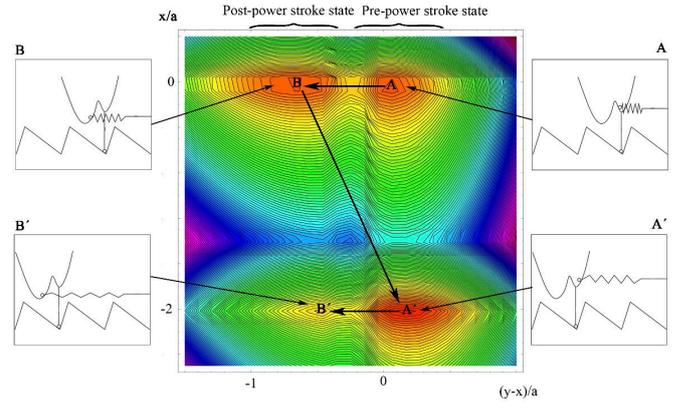

**Figure 50.** Contour plot of the effective energy $v(x, y; z_0)$ at $z_0 = 0$. Inserts illustrate the states of various mechanical subunits.

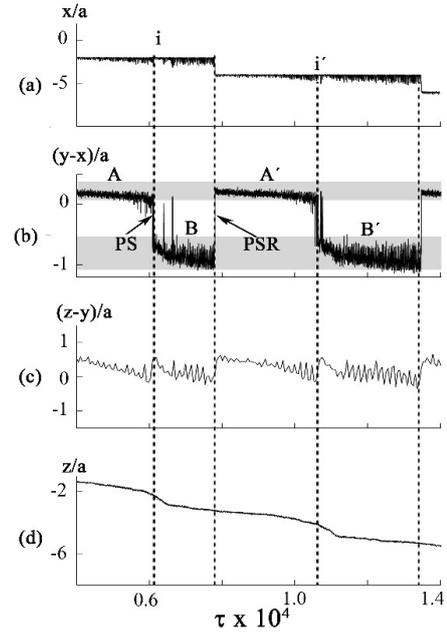

**Figure 51.** The numerical simulation of the time histories for different mechanical units in a load clamp simulation at zero external force: (a) the behavior of the myosin catalytic domain; (b) the behavior of the power stroke element (snap-spring); (c) the behavior of the elastic element; (d) the total displacement of the backbone.

Observe that the position of the backbone can be considered stationary during the recharging of the power stroke. In this situation, the key-factor for the possibility of recharging (after the variable $x$ has overcome the barrier in the periodic potential) is that the total energy $v(x, y)$ has a minimum when the snap-spring is in the pre-power stroke state. The corresponding analytical condition is $(Q/v_0) > (\lambda_1 L)/a$ which places an important constraint on the choice of parameters [269].

A direct comparison of the simulated mechanical cycle with the Lymn–Taylor cycle (see Fig. 2) shows that while the two attached configurations are represented in this model adequately, the detached configurations appear only as transients. In fact, one can see that the (slow) transition $B \to A'$ represents a combined description of the detachment, of the power stroke recharge and then of another attachment. Since in the actual biochemical cycle such a transition are described by at least two distinct chemical states, the ratchet driven model is only in partial agreement with biochemical data.

### 4.2. Contractions driven by the power stroke

We now consider a possibility that acto-myosin contractions are propelled directly through a conformational change. The model where the power-stroke is the only active mechanism driving muscle contraction was developed in Ref. [244].

To justify such change of the modeling perspective, we recall that in physiological literature active force generation is largely attributed to the power-stroke which is perceived as a part of active rather than passive machinery [153]. This opinion is supported by observations that both the power-stroke and the reverse-power-stroke can be induced by ATP even in the absence of actin filaments [71], that contractions can be significantly inhibited by antibodies which impair lever arm activity [283], that sliding velocity in mutational myosin forms depends on the lever arm length [192] and that the directionality can be reversed as a result of modifications in the lever arm domain [284; 285].

Although the simplest models of Brownian ratchets neglect the conformational change in the head domain, some phases of the attachment-detachment cycle have been interpreted as a power-stroke viewed as a part of the horizontal shift of the myosin head [144; 286]. In addition, ratchet models were considered with the periodic spatial landscape supplemented by a reaction coordinate, representing the conformational change [287; 288]. In all these models, however, the power stroke was viewed as a secondary element and contractions could be generated even with the disabled power stroke. The main functionality of the power-stroke mechanism was attributed to fast force recovery which could be activated by loading but was not directly ATP-driven [74; 99; 289].

The apparently conflicting viewpoint that the power-stroke mechanism consumes metabolic energy remains, however, the underpinning of the phenomenological chemo-mechanical models that assign active roles to both the attachment-detachment and the power-stroke [86; 102]. These models pay great attention to structural details and in their most comprehensive versions faithfully reproduce the main experimental observations [68; 115; 290].

In an attempt to reach a synthetic description, several hybrid models, allowing chemical states to coexist with springs and forces, have been also proposed, see Refs. [112; 152; 291].

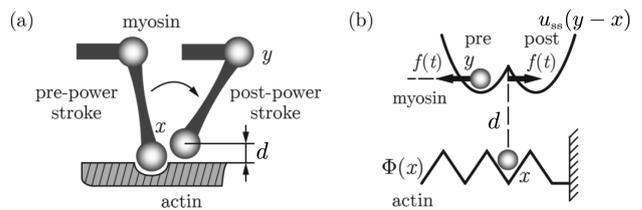

**Figure 52.** (a) An illustration of the steric effect associated with the power-stroke; (b) sketch of the mechanical model. Adapted from Ref. [244].

These models, however, still combine continuous dynamics with jump transitions which makes the precise identification of structural analogs of the chemical steps and the underlying micro-mechanical interactions challenging [154].

*4.2.1. The model.* Here, following Ref. [244], we sketch a mechanistic model of muscle contractions where power stroke is the only active agent. To de-emphasize the ratchet mechanism discussed in the previous section, we simplify the real picture and represent actin filaments as passive, non-polar tracks. The power-stroke mechanism is represented again by a symmetric bi-stable potential and the ATP activity is modeled as a center-symmetric correlated force with zero average acting on the corresponding configurational variable.

A schematic representation of the model for a single cross-bridge is given in Fig. 52(b), where $x$ is the observable position of a myosin catalytic domain, $y - x$ is the internal variable characterizing the phase configuration of the power stroke element and $d$ is the distance between the myosin head and the actin filament. Through the variable $d$ we can take into account that when the lever arm swings, the interaction of the head with the binding site weakens, see Fig. 52(a). The implied steric rotation-translation coupling in ratchet models has been previously discussed in Refs. [154; 292; 293].

We write the energy of a single cross-bridge in the form

$$\hat{G}(x, y, d) = d\,\Phi(x) + u_{ss}(y - x), \quad (4.5)$$

where $\Phi(x)$ is a non-polar periodic potential representing the binding strength of the actin filament and $u_{ss}(y - x)$ is a symmetric double-well potential describing the power-stroke element, see Fig. 49. The coupling between the state of the power-stroke element $y - x$ and the spatial position of the motor $x$ is implemented through the variable $d$. In the simplest version of the model $d$ is assumed to be a function of the state of the power-stroke element

$$d(x, y) = \Psi(y - x). \quad (4.6)$$

To mimic the underlying steric interaction, we assume that when a myosin head executes the power-stroke, it moves away from the actin filament and therefore the control function $\Psi(y - x)$ progressively switches off the actin potential, see Fig. 52(b). Similarly, as the power-stroke is recharging, the myosin head moves progressively closer to the actin filament and therefore the function $\Psi(y - x)$ should be bringing the actin potential back into the bound configuration.

In view of (4.6), we can eliminate the variable $d$ and introduce the redressed potential $G(x, y) = \hat{G}[x, y, \Psi(y - x)]$.

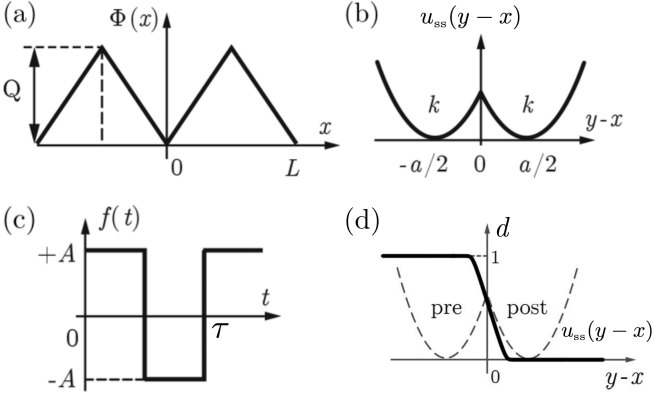
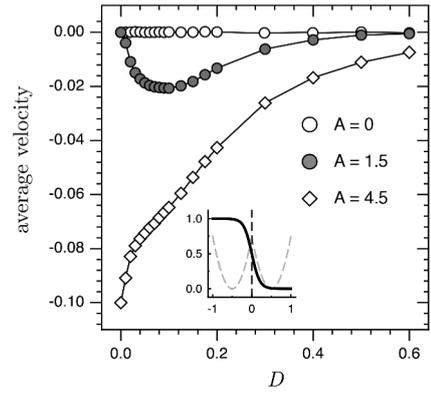

**Figure 53.** The functions $\Phi$, $u_{ss}$, $f$ and the coupling function $\Phi$ used in numerical experiments. Analytic expressions for (a),(b) and (c) are given by Eqs. [(4.1),(4.2) and (4.8)], respectively. Adapted from Ref. [244].

**Figure 54.** The dependence of the average velocity $v$ on temperature $D$ and the amplitude of the ac signal $A$. The pre- and post-power-stroke states are labeled in such a way that the purely mechanical ratchet would move to the left. Adapted from Ref. [244].

Then the overdamped stochastic dynamics can be described by the system of dimensionless Langevin equations

$$\dot{x} = -\partial_x G(x, y) - f(t) + \sqrt{2D}\xi_x(t)$$
$$\dot{y} = -\partial_y G(x, y) + f(t) + \sqrt{2D}\xi_y(t). \quad (4.7)$$

Here $\xi(t)$ is the standard white noise with $\langle \xi_i(t) \rangle = 0$, and $\langle \xi_i(t)\xi_j(s) \rangle = \delta_{ij}\delta(t-s)$ and $D$ is a dimensionless measure of temperature; for simplicity the viscosity coefficients are assumed to be the same for variables $x$ and $y$. The time dependent force couple $f(t)$ with zero average represents a correlated component of the noise. In the computational experiments a periodic extension of the symmetric triangular potential $\Phi(x)$ with amplitude $Q$ and period $L$ was used, see Fig. 53(a). The symmetric potential $u_{ss}(y-x)$ was taken to be bi-quadratic with the same stiffness $\kappa$ in both phases and the distance between the bottoms of the wells denoted by $a$, see Fig. 53(b). The correlated component of the noise $f(t)$ was described by a periodic extension of a rectangular shaped function with amplitude $A$ and period $\tau$, Fig. 53(c). Finally, the steric control ensuring the gradual switch of the actin potential is described by a step function

$$\Psi(s) = (1/2)[1 - \tanh(s/\varepsilon)], \quad (4.8)$$

where $\varepsilon$ is a small parameter, see Fig. 53(d).

The first goal of any mechanical model of muscle contraction is to generate a systematic drift $v = \lim_{t\to\infty}\langle x(t)\rangle/t$ without applying a biasing force. The dependence of the average velocity $v$ on the parameters of the model is summarized in Fig. 54. It is clear that the drift in this model is exclusively due to $A \neq 0$. When $A$ is small, the drift velocity shows a maximum at finite temperatures which implies that the system exhibits stochastic resonance [294]. At high amplitudes of the ac driving, the motor works as a purely mechanical ratchet and the increase of temperature only worsens the performance [136; 137; 143].

One can say that the system (4.7) describes a power-stroke-driven ratchet because the correlated noise $f(t)$ acts on the relative displacement $y - x$. It effectively "rocks" the bi-stable potential and the control function $\Psi(y-x)$ converts such "rocking" into the "flashing" of the periodic potential $\Phi(x)$.

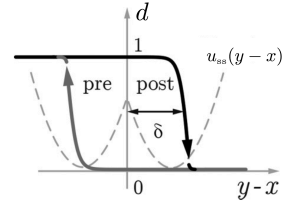

**Figure 55.** The hysteresis operator $\widehat{\Psi}\{y(t) - x(t)\}$ linking the degree of attachment $d$ with the previous history of the power-stroke configuration $y(t) - x(t)$. Adapted from Ref. [244].

It is also clear that the symmetry breaking in this problem is imposed exclusively by the asymetry of the coupling function $\Psi(y-x)$. Various other types of rocked-pulsated ratchet models have been studied in Refs. [295; 296].

The idea that the source of non-equilibrium in Brownian ratchets is resting in internal degrees of freedom [297; 298] originated in the theory of processive motors [299–302]. For instance, in the description of dimeric motors it is usually assumed that ATP hydrolysis induces a conformational transformation which then affects the position of the motor legs [303]. Here the same idea is used to describe a non-processive motor with a single leg that remains on track due to the presence of a thick filament. By placing emphasis on active role of the conformational change in non-processive motors the model brings closer the descriptions of porters and rowers as it was originally envisaged in Ref. [304].

*4.2.2. Hysteretic coupling* The analysis presented in Ref. [244] has shown that in order to reproduce the whole Lymn–Taylor cycle, the switchings in the actin potential must take place at different values of the variable $y - x$ depending on the direction of the conformational change. In other words, we need to replace the holonomic coupling (4.6) by the memory operator

$$d\{x, y\} = \widehat{\Psi}\{y(t) - x(t)\} \quad (4.9)$$

whose output depends on whether the system is on the "striking" or on the "recharging" branch of the trajectory, see Fig. 55. Such memory structure can be also described by a rate independent differential relation of the form $\dot{d} = Q(x, y, z)\dot{x} + R(x, y, d)\dot{y}$, which makes the model non-holonomic.

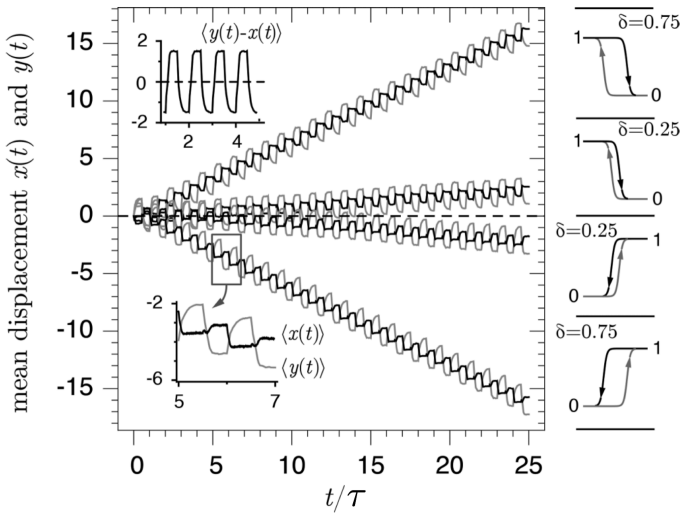 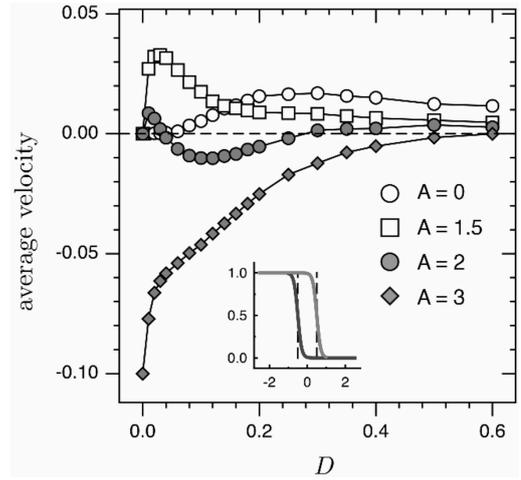

**Figure 56.** Stationary particle trajectories in the model with the hysteretic coupling (4.9). Parameters are: $D = 0.02$ and $A = 1.5$. Adapted from Ref. [244].

**Figure 57.** The dependence of the average velocity $v$ on temperature $D$ in the hysteretic model with $\delta = 0.5$. Adapted from Ref. [244].

Using (4.9) we can rewrite the energy of the system as a functional of its history $y(t)$ and $x(t)$

$$G\{x, y\} = \widehat{\Psi}\{y(t) - x(t)\}\Phi(x) + u_{ss}(y - x). \quad (4.10)$$

In the Langevin setting (4.7), the history dependence may mean that the underlying microscopic stochastic process is non-Markovian (due to, say, configurational pinning [305]), or that there are additional non-thermalized degrees of freedom that are not represented explicitly, see Ref. [306]. In general, it is well known that the realistic feedback implementations always involve delays [307].

To simulate our hysteretic ratchet numerically we used two versions of the coupling function (4.8) shifted by $\delta$ with the branches $\Psi(y - x \pm \delta)$ blended sufficiently far away from the hysteresis domain, see Fig. 55. Our numerical experiments show that the performance of the model is not sensitive to the shape of the hysteresis loop and depends mostly on its width characterized by the small parameter $\delta$.

In Fig. 56 we illustrate the "gait" of the ensuing motor. The center of mass advances in steps and during each step the power-stroke mechanism gets released and then recharged again, which takes place concurrently with attachment-detachment. By coupling the attached state with either pre- or post-power-stroke state, we can vary the directionality of the motion. The average velocity increases with the width of the hysteresis loop which shows that the motor can extract more energy from the coupling mechanism with longer delays.

The results of the parametric study of the model are summarized in Fig. 57. The motor can move even in the absence of the correlated noise, at $A = 0$, because the non-holonomic coupling (4.10) breaks the detailed balance by itself. At finite $A$ the system can use both sources of energy (hysteretic loop and ac noise) and the resulting behavior is much richer than in the non-hysteretic model, see Ref. [244] for more details.

*4.2.3. Lymn–Taylor cycle.* The mechanical "stroke" in the space of internal variables $(d, y - x)$ can be now compared with the Lymn–Taylor acto-myosin cycle [59] shown in Fig. 2 and in the notations of this Section in Fig. 58(a).

We recall that the chemical states constituting the Lymn–Taylor cycle have been linked to the structural configurations (obtained from crystallographic reconstructions): $A$ (attached, pre-power-stroke $\rightarrow$ *AM-ADP-Pi*), $B$ (attached, post-power-stroke $\rightarrow$ *AM-ADP*), $C$ (detached, post-power-stroke $\rightarrow$ *M-ATP*), $D$ (detached, pre-power-stroke $\rightarrow$ *M-ADP-Pi*). In the discussed model the jump events are replaced by continuous transitions and the association of chemical states with particular regimes of stochastic dynamics is not straightforward.

In Fig. 58(b), we show a fragment of the averaged trajectory of a steadily advancing motor projected on the $(x, y - x, 0)$ plane. In Fig. 58(c) the same trajectory is shown in the $(x, y - x, d)$ plane with fast advances in the $d$ direction intentionally schematized as jumps. By using the same letters $A, B, C, D$ as in Fig. 58(a) we can visualize a connection between the chemical/structural states and the transient mechanical configurations of the advancing motor.

Suppose, for instance, that we start at point $A$ corresponding to the end of the negative cycle of the ac driving $f(t)$. The system is in the attached, pre-power-stroke state and $d = 1$. As the sign of the force $f(t)$ changes, the motor undergoes a power-stroke and reaches point $B$ while remaining in the attached state. When the configurational variable $y - x$ passes the detachment threshold, the myosin head detaches which leads to a transition from point $B$ to $B'$ on the plane $d = 0$. Since the positive cycle of the force $f(t)$ continues, the motor completes the power-stroke by moving from $B'$ to point $C$. At this moment, the rocking force changes sign again which leads to recharging of the power-stroke mechanism in the detached state, described in Fig. 58(a) as a transition from $C$ to $D$. In point $D$, the variable $y - x$ reaches the attachment threshold. The myosin head reattaches and the system moves to point $D'$ where $d = 1$ again. The recharging continues in the attached state as the motor evolves from $D'$ to a new state $A$, shifted by one period.

One can see that the chemical states constituting the minimal enzyme cycle can be linked to the mechanical configurations traversed by this stochastic dynamical system. The detailed mechanical picture, however, is more complicated than in the prototypical Lymn–Taylor scheme. In some stages of the cycle one can use the Kramers approximation to build a description in terms of a discrete set of chemical reactions, however, the number of such reactions should be larger than in the minimal Lymn–Taylor model.

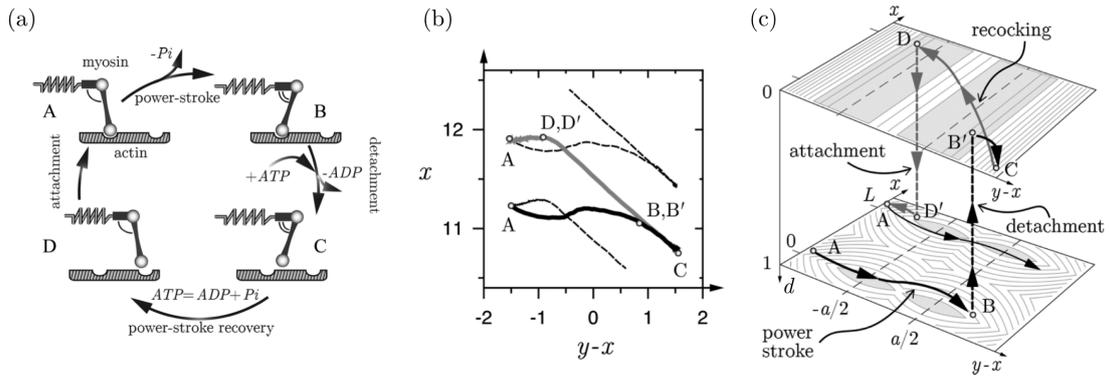

**Figure 58.** (a) Schematic illustration of the four-step Lymn–Taylor cycle in the notations of this Section. (b) A steady-state cycle in the hysteretic model projected on the $(x, y-x)$ plane; color indicates the sign of the rocking force $f(t)$: black if $f(t) > 0$ and gray if $f(t) < 0$; (c) Representation of the same cycle in the $(d, x, y-x)$ space with identification of the four chemical states A, B, C, D constituting the Lymn–Taylor cycle shown in (a). The level sets represent the energy landscape $G$ at $d = 0$ (detached state) and $d = 1$ (attached state). The parameters are: $D = 0.02$, $A = 1.5$, and $\delta = 0.75$. Adapted from Ref. [244].

In conclusion, we mention that the identification of the chemical states, known from the studies of the prototypical catalytic cycle in solution, with mechanical states, is a precondition for the bio-engineering reproduction of a wide range of cellular processes. In this sense, the discussed schematization of the contraction phenomenon can be viewed as a step towards building engineering devices imitating actomyosin enzymatic activity.

*4.2.4. Force-velocity relations.* The next question is how fast such motor can move against an external cargo. To answer this question we assume that the force $f_{\text{ext}}$ acts on the variable $y$ which amounts to tilting of the potential (4.10) along the $y$ direction

$$G\{x, y\} = \widehat{\Psi}\{y(t) - x(t)\}\Phi(x) + u_{\text{ss}}(y - x) - yf_{\text{ext}}. \quad (4.11)$$

A stochastic system with energy (4.11) was studied numerically in Ref. [244] and in Fig. 59 we illustrate the obtained force-velocity relations. The quadrants in the $(f_{ext}, v)$ plane where $R = f_{ext}v > 0$ describe dissipative behavior. In the other the other two quadrants, where $R = f_{ext}v < 0$, the system shows anti-dissipative behavior.

Observe that at low temperatures the convexity properties of the force-velocity relations in active pushing and active pulling regimes are different. In the case of pulling the typical force-velocity relation is reminiscent of the Hill's curve describing isotonic contractions, see Ref. [79]. In the case of pushing, the force-velocity relation can be characterized as convex-concave and such behavior has been also recorded in muscles, see Refs. [161; 308; 309]. The asymmetry is due to the dominance of different mechanisms in different regimes. For instance, in the pushing regimes, the motor activity fully depends on ac driving and at large amplitudes of the driving the system performs as a mechanical ratchet. Instead, in the pulling regimes, associated with small amplitudes of external driving, the motor advances because of the delayed feedback. Interestingly, the dissimilarity of convexity properties of the force-velocity relations in pushing and pulling regimes has been also noticed in the context of cell motility where actomyosin contractility is one of the two main driving forces, see Ref. [310].

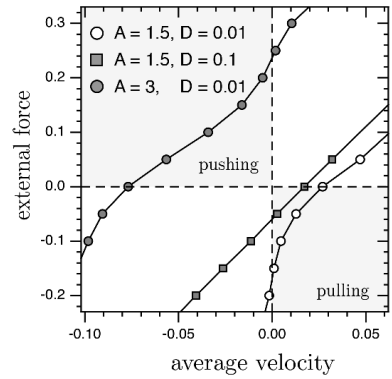

**Figure 59.** The force-velocity relation in the model with hysteretic coupling at different amplitudes of the ac driving $A$ and different temperatures $D$. The hysteresis width is $\delta = 0.5$. Adapted from Ref. [244].

## 5. Descending limb

In this Section, following Ref. [238], we briefly address one of the most intriguing issues in mesoscopic muscle mechanics: an apparently stable behavior on the "descending limb" which is a section of the force-length curve describing isometrically tetanized muscle [17–19; 39].

As we have seen in the previous Sections, the active force $f$ generated by a muscle in a hard (isometric) device depends on the number of pulling cross-bridge heads. The latter is controlled by the filament overlap which may be changed by the (pre-activation) passive stretch $\Delta\ell$. A large number of experimental studies have been devoted to the measurement of the isometric tetanus curve $f(\Delta\ell)$, see Fig. 60 and Fig. 3.

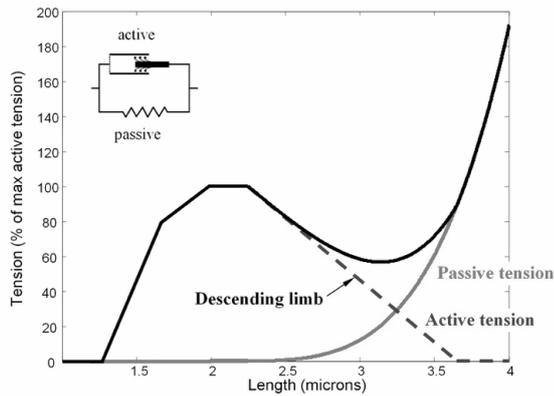

**Figure 60.** Schematic isometric tetanus with a descending limb. Adapted from Ref. [238].

Since the stretch beyond a certain limit would necessarily decrease the filament overlap, the active component of $f(\Delta\ell)$ must contain a segment with a negative slope known as the "descending limb" [78; 311–315]. The negative stiffness associated with active response is usually corrected by the positive stiffness provided by passive crosslinkers that connect actin and myosin filaments. However, for some types of muscles the total force-length relation $f(\Delta\ell)$ describing active and passive elements connected in parallel, still has a range where force decreases with elongation. It is this overall negative stiffness that will be the focus of the following discussion.

If the curve $f(\Delta\ell)$ is interpreted as a description of the response of the "muscle material" shown in Fig. 61, the softening behavior associated with negative overall stiffness should lead to localization instability and the development of strain inhomogeneities [227; 316]. In terms of the observed quantities, the instability would mean that any initial imperfection would cause a single myosin filament to be pulled away from the center of the activated half-sarcomeres.

Some experiments seem to be indeed consistent with non-uniformity of the Z-lines spacing, and with random displacements of the thick filaments away from the centers of the sarcomeres [225; 312; 314; 317; 318]. The nontrivial half-sarcomeres length distribution can be also blamed for the observed disorder and skewing [319; 320]. The link between non-affine deformation and the negative stiffness is also consistent with the fact that the progressive increase of the range of dispersion in half-sarcomere lengths, associated with a slow rise of force during tetanus (creep phase), was observed mostly around the descending limb [321–323], even though the expected ultimate strain localization leading to failure was not recorded.

A related feature of the muscle response on the descending limb is the non-uniqueness of the isometric tension, which was shown to depend on the pathway through which the elongation is reached. Experiments demonstrate that when a muscle fiber is activated at a fixed length and then suddenly stretched while active, the tension first rises and then falls without reaching the value that the muscle generates when stimulated isometrically [320; 324–330]. The difference between tetani subjected to such post-stretch and the corresponding isometric tetani reveals a positive instantaneous stiffness on the descending limb. Similar phenomena have been observed during sudden shortening of the active muscle fibers: if a muscle is allowed to shorten to the prescribed length it develops less tension than during direct tetanization at the final length.

All these puzzling observations have been discussed extensively in the literature interpreting half-sarcomeres as softening elastic springs [54; 78; 331–334]. The fact of instability on the descending limb for such spring chain was realized already by Hill [331] and various aspects of this instability were later studied in Refs. [332; 335]. It is broadly believed that a catastrophic failure in this system is warranted but is not observed because of the anomalously slow dynamics [313; 320; 336–338]. In a dynamical version of the model of a chain with softening springs, each contractile component is additionally bundled with a dashpot characterized by a realistic (Hill-Katz) force-velocity relation [226; 313; 320; 336–339]. A variety of numerical tests in such dynamic setting demonstrated that around a descending limb the half-sarcomeres configuration can become non uniform but at the time scale which is unrealistically long. Such over-damped dynamic model was shown to be fully compatible with the residual force after stretch on the descending limb, and the associated deficit of tension after shortening.

These simulations, however, left unanswered the question about the fundamental origin of the multi-valudness of the muscle response around the descending limb. For instance, it is still debated whether such non-uniqueness is a property of individual half-sarcomeres or a collective property of the whole chain. It is also apparently unclear how the local (microscopic) inhomogeneity of a muscle myofibril can coexist with the commonly accepted idea of a largely homogenous response at the macro-level.

To address these questions we revisit here the one-dimensional chain model with softening springs reinforced by parallel (linear) elastic springs, see Fig. 61 and 62. A formal analysis [238], following a similar development in the theory of shape memory alloys [227], shows that this mechanical system has an exponentially large (in $N$) number of configurations with equilibrated forces, see an illustration for small $N$ in Fig. 63 and our goal will be to explore the consequences of the complexity of the properly defined energy landscape.

### 5.1. Pseudo-elastic energy.

The physical meaning of the energy associated with the parallel passive elements is clear but the challenge is to associate an energy function with active elements. In order to generate active force, motors inside the active element receive and dissipate energy, however, this not the energy we need to account for.

As we have already seen, active elements posses their own passive mechanical machinery which is loaded endogenously by molecular motors. Therefore some energy is stored in these passive structures. For instance, we can account for the elastic energy of attached springs and also consider the energy of de-bonding. A transition from one tetanized state to another tetanized state, leads to the change in the stored energy of these passive structures. Suppose that to make an elongation $d\ell$ along the tetanus, the external force $f(\ell)$ must perform the work $f d\ell = dW$ where $W(\ell)$ is the energy of the passive structures that accounts not only for elastic stretching but also for inelastic effect associated with the changes in the number of attached cross-bridges.

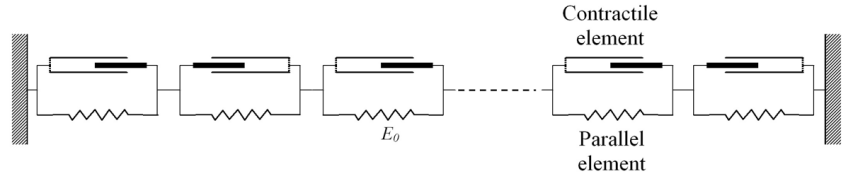

**Figure 61.** The model of a muscle myofibril. Adapted from Ref. [238].

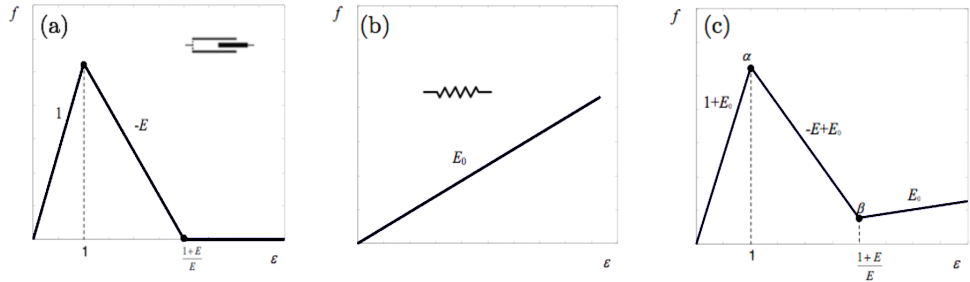

**Figure 62.** Non-dimensional tension-elongation relations for the active element (a), for the passive elastic component (b) and for the bundle (c). Adapted from Ref. [238].

By using the fact that the isometric tetanus curve $f(\ell)$ has a up-down-up structure we can conclude that the effective energy function $W(\ell)$ must have a double-well structure. If we subtract the contribution due to parallel elasticity $W_p(\ell)$, we are left with the active energy $W_a(\ell)$, which will then have the form of a Lennard-Jones potential. Shortening below the inflection point of this potential would lead to partial "neutralization" of cross-bridges, and as a result the elastic energy of contributing pullers progressively diminishes. Instead, we can assume that when the length increases beyond the inflection point (point of optimal overlap), the system develops damage (debonding) and therefore the energy increases. After all bonds are broken, the energy of the active element does not change any more and the generated force becomes equal to zero.

### 5.2. Local model.

Consider now a chain of half sarcomeres with nearest neighbor interactions and controlled total length, see Fig. 61. Suppose that the system selects mechanical configurations where the energy invested by pullers in loading the passive sub-structures is minimized. The energy minimizing configurations will then deliver an optimal trade-off between elasticity and damage in the whole ensemble of contractile units. This assumption is in agreement with the conventional interpretation of how living cells interact with an elastic environment. For instance, it is usually assumed that active contractile machinery inside a cell rearranges itself in such a way that the generated elastic field in the environment minimizes the elastic energy [340; 341].

The analysis of the zero temperature chain model for a myofibril whose series elements are shown in Fig. 62 confirms that the ensuing energy landscape is rugged, see Ref. [238]. The possibility of a variety of evolutionary paths in such a landscape creates a propensity for history dependence, which, in turn, can be used as an explaination of both the "permanent extra tension" and the "permanent deficit of tension" observed in the areas adjacent to the descending limb. The domain of metastability on the force-length plane, see Fig. 63, is represented by a dense set of stable branches with a fixed degree of inhomogeneity. Note that in this system the negative overall slope of the force-length relation along the global minimum path can be viewed as a combination of a large number of micro-steps with positive slopes. Such "coexistence" of the negative averaged stiffness with the positive instantaneous stiffness, first discussed in Ref. [332], can be responsible for the stable performance of the muscle fiber on the descending limb.

Observe, however, that the strategy of global energy minimization contradicts observations because the reported negative overall stiffness is incompatible with the implied convexification of the total energy. Moreover, the global minimization scenario predicts considerable amount of vastly over-stretched (popped) half-sarcomeres that have not been seen in experiments. We are then left with a conclusion that along the isometric tetanus at least some of the active, non-affine configurations correspond to local rather than global minima of the stored energy.

A possible representation of the experimentally observed tetanus curve as a combination of local and global minimization segments is presented by a solid thick line in Fig. 63. In view of the quasi-elastic nature of the corresponding response, it is natural to associate the ascending limb of the tetanus curve at small levels of stretch with the homogeneous (affine) branch of the global minimum path (segment AB in Fig. 63). Assume that around the point where the global minimum configuration becomes non-affine (point B in Fig. 63), the system remains close to the global minimum path. Then, the isometric tetanus curve forms a plateau separating ascending and descending limbs (segment between points B and C in Fig. 63). Such plateau is indeed observed in experiments on myofibrils and is known to play an important physiological role ensuring robustness of the active response. We can speculate that a limited mixing of "strong" and "weak" (popped) half-sarcomeres responsible for this plateau can be confined close to the ends of a myofibril while remaining almost invisible in the bulk of the sample.

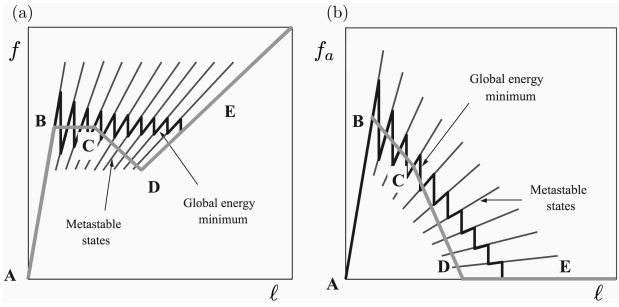

**Figure 63.** The structure of the set of metastable branches of the tension-elongation relation for $N = 10$. Here $f$ is the total tension (a) and $f_a$ is the active tension (b). The thick gray line represents the anticipated tetanized response. Adapted from Ref. [238].

To account for the descending limb, we must assume that as the length of the average half-sarcomere increases beyond the end of the plateau (point C in Fig. 63), the tetanized myofibril can no longer reach the global minimum of the stored energy. To match observations we assume that beyond point C in Fig. 63 the attainable metastable configurations are characterized by the value of the active force, which deviates from the Maxwell value and becomes progressively closer to the value generated by the homogeneous configurations as we approach the state of no overlap (point D). The numerical simulations show [238] that the corresponding non-affine configurations can be reached dynamically as a result of the instability of a homogeneous state. One may argue that such, almost affine metastable configurations, may be also favored due to the presence of some additional mechanical signaling, which takes a form of inter-sarcomere stiffness or next to nearest neighbor ($NNN$) interaction. As the point D in Fig. 63 is reached, all cross-bridges are detached and beyond this point the myofibril is supported exclusively by the passive parallel elastic elements (segment DE).

Since all the metastable non-affine states involved in this construction have an extended range of stability, the application of a sudden deformation will take the system away from the isometric tetanus curve BCD in Fig. 63. It is then difficult to imagine that the isometric relaxation, following such an eccentric loading, will allow the system to stabilize again exactly on the curve BCD. Such "metastable" response would be consistent with residual force enhancement observed not only around the descending limb but also above the optimal (physiological) plateau and even around the upper end of the ascending limb. It is also consistent with the observations showing that the residual force enhancement after stretch is independent of the velocity of the stretch, that it increases with the amplitude of the stretch and that it is most pronounced along the descending limb.

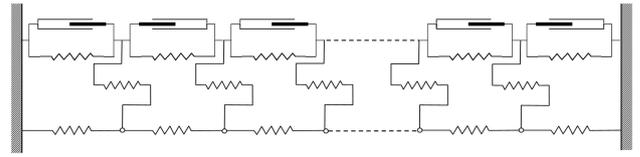

**Figure 64.** Schematic representation of the structure of a half-sarcomere chain surrounded by the connecting tissue. Adapted from Ref. [238].

### 5.3. Nonlocal model.

While the price of stability in this system appear to be the emergence of the limited microscopic non-uniformity in the distribution of half sarcomere lengths, we now argue that it may be still compatible with the macroscopic (averaged) uniformity of the whole myofibril [319]. To support this statement we briefly discuss here a model of a myofibril which involves long range mechanical signaling between half-sarcomeres via the surrounding elastic medium, see Ref. [238].

The model is illustrated in Fig. 64. It includes two parallel elastically coupled chains. One of the chains, containing double well springs, is the same as in the local model. The other chain contains elements mimicking additional elastic interactions in the myofibril of possibly non-one-dimensional nature; it is assumed that the corresponding shear (leaf) springs are linearly elastic.

The ensuing model is nonlocal and involves competing interactions: the double-well potential of the snap-springs favors sharp boundaries between the "phases", while the elastic foundation term favors strain uniformity. As a result of this competition the energy minimizing state can be expected to deliver an optimal trade off between the uniformity at the macro-scale and the non-uniformity (non-affinity) at the micro-scale.

The nonlocal extension of the chain model lacks the permutation degeneracy and generates peculiar microstructures with fine mixing of shorter half sarcomeres located on the ascending limb of the tension-length curve and longer half sarcomeres supported mostly by the passive structures [238]. The mixed configurations represent periodically modulated patterns that are undistinguishable from the homogeneous deformation if viewed at a coarse scale. The descending limb can be again interpreted as a union of positively sloped steps that can be now of vastly different sizes. It is interesting that the discrete structure of the force-length curve survives in the continuum limit, which instead of smoothening makes it extremely singular. More specifically, the variation of the degree of non-uniformity with elongation along the global energy minimum path exhibits a complete devil's staircase type behavior first identified in a different but conceptually related system [342], see Fig. 65 and Ref. [238] for more details.

To make the nonlocal model compatible with observations, one should again abandon the global minimisation strategy and associate the descending limb with metastable (rather than stable) states. In other words, one needs to apply an auxiliary construction similar to the one shown in Fig. 63 for the local model, which anticipates an outcome produced by a realistic kinetic model of tetanization.

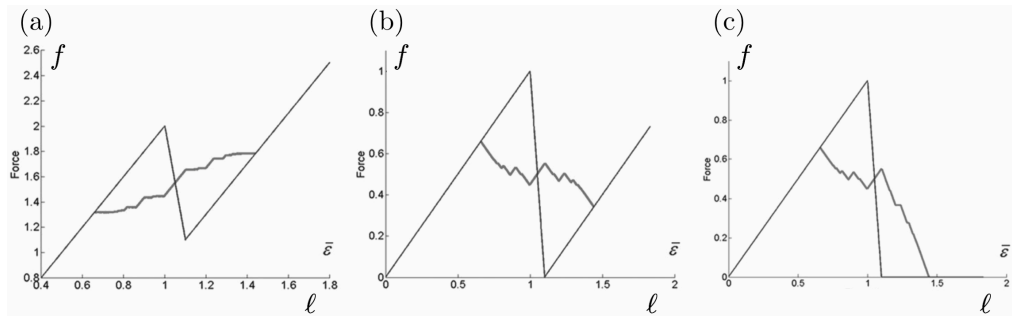

**Figure 65.** (a) The force-length relation along the global energy minimum path in the continuum limit for the model shown in Fig. 64. (b) The force-length relation along the global energy minimum path with the contribution due to connecting tissue subtracted. (c) The active force-length relation along the global energy with the contribution due to connecting tissue and sarcomere passive elasticity subtracted. Adapted from Ref. [238].

## 6. Non-muscle applications

The prototypical nature of the main model discussed in this review (HS model, a parallel bundle of bistable units in passive or active setting ) makes it relevant far beyond the skeletal muscle context. It provides the most elementary description of molecular devices capable of transforming in a Brownian environment a continuous input into a binary, all-or-none output that is crucial for the fast and efficient stroke-like behavior. The capacity of such systems to flip in a reversible fashion between several metastable conformations is essential for many processes in cellular physiology, including cell signaling, cell movement, chemotaxis, differentiation, and selective expression of genes [343; 344]. Usually, both the input and the output in such systems, known as allosteric, are assumed to be of biochemical origin. The model, dealing with mechanical response and relying on mechanical driving, complements biochemical models and presents an advanced perspective on allostery in general.

The most natural example of the implied hypersensitivity concerns the transduction channels in hair cells [345]. Each hair cell contains a bundle of $N \approx 50$ stereocilia which are mechanically stimulated by the vibrations in the inner ear. The stereocilia possess transduction channels closed by "gating springs" which can open (close) in response to a positive (negative) shear strain imposed on the cilia from outside.

The broadly accepted model of this phenomenon [119] views the hair bundle as a set of $N$ bistable springs arranged in parallel. It is identical to the HS model if the folded (unfolded) configurations of cross-bridges are identified with the closed (opened) states of the channels. The applied loading, which tilts the potential and biases in this way the distribution of closed and open configurations, is treated in this model as a hard device version of HS model. Experiments, involving a mechanical solicitation of the hair bundle through an effectively rigid glass fiber, showed that the stiffness of the hair bundle is negative around the physiological functioning point of the system [120], which is fully compatible with the predictions of the HS model.

A similar analogy can be drawn between the HS model and the models of collective unzipping for adhesive clusters [7; 12; 118; 341; 346]. At the micro-scale we again encounter $N$ elements representing, for instance, integrins or cadherins, that are attached in parallel to a common, relatively rigid pad. The two conformational states, which can be described by a single spin variable, are the bound and the unbound configurations.

The binding-unbinding phenomena in a mechanically biased system of the HS type are usually described by the Bell model [117], which is a soft device analog of the HS model with $\kappa_0 = \infty$. In this model the breaking of an adhesive bond represents an escape from a metastable state and the corresponding rates are computed by using Kramers' theory [341; 347] as in the HS model. In particular, the rebinding rate is often assumed to be constant [263; 348], which is also the assumption of HS for the reverse transition from the post- to the pre-power-stroke state. More recently, Bell's model was generalized through the inclusion of ligand tethers, bringing a finite value to $\kappa_0$ and using the master equation for the probability distribution of attached units [118; 348].

The main difference between the Bell-type models and the HS model is that the detached state cannot bear force while the unfolded conformation can. As a result, while the cooperative folding-unfolding (ferromagnetic) behavior in the HS model is possible in the soft device setting [99], similar cooperative binding-unbinding in the Bell model is impossible because the rebinding of a fully detached state has zero probability. To obtain cooperativity in models of adhesive clusters, one must use a mixed device, mimicking the elastic backbone and interpolating between soft and hard driving [118; 179; 197; 341].

Muscle tissues maintain stable architecture over long periods of time. However, it is also feasible that transitory muscle-type structures can be assembled to perform particular functions. An interesting example of such assembly is provided by the SNARE proteins responsible for the fast release of neurotransmitors from neurons to synaptic clefts. The fusion of synaptic vesicles with the presynaptic plasma membrane [349; 350] is achieved by mechanical zipping of the SNARE complexes which can in this way transform from opened to closed conformation [351].

To complete the analogy, we mention that individual SNAREs participating in the collective zipping are attached to an elastic membrane that can be mimicked by an elastic or even rigid backbone [352]. The presence of a backbone mediating long-range interactions allows the SNAREs to cooperate in fast and efficient closing of the gap between the vesicle and the membrane. The analogy with muscles is corroborated by the fact that synaptic fusion takes place at the same time scale as the fast force recovery (1 ms) [353].

Yet another class of phenomena that can be rationalized within the HS framework is the ubiquitous flip-flopping of macro-molecular hairpins subjected to mechanical loading [187; 188; 196; 199]. We recall that in a typical experiment of this type, a folded (zipped) macromolecule is attached through compliant links to micron-sized beads trapped in optical tweezers. As the distance between the laser beams is increased, the force applied to the molecule rises up to a point where the subdomains start to unfold. An individual unfolding event may correspond to the collective rupture of $N$ molecular bonds or an unzipping of a hairpin. The corresponding drops in the force accompanied by an abrupt increase in the total stretch can lead to an overall negative stiffness response [186; 199; 203]. Other molecular systems exhibiting cooperative unfolding include protein $\beta$-hairpins [354] and coiled coils [209]. The backbone dominated internal architecture in all these systems leads to common mean-field type mechanical feedback exploited by the parallel bundle model [355; 356].

Realistic examples of unfolding in macromolecules may involve complex "fracture" avalanches [357] that cannot be modeled by using the original HS model. However, the HS theoretical framework is general enough to accommodate hierarchical meta-structures whose stability can be also biased by mechanical loading. The importance of the topology of interconnections among the bonds and the link between the collective nature of the unfolding and the dominance of the HS-type parallel bonding have been long stressed in the studies of protein folding [358]. The broad applicability of the HS mechanical perspective on collective conformational changes is also corroborated by the fact that proteins and nucleic acids exhibit negative stiffness and behave differently in soft and hard devices [209; 359; 360].

The ensemble dependence in these systems suggests that additional structural information can be obtained if the unfolding experiments are performed in the mixed device setting. The type of loading may be affected through the variable rigidity of the "handles" [361; 362] or the use of an appropriate feedback control that can be modeled in the HS framework by a variable backbone elasticity.

As we have already mentioned, collective conformational changes in distributed biological systems containing coupled bistable units can be driven not only mechanically, by applying forces or displacements, but also biochemically by, say, varying concentrations or chemical potentials of ligand molecules in the environment [363]. Such systems can become ultrasensitive to external stimulations as a result of the interaction between individual units undergoing conformational transformation which gives rise to the phenomenon of conformational spread [344; 364].

The switch-like input-output relations are required in a variety of biological applications because they ensure both robustness in the presence of external perturbations and ability to quickly adjust the configuration in response to selected stimuli [343; 365]. The mastery of control of biological machinery through mechanically induced conformational spread is an important step in designing efficient biomimetic nanomachines [195; 366; 367]. Since interconnected devices of this type can be arranged in complex modular metastructures endowed with potentially programmable mechanical properties, they are of particular interest for micro-enginnering of energy harvesting devices [13].

To link this behavior to the HS model, we note that the amplified dose response, characteristic of allostery, is analogous to the sigmoidal stress response of the paramagnetic HS system where an applied displacement plays the role of the controlled input of a ligand. Usually, in allosteric protein systems, the ultrasensitive behavior is achieved as a result of nonlocal interactions favoring all-or-none types of responses; moreover, the required long-range coupling is provided by mechanical forces acting inside membranes and molecular complexes. In the HS model such coupling is modeled by the parallel arrangement of elements, which preserves the general idea of nonlocality. Despite its simplicity, the appropriately generalized HS model [99] captures the main patterns of behavior exhibited by apparently purely chemical systems, including the possibility of a critical point mentioned in Ref. [363].

## 7. Conclusions

In contrast to inert matter, mechanical systems of biological origin are characterized by structurally complex network architecture with domineering long-range interactions. This leads to highly unusual mechanical properties in both statics and dynamics. In this review we identified a particularly simple system of this type, mimicking a muscle half-sarcomere, and systematically studied its peculiar mechanics, thermodynamics and kinetics.

In the study of passive force generation phenomena our starting point was the classical model of Huxley and Simmons (HS). The original prediction of the possibility of negative stiffness in this model remained largely unnoticed. For 30 years the HS model was studied exclusively in the hard device setting which concealed the important role of cooperative effects. A simple generalization of the HS model for the mixed device reveals many new effects, in particular the ubiquitous presence of coherent fluctuations.

Among other macroscopic effects exhibited by the generalized HS model are the non-equivalence of the response in soft and hard devices and the possibility of negative susceptibilities. These characteristics are in fact typical for nonlinear elastic materials in 3D at zero temperature. Thus, the relaxed energy of a solid material must be only quasi-convex which allows for non-monotone stress strain relations and different responses in soft and hard devices [368]. Behind this behavior is the long range nature of elastic interactions which muscle tissues appear to be emulating in 1D.

For a long time it was also not noticed that the original parameter fit by HS placed skeletal muscles almost exactly in the critical point. Such criticality is tightly linked to the fact that the number of cross-bridges in a single half sarcomere is of the order of 100. This number is now shown to be crucial to ensure mechanical ultra sensitivity that is not washed out by finite temperature and it appears quite natural that muscle machinery is evolutionaty tuned to perform close to a critical point. This assumption is corroborated by the observation that criticality is ubiquitous in biology from the functioning of auditory system [120] to the macroscopic control of upright standing [369; 370].

The mechanism of fine tuning to criticality can be understood if we view the muscle fiber as a device that can actively modify its rigidity. To this end the system should be able to generate a family of stall states parameterized by the value of the meso-scopic strain. A prototypical model reviewed in this paper shows that by controlling the degree of non-equilibrium in the system, one can indeed stabilize apparently unstable or marginally stable mechanical configurations, and in this way modify the structure of the effective energy landscape (when it can be defined). The associated pseudo-energy wells with resonant nature may be crucially involved in securing robustness of the near critical behavior of the muscle system. Needless to say that the mastery of tunable rigidity in artificial conditions can open interesting prospects not only in biomechanics [371] but also in engineering design incorporating negative stiffness [372] or aiming at synthetic materials involving dynamic stabilization [373; 374].

In addition to the stabilization of passive force generation, we also discussed different modalities of how a power-stroke-driven machinery can support active muscle contraction. We have shown that the use of a hysteretic design for the power-stroke motor allows one to reproduce mechanistically the complete Lymn–Taylor cycle. This opens a way towards dynamic identification of the chemical states, known from the studies of the prototypical catalytic reaction in solution, with particular transient mechanical configurations of the acto-myosin complex.

At the end of this review we briefly addressed the issue of ruggedness of the global energy landscape of a tetanized muscle myofibril. The domain of metastability on the force-length plane was shown to be represented by a dense set of elastic responses parameterized by the degree of cross-bridge connectivity to actin filaments. This observation suggests that the negative overall slope of the force-length relation may be a combination of a large number of micro-steps with positive slopes.

In this review we focused almost exclusively on the results obtained in our group and mentioned only peripherally some other related work. For instance, we did not discuss a vast body of related experimental results, e.g. Refs. [116; 166; 375; 376]. Among the important theoretical work that we left outside, are the results on active collective dynamics of motors [12; 377–379]. Interesting attempts of building alternative models of muscle contraction [56; 380] and of creating artificial devices imitating muscle behavior [195] were also excluded from the scope of this paper. Other important omissions concern the intriguing mechanical behavior of smooth [381; 382] and cardiac [383–387] muscles.

Despite the significant progress in the understanding of the microscopic and mesoscopic aspects of muscle mechanics, achieved in the last years, many fundamental problems remain open. Thus, the peculiar temperature dependence of the fast force recovery [207; 388] has not been systematically studied, despite some recent advances [121; 180]. A similarly important challenge presents the delicate asymmetry between shortening and stretching, which may require the account of the second Myosin head [94]. Left outside most of the studies is the short-range coupling between cross-bridges due to filaments extensibility [76], the inhomogeneity of the relative displacement between myosin and actin filaments, and more generally the possibility of a non-affine displacements in the system of interacting cross bridges. Other under-investigated issues include the mechanical role of additional conformational states [74] and the functionality of parallel elastic elements [389].

We anticipate that more efforts will be also focused on the study of contractional instabilities and actively generated internal motions [148] that should lead to the understanding of the self-tuning mechanism bringing sarcomeric systems towards criticality [99; 390; 391]. Criticality implies that fluctuations become macroscopic, which is consistent with observations at stall force conditions. The proximity to the critical point allows the system to amplify interactions, ensure strong feedback, and achieve considerable robustness in front of random perturbations. In particular, it is a way to quickly and robustly switch back and forth between highly efficient synchronized stroke and stiff behavior in the desynchronized state [99].

## Acknowledgments

We thank J.-M. Allain, L. Marcucci, I. Novak, R. Sheska and P. Recho for collaboration in the projects reviewed in this article. We are also grateful to V. Lombardi group, D. Chapelle, P. Moireau, T. Lelièvre and P. Martin for numerous inspiring discussions. The PhD work of M.C. was supported by the Monge Fellowship from Ecole Polytechnique. L. T. was supported by the French Governement under the Grant No. ANR-10-IDEX-0001-02 PSL.